\documentstyle[12pt,epsfig]{article}

\renewcommand{\theequation}{\arabic{section}.\arabic{equation}}

\def\hybrid{\topmargin -20pt    \oddsidemargin 0pt
        \headheight 0pt \headsep 0pt
        \textwidth 6.25in       
        \textheight 9.5in       
        \marginparwidth .875in
        \parskip 5pt plus 1pt   \jot = 1.5ex}
\hybrid
\def\be{\begin{equation}}       \def\eq{\begin{equation}}
\def\ee{\end{equation}}         \def\eqe{\end{equation}}

\def\bea{\begin{eqnarray}}      \def\eqa{\begin{eqnarray}}
\def\ena{\end{eqnarray}}        \def\eea{\end{eqnarray}}
                                \def\eqae{\end{eqnarray}}

\def\ba{\begin{array}}
\def\ea{\end{array}}
\def\unit{1 \hskip-.3em \raise2pt\hbox{$ \scriptstyle |$ } }

\def\a{\alpha}
\def\b{\beta}
\def\c{\chi} 
\def\d{\delta}
\def\e{\epsilon}           
\def\f{\phi}               
\def\g{\gamma}

\def\j{\psi}
\def\k{\kappa}                    
\def\l{\lambda}
\def\m{\mu}
\def\n{\nu}
\def\o{\omega}  
\def\q{\theta}                    
\def\r{\rho}                                     
\def\s{\sigma}                                   
\def\t{\tau}

\def\D{\Delta}
\def\F{\Phi}
\def\G{\Gamma}

\def\L{\Lambda}
\def\O{\Omega}  
\def\P{\Pi}

\def\S{\Sigma}

\def\del{\partial}

\def\cm{{\cal M}}

\def\half{{1 \over 2}}

\def\bop#1{\setbox0=\hbox{$#1M$}\mkern1.5mu
	\vbox{\hrule height0pt depth.04\ht0
	\hbox{\vrule width.04\ht0 height.9\ht0 \kern.9\ht0
	\vrule width.04\ht0}\hrule height.04\ht0}\mkern1.5mu}
\def\pa{\partial}                              
\def\ddg{\sp\ddagger} 

\def\>{\rangle} 

\def\<{\langle} 
\def\Dsl{D \hskip-.6em \raise1pt\hbox{$ / $ } }

\def\leftrightarrowfill{$\mathsurround=0pt \mathord\leftarrow \mkern-6mu
       \cleaders\hbox{$\mkern-2mu \mathord- \mkern-2mu$}\hfill
       \mkern-6mu \mathord\rightarrow$}
\def\dvec#1{\vbox{\ialign{##\crcr
       \leftrightarrowfill\crcr\noalign{\kern-1pt\nointerlineskip}
       $\hfil\displaystyle{#1}\hfil$\crcr}}}          
\def\hook#1{{\vrule height#1pt width0.4pt depth0pt}}
\def\leftrighthookfill#1{$\mathsurround=0pt \mathord\hook#1
       \hrulefill\mathord\hook#1$}
\def\underhook#1{\vtop{\ialign{##\crcr                 
       $\hfil\displaystyle{#1}\hfil$\crcr
       \noalign{\kern-1pt\nointerlineskip\vskip2pt}
       \leftrighthookfill5\crcr}}}
\def\smallunderhook#1{\vtop{\ialign{##\crcr      
       $\hfil\scriptstyle{#1}\hfil$\crcr
       \noalign{\kern-1pt\nointerlineskip\vskip2pt}
       \leftrighthookfill3\crcr}}}

\def\sfrac#1#2{{\vphantom1\smash{\lower.5ex\hbox{\small$#1$}}\over
       \vphantom1\smash{\raise.4ex\hbox{\small$#2$}}}} 
\def\bfrac#1#2{{\vphantom1\smash{\lower.5ex\hbox{$#1$}}\over
       \vphantom1\smash{\raise.3ex\hbox{$#2$}}}}      
\def\afrac#1#2{{\vphantom1\smash{\lower.5ex\hbox{$#1$}}\over#2}}  
\def\on#1#2{{\buildrel{\mkern2.5mu#1\mkern-2.5mu}\over{#2}}}
\def\ddt#1{\on{\hbox{\LARGE .\kern-2pt.}}#1}             
\def\tdt#1{\on{\hbox{\LARGE .\kern-2pt.\kern-2pt.}}#1}   

\def\boxes#1{
       \newcount\num
       \num=1
       \newdimen\downsy
       \downsy=-1.5ex
       \mskip-2.8mu
       \bo
       \loop
       \ifnum\num<#1
       \llap{\raise\num\downsy\hbox{$\bo$}}
       \advance\num by1
       \repeat}
\def\boxup#1#2{\newcount\numup
       \numup=#1
       \advance\numup by-1
       \newdimen\upsy
       \upsy=.75ex
       \mskip2.8mu
       \raise\numup\upsy\hbox{$#2$}}

\newskip\humongous \humongous=0pt plus 1000pt minus 1000pt

\newif\ifdtup

\def\to{\rightarrow}

\def\eff{e^{\phi-\bar{\phi}}}
\def\effb{e^{\bar{\phi}-\phi}}

\def\dda{\dot{\alpha}} 
\def\ddb{\dot{\beta}}

\def\ddd{\dot{\delta}}
\def\dde{\dot{\epsilon}}

\def\ddg{\dot{\gamma}}

\def\ddm{\dot{\m}}
\def\ddn{\dot{\n}}

\def\ddt{\dot{\t}}

\def\ta{\tilde{\a}} 
\def\tta{\tilde{\a}} 
\def\tb{\tilde{\b}}
\def\ttb{\tilde{\b}}

\def\td{\tilde{\d}}
\def\te{\tilde{\e}}

\def\tg{\tilde{\g}}
\def\ttg{\tilde{\g}}

\def\ome{\omega}

\def\pa{\partial}
\def\del{\nabla}
\def\delbar{\bar{\nabla}}

\def\bbk{\bar{\kappa}}
\def\bbe{\bar{\epsilon}}
\def\kb{\bar{k}}
\def\Kb{\bar{K}}
\def\zb{\bar{z}}
\def\wb{\bar{w}}
\def\pb{\bar{p}}
\def\fb{\bar{\f}}
\def\Fb{\bar{\F}}
\def\Mb{\bar{M}}
\def\Lb{\bar{L}}
\def\Lab{\bar{\L}}
\def\Kb{\bar{K}}
\def\jb{\bar{\j}}
\def\qb{\bar{\q}}
\def\Sib{\bar{\S}}
\def\Pb{\bar{\P}}

\def\hL{\hat{L}} 
\def\hM{\hat{M}}
\def\hj{\hat{\j}} 
\def\hf{\hat{\f}}
\def\hMb{\bar{\hat{M}}}
\def\hfb{\bar{\hat{\f}}}

\def\dif{\partial}
\def\difb{\bar{\partial}}
\def\dbar{\bar{\partial}}
\def\nonu{\nonumber \\{}}
\def\half{{1 \over 2}}
\def\trr{{\rm Tr}}

\begin{document}

\thispagestyle{empty}
\begin{flushright}
{\sc ITP-SB}-96-40
\end{flushright}
\vspace{1cm}
\setcounter{footnote}{0}
\begin{center}
{\LARGE{Covariant Computation of the Low Energy Effective Action of
the Heterotic Superstring}}\\[14mm]

\sc{Jan de Boer\footnote{e-mail: deboer@insti.physics.sunysb.edu, address
after September 1st: Department of Physics, University of California at
Berkeley, 366 LeConte Hall, Berkeley, CA 94720-7300}
    and Kostas Skenderis\footnote{e-mail: kostas@insti.physics.sunysb.edu,
address after September 1st: Instituut voor Theoretische Fysica, 
Katholieke Universiteit Leuven, Celestijnenlaan 200D, B-3001 Leuven, Belgium},
    }\\[5mm]
{\it Institute for Theoretical Physics\\
State University of New York at Stony Brook\\
Stony Brook, NY 11794-3840, USA}\\[10mm]
{\sc Abstract}\\[2mm]
\end{center}
We derive the low energy effective action of the heterotic
superstring in superspace. This is achieved by coupling 
the covariantly quantized Green-Schwarz superstring of Berkovits
to a curved background and requiring that the sigma model has
superconformal invariance at tree level and at
one loop in $\a'$. Tree level superconformal invariance
yields the complete supergravity algebra, and one-loop
superconformal invariance the equations of motion of the low
energy theory.  The resulting low energy theory is 
old-minimal supergravity coupled to a tensor multiplet.
The dilaton is part of the compensator multiplet.

\vfill

\section{Introduction}
\setcounter{equation}{0}

Low energy effective actions play an important role in the study of
string theory, and not only for phenomenological applications. They also
provide important pieces of evidence for the existence of various
dual descriptions of string theories. One can consider
low energy effective actions in their own right as specific examples
of locally supersymmetric field theories. The structure of these
low energy effective field theories is constrained not only by
the requirement of local supersymmetry, but also by the fact that they
represent low energy effective theories of some string theory. It is
clearly desirable to obtain as much understanding of these constraints
as possible. One of the most powerful ways to obtain such constraints
is by analyzing the two-dimensional sigma model that describes
the string theory coupled to the background fields of the low
energy effective field theory. Unfortunately, until recently no
sigma model description was known which was formulated directly
in terms of a target superspace and which had manifest local
target space supersymmetry (this problem is also known as 
`the covariant quantization of the superstring'). Without such a 
sigma model description, one can only work in components and not
determine the off-shell description that is selected
by string theory. In addition, it is in general quite difficult to
obtain the fermionic part of the effective action without a 
manifestly supersymmetric sigma model. In the RNS formalism
for example, the
fermionic terms can only be obtained by a calculation of
S-matrix scattering amplitudes.

It is important to know the precise off-shell 
description, as it provides constraints on the allowed types
of matter couplings in the low energy effective action. Furthermore,
it provides constraints on the structure of 
higher order $\alpha'$ corrections, and it provides a natural
framework to study the interplay between $T$-duality and
supersymmetry, supersymmetry breaking, and non-renormalization 
theorems. 

The sought for sigma model description of string theory was obtained
recently by Berkovits \cite{ber1,ber2,ber3}. In the case of
heterotic string theory compactified to four dimensions, it is
a sigma model with $N=(2,0)$ supersymmetry on the world-sheet.
The coupling of the theory to background fields was discussed in
\cite{bersie}, where also the form of the low-energy effective
action was proposed, based on indirect arguments. In this paper
we perform a detailed study of this sigma model. We show how
the tree-level analysis of the sigma model yields a complete
set of supergravity constraints, and how this determines, by
means of the Bianchi identities, the supergravity algebra. 
Next, we study the sigma model at one loop, and find the
equations of motion of the supergravity theory, all in superspace.
We determine the action that reproduces these field equations,
and thereby establish that the theory that describes the four
dimensional fields of the compactification of the heterotic
string is old-minimal supergravity 
coupled to a tensor multiplet. The dilaton is part of the
chiral and anti-chiral compensator fields \cite{wa2,wa3}, and not, as is
often claimed, part of the tensor multiplet. This affects,
in the presence of charged matter fields,
the structure of the effective action, especially at higher
string loops. 

The outline of this paper is as follows. In section two we review
the new covariant sigma model description for the heterotic string.
We first describe the sigma model in flat space, in which case it
is an exact conformal field theory. Next we couple the
sigma model to a curved background.
To do this we first rewrite the action in the flat background 
in a manifestly supersymmetric way, and then promote the flat vielbeins
and antisymmetric tensor fields into arbitrary unconstrained fields.
The sigma model is then by construction 
target space super-reparametrization invariant. 
Moreover, when expanded around
a flat background it contains all massless vertex operators of the heterotic
superstring. 
For simplicity we set both the compactification dependent 
and the Yang-Mills background fields equal to zero. 
Our method can easily be extended to 
include these fields as well. 
Next, we add a Fradkin-Tseytlin term to the sigma model to
accomodate the dilaton. The structure of the Fradkin-Tseytlin term,
proposed in \cite{bersie}, forces the dilaton to be part of a
chiral and anti-chiral target space superfield, and shows that it cannot
sit in the same multiplet as the antisymmetric tensor field.

The sigma-model, thus constructed, has a local target space 
Lorentz and $U(1)$ invariance. In section 3 we show how to
maintain this symmetry at the 
quantum level using the background field method. 
We develop a background field expansion that it is covariant with respect 
to both the Lorentz and the $U(1)$ symmetry. One of the interesting
features of the sigma model is that it has a chiral boson $\rho$ whose
kinetic term is $\a'$ independent, and which has to be treated exactly.
This can be achieved by means of a field redefinition, and
allows us to develop a hybrid perturbation theory
where we treat all fields perturbatively except the field $\r$ that 
we treat exactly. We use this perturbation theory to analyze the 
sigma model at tree level. Due to the presence of the chiral boson
$\rho$, a Poisson bracket analysis of the theory would not
be sufficient, and one has to analyze tree diagrams in perturbation
theory instead. From the tree diagrams we obtain a set of constraints,
which, when supplemented with a maximal set of conventional constraints,
provides a complete set of constraints for conformal supergravity
coupled to a tensor multiplet. This, as well as the complete
supergravity algebra, follows from the solution to the Bianchi
identities. 

In section 4, we check the superconformal invariance of the sigma
model at one loop, by
computing the OPE's of the $N=2$ generators in perturbation
theory. We evaluate all one-loop graphs in momentum space
using a version of dimensional regularization suitable for this
sigma model. We find that the non-holomorphic terms in the OPE's
yield the field equations of the low energy effective field theory.
The extra holomorphic terms that appear in the calculation but should
not be there can all be removed using suitable one-loop counterterms
and a modification of the background field expansion. The latter
is, unfortunately, rather ad hoc, and we have not yet understood
whether this has a geometrical interpretation or is an artifact
of the scheme we use. We conclude this section by a discussion
of sigma model anomalies, and show how they can be cancelled by a superspace
version of the Green-Schwarz mechanism. Also, we briefly indicate
which part of the calculation is reproduced by a standard beta-function
calculation. Such a calculation would be the most efficient way
to obtain some results at two loops.

Having obtained the field equations we then look in section 5
for a supergravity 
action that reproduces the same field equations. 
This is a rather tedious exersice in supergravity theory. 
Although it is straightforward to write down candidate actions, to find
out what their field equations are is not an easy task. 
This is because the supergravity fields are
constrained. Hence, when  varying the action to obtain the
field equations the variations should respect the constraints. 
To achieve this we express the variations in terms of unconstrained
prepotentials.  Our final result is that the low energy effective 
theory of the heterotic superstring is old-minimal supergravity coupled to a 
tensor multiplet.  

In section six we present our conclusions and some possible applications
of our results. The superspace conventions we used throughout the paper
are summarized in appendix A, and appendix B provides details of the
solution of the Bianchi identities presented in the main text.

\section{The covariant description of 4d compactifications of the heterotic
string}
\setcounter{equation}{0}

\subsection{Flat space}

In this section we review the construction and some of the properties
of the two-dimensional sigma model which is manifestly space-time
supersymmetric and describes 4d compactifications of the heterotic string.
This covariant description originated in \cite{ber1} and was further
developed in \cite{ber2}, \cite{ber3} and \cite{bersie}. For a review,
see \cite{bermore}.

An important distinction between the covariant formulation and 
the RNS or Green-Schwarz formulations of the heterotic string is that
in the covariant formalism the heterotic string arises as a critical
$(2,0)$ string theory rather than a $(1,0)$ one. 
In a flat 4d background, the 4d-part of the heterotic string is
in the $(2,0)$ superconformal gauge 
desribed by the following action\footnote{We use an Euclidean world-sheet,
and the measure $d^2 z$ denotes $dxdy/\pi$, with $z=x+iy$. The advantage
of this convention is that we never see explicit factor of $\pi$. In 
particular the following ditributional identity holds:
$\difb_{\bar{z}} (z-w)^{-1} = \d^{(2)}(z-w)$, where 
$\d^{(2)}(z-w)$ is the delta function with respect to the measure $d^2 z$.}
\be \label{act1}
S=\frac{1}{\alpha '} \int d^2 z (
\frac{1}{2} \dif x^m \difb x^m 
+ p_{\a} \difb \theta^{\a}
+ p_{\dda} \difb \bar{\theta}^{\dda}
-\frac{\alpha'}{2} \difb \rho (\dif \rho + a_z) )
\ee
where $z^M \equiv
(x^m,\theta^{\a},\bar{\theta}^{\dda})$ are the coordinates of
flat 4d superspace, the $p$'s are the conjugate fields of the $\q$'s
and $\rho$ is a boson with a `wrong' sign for the kinetic term.
We omitted here the $32-2r$ heterotic fermions and the $c=(9,6+r)$
internal
superconformal field theory coming from the Calabi-Yau sector, as
we will not consider background fields coming from these sectors
in this paper. However, our methods can be easily extended to include
these fields as well.
Notice the explicit factor of $\alpha'$ in front of $\rho$, from which
it follows that the $\rho$-sector of the theory is completely 
$\alpha'$-independent. One of the consequences this will have for us 
later is that in order to obtain results at a fixed order in $\alpha'$,
we need to treat the $\rho$-sector exactly rather than perturbatively.
The $U(1)$ gauge field component $a_z$ will be viewed as a Lagrange
multiplier imposing the constraint $\difb\rho=0$, so that $\rho$ becomes 
a chiral boson. If we couple the theory to $(2,0)$ world-sheet
supergravity, the theory will contain both components $a_z$ and $a_{\bar{z}}$
of a $U(1)$ gauge field. One might think that the gauge fixed theory should
therefore have both a left and right moving $U(1)$ symmetry, but this
is not quite true, since the two $U(1)$ gauge fields gauge only one
symmetry ($\rho\rightarrow \rho+{\rm constant}$), so that only one component
of the gauge field can be gauged away. The other remains as a Lagrange
multiplier (modulo global issues, which we ignore in the remainder).

From the action (\ref{act1}) one derives the following OPE's
\bea
\dif x_{\a\dda}(z) \dif x_{\b\ddb}(w) & = & 
\frac{- \alpha' C_{\a\b} C_{\dda \ddb} }{(z-w)^2} \nonu
p_{\a}(z) \theta^{\b}(w) & = & \frac{\alpha' \delta_{\a}{}^{\b} }{(z-w)}
\nonu
{p}_{\dda}(z) \bar{\theta}^{\ddb}(w) & = &
 \frac{\alpha' \delta_{\dda}{}^{\ddb} }{(z-w)}
\nonu
\dif \rho(z) \dif \rho(w) & = & \frac{1}{(z-w)^2}
\eea
It has a $(2,0)$ superconformal
symmetry on the world-sheet with generators
\bea \label{generators1}
T & = & \frac{1}{\alpha'} \left(
 -\frac{1}{2} \dif x^{\a\dda} \dif x_{\a\dda}  - p_{\a} \dif \theta^{\a}
- {p}_{\dda} \pa \bar{\theta}^{\dda} 
+ \frac{\alpha'}{2} \dif \rho \dif \rho
\right) \nonu
G & = &  \frac{1}{i \alpha' \sqrt{8\alpha'}} e^{i\rho} d^\a d_\a \nonu
\bar{G} & = &  \frac{1}{i \alpha' \sqrt{8\alpha'}} e^{-i\rho}
 d^{\dda} d_{\dda} \nonu
J & = & -i \dif \rho
\eea
where we defined\footnote{For simplicity, we will abbreviate $d^{\dda}
d_{\dda}$ and $\nabla^{\dda} \nabla_{\dda}$ by $\bar{d}^2$ and
$\bar{\nabla}^2$.}
\bea
d_{\a} & = & p_{\a} + i \bar{\theta}^{\dda} \dif x_{\a\dda}
+\frac{1}{2} \bar{\theta}^2 \dif \theta_{\a} 
-\frac{1}{4} \theta_{\a} \dif (\bar{\theta}^2) \nonu
d_{\dda} & = & {p}_{\dda} + i \bar{\theta}^{\a} \dif x_{\a\dda}
+\frac{1}{2} \theta^2 \dif \bar{\theta}_{\dda} 
-\frac{1}{4} \bar{\theta}_{\dda} \dif (\theta^2)
\eea
The generators in (\ref{generators1})
satisfy the usual $N=2$ superconformal algebra
\bea \label{n2algebra}
T(z) T(w) & = & \frac{c/2}{(z-w)^4} + 
\frac{2 T(w)}{(z-w)^2} + \frac{\dif T(w)}{(z-w)} \nonu
T(z) G(w) & = & \frac{\frac{3}{2} G(w) }{(z-w)^2} + 
\frac{\dif G (w)}{(z-w)} \nonu
T(z) \bar{G}(w) & = & \frac{\frac{3}{2} \bar{G}(w) }{(z-w)^2} + 
\frac{\dif \bar{G} (w)}{(z-w)} \nonu
J(z) G(w) & = & \frac{G(w)}{(z-w)} \nonu
J(z) \bar{G}(w) & = & \frac{-\bar{G}(w)}{(z-w)} \nonu
J(z) J(w) & = & \frac{c/3}{(z-w)^2} \nonu
G(z) \bar{G}(w) & = & \frac{2c/3}{(z-w)^3} + \frac{2J(w)}{(z-w)^2} + 
 \frac{2 T(w) + \dif J(w)}{(z-w)}
\eea
with $c=-3$. This central charge is exactly what we need, since
combined with the $c=+9$ $N=2$ algebra in the Calabi-Yau sector
the total $N=2$ algebra has central charge $c=+6$, which is 
precisely the central charge needed to obtain 
a critical $N=2$ string ($6=26-11-11+2$). In contrast to the RNS
string where only an $N=1$ subalgebra of the $N=2$ algebra of
the internal Calabi-Yau sector appears as a local symmetry algebra,
here we need the full $N=2$ algebra.

In verifying (\ref{n2algebra}) one has to be a bit careful. For example,
to verify that $G$ has a regular OPE with itself, one needs to check
that the normal ordered product of $(d)^2$ with itself vanishes.
If this would be nonzero, it would (due to the factor $e^{i\rho}$ in $G$)
appear as a first order pole in the OPE of $G$ with $G$. The easiest
way to verify (\ref{n2algebra}) is to use the following
intermediate results
\bea
d_{\a}(z) d_{\b}(w) & = & {\rm regular} \nonu
d_{\dda}(z) d_{\ddb}(w) & = & {\rm regular} \nonu
d_{\a}(z) d_{\ddb}(w) & = & \frac{2i \alpha'
 \Pi_{\a\ddb} }{(z-w)} \nonu
d_{\a}(z) \Pi_{\b\ddb}(w) & = & 
\frac{-2 \alpha' i C_{\a\b} \dif \bar{\theta}_{\ddb}}{(z-w)} \nonu
d_{\dda}(z) \Pi_{\b\ddb}(w) & = & 
\frac{-2 \alpha' i C_{\dda\ddb} \dif \theta_{\b}}{(z-w)} \nonu
d^2(z) d_{\dda}(w) & = & -\frac{8 (\alpha ')^2  \dif 
\bar{\theta}_{\dda}(w)}{(z-w)^2} + 
\frac{4\alpha ' i d^{\a} \Pi_{\a\dda}(w)}{(z-w)} \nonu
\bar{d}^2(z) {d}_{\a}(w) & = & -\frac{8 (\alpha ')^2  \dif 
\theta_{\a}(w)}{(z-w)^2} + 
\frac{4\alpha ' i d^{\dda} \Pi_{\a\dda}(w)}{(z-w)} 
\eea
where $d^2=d^\a d_\a$, $\bar{d}^2=d^{\dda} d_{\dda}$ and we defined
\be \label{eq01}
\Pi_{\a\dda} = \dif x_{\a\dda} - i \theta_{\a} \dif \bar{\theta}_{\ddb}
+ i \dif \theta_{\a} \bar{\theta}_{\ddb}
\ee
in terms of which $T$ becomes
\be
  \frac{1}{\alpha'} \left(
 -\frac{1}{2} \Pi^{\a\dda} \Pi_{\a\dda}  - d_{\a} \dif \theta^{\a}
 - d_{\dda} \dif \bar{\theta}^{\dda} 
  + \frac{\alpha'}{2} \dif \rho \dif \rho
\right) 
\ee
The advantage of the variables $d$ and $\Pi$ over $p$ and $x$ is that
they commute with the target space supersymmetry generators
\bea
q_{\a} & = & \oint \frac{dz}{2\pi i} \left(
p_{\a} - i \bar{\theta}^{\dda} \dif x_{\a\dda}
+\frac{1}{4} \theta_{\a} \dif (\bar{\theta}^2)
\right) \nonu
{q}_{\dda} & = & 
\oint \frac{dz}{2\pi i} \left(
\bar{p}_{\dda} - i \theta^{\a} \dif x_{\a\dda}
+\frac{1}{4} \bar{\theta}_{\dda} \dif (\theta^2)
\right)
\eea
In general, we can define variables $\Pi^A, \bar{\Pi}^A$ using
vielbeins $E_M{}^A$ to convert curved into flat indices
\be  \label{def:vielbein}
\Pi^A = \dif z^M E_M{}^A, \qquad \bar{\Pi}^A = \difb z^M E_M{}^A.
\ee
$\Pi^{\a\dda}$ reduces to (\ref{eq01}) when $E_M{}^A$ is
the vielbein of flat superspace, which is one on the diagonal and
has off-diagonal components
\be \label{eq02}
E_{\ddm}{}^{\a\dda} = i \delta_{\ddm}{}^{\dda} \theta^{\a},
\qquad E_{\m}{}^{\a\dda} = i \delta_{\m}{}^{\a} \bar{\theta}^{\dda} .
\ee
The action also becomes manifestly target space
supersymmetric when written in terms of $\Pi^A,\bar{\Pi}^A$,
\be \label{act2}
S=\frac{1}{\alpha '} \int d^2 z (
\frac{1}{2} \Pi^{\a\dda} \bar{\Pi}_{\a\dda}
+ d_{\a} \bar{\Pi}^{\a}
+ d_{\dda} \bar{\Pi}^{\dda}
+\frac{1}{2} \bar{\Pi}^A \Pi^B B_{BA}
-\frac{\alpha'}{2} \difb \rho (\dif \rho + a_z))
\ee
where we introduced an anti-symmetric tensor field $B_{BA}$ whose
only non-zero components are
\be \label{eq03}
\begin{array}{cc} B_{\a\dda,\ddb} = i C_{\ddb\dda} \theta_{\a} &
                  B_{\a\dda,\b} = i C_{\b\a} \bar{\theta}_{\dda} \\
                  B_{\ddb,\a\dda} = -i C_{\ddb\dda} \theta_{\a} &
                  B_{\b,\a\dda} = -i C_{\b\a} \bar{\theta}_{\dda} 
\end{array}
\ee

Covariant derivatives are in flat superspace given by
$\nabla_A=E_A{}^M \dif_M$, where the inverse vielbein $E_A{}^M$ has
one on the diagonal and off-diagonal components
\be E_{\dda}{}^{\m\ddm} = -i \delta_{\dda}{}^{\ddm} \theta^{\m},
\qquad E_{\a}{}^{\m\ddm} = -i \delta_{\a}{}^{\m} \bar{\theta}^{\ddm},
\ee
as follows from (\ref{eq02}). From this we deduce that the 
non-zero torsion of flat superspace is in our conventions given by
\be T_{\a,\dda}{}^{\b\ddb}=-2i \delta_{\a}{}^{\b} \delta_{\dda}{}^{\ddb}.
\label{flattor}
\ee
Associated to the anti-symmetric tensor field in (\ref{eq03}) is
a non-zero field strength $H_{ABC}$. With curved indices it
is given by $H_{MNP}=\frac{1}{4} \partial_{[M} B_{NP)}$ which
in terms of flat indices becomes
\be H_{ABC} = \frac{1}{4}  \nabla_{[A} B_{BC)} -\frac{1}{4}
T_{[AB}{}^D B_{|D|C)}.
\ee
The only non-zero components of $H$ are
\be
H_{\a,\dda,\b\ddb}=-i C_{\a\b} C_{\dda\ddb}
\label{flath}
\ee
and its permutations.

\subsection{Relation with RNS and Green-Schwarz formalism}

For completeness, we briefly indicate the relation between this
formalism and the usual RNS and Green-Schwarz formalisms\cite{ber3}.

The usual light-cone variables of the four-dimensional
Green-Schwarz superstring appear after bosonizing a pair
$(p_{\a},\theta^{\a})$ for $\a=1$ or $\a=2$, or a pair
$({p}_{\dda},\bar{\theta}_{\dda})$, and by imposing the
constraints that follow from the $N=2$ algebra. The stress-energy
tensor $T$ can be used to gauge $\dif(x^0+x^3)$ to $1$, and
the constraint $T=0$ can be used to eliminate $x^0-x^3$. The $U(1)$
current $J$ can be used to gauge $\rho=i\sigma$, where $\sigma$ is
the boson used to bosonize a pair $(p,\theta)$ via $p=e^{-i\s},\theta=
e^{i\s}$, and subsequently $J=0$ is used to take $\dif\rho=0$. Next,
$G$ and $\bar{G}$ can be used to gauge a $\theta$ and a $\bar{\theta}$
equal to zero, and $G=\bar{G}=0$ can be used to eliminate the
two corresponding $p$'s, so that all that remains is $x^1,x^2$ and
one pair $(p,\theta)$, which are the light-cone Green-Schwarz variables.

The relation with the RNS formalism starts by embedding the $N=1$
RNS string in a critical $N=2$ string following \cite{berva}.
The cohomologies are the same in the two cases, which can easily be seen
from the fact that the two BRST operators are related by a similarity
transformation \cite{ohpe}. Subsequently, one needs to perform a 
field redefinition and a further unitary transformation to arrive
at the $N=2$ string described in the previous paragraph. A nice
check on this procedure is that only those fields in the RNS string
that survive the GSO projection can be transformed into a single
valued field in the $N=2$ string, which is what one would expect in
view of the manifest target space supersymmetry of the latter.

\subsection{The $\sigma$-model in a curved background}
\label{subsect13}

Our next task is to formulate the sigma model in (\ref{act1}) in
a curved background \cite{bersie}. The main requirements this formalism should
be subject to are: (i) it should be manifestly world-sheet covariant,
(ii) it should be target space supersymmetric, (iii) when expanded
to first order around a flat background one should recover the
massless vertex operators of a flat background and (iv) the complete
set of massless physical states of the heterotic superstring should be
present.

The massless vertex operators were discussed in \cite{ber3} and can be
obtained, for example, by transforming the
vertex operators in the RNS formalism to ones for the $N=2$ string
discussed here, as explained in the previous section. They have the
form 
\be \label{mvec}
V=\int d^2 z \{ \bar{G}, [ G, V_{\a\dda} ] \} \bar{\Pi}^{\a\dda}
\ee
where $G=\oint \frac{dz}{2\pi i} G(z) = G_{-1/2}$ and similarly
$\bar{G}$ are the $N=2$ world-sheet supersymmetry generators,
and they have precisely the form as one would expect for a theory
with $N=(2,0)$ world-sheet supersymmetry. Furthermore, from the
requirement that the vertex operator produces a state in the BRST cohomology 
it follows that $V$ must be an $N=2$ primary of conformal weight
zero and $U(1)$ charge zero, i.e. $V$ should have only single
poles in the OPE with $T,G,\bar{G}$ and have a regular OPE with $J$.
To analyze this condition, consider the following OPE's of $T,G,\bar{G}$
with an arbitrary function $M$ of $x_{\a\dda}$ and 
$\theta^{\a},\bar{\theta}^{\dda}$
\bea \label{mopes}
d_{\a}(z) M(w) & = & \frac{\alpha' \nabla_{\a} M(w) }{(z-w)} \nonu
d_{\dda}(z) M(w) & = & \frac{\alpha' 
  \nabla_{\dda} M(w) }{(z-w)} \nonu
T(z) M(w) & = & 
\frac{-\frac{1}{2} \alpha' \nabla^{\a\dda} \nabla_{\a\dda} M(w)}{(z-w)^2}
 + \frac{\dif M(w)}{(z-w)} \nonu
G(z) M(w) & = & \frac{1}{i\sqrt{8\alpha '} }
\left(
\frac{\alpha' e^{i\rho} \nabla^{\gamma} \nabla_{\gamma} M(w) }{(z-w)^2} 
+ \frac{2 e^{i \rho} d^{\gamma} \nabla_{\gamma} M(w) }{(z-w)} 
\right) \nonu
\bar{G}(z) M(w) & = & \frac{1}{i\sqrt{8\alpha '} }
\left(
\frac{\alpha' e^{-i\rho} \nabla^{\dot{\gamma}}
 \nabla_{\dot{\gamma}} M(w) }{(z-w)^2} 
+ \frac{2 e^{-i \rho} d^{\dot{\gamma}} 
\nabla_{\dot{\gamma}} M(w) }{(z-w)} 
\right) \nonu
\eea
where we as an intermediate step also included the OPE of the $d$'s with $M$.
From these equations we read off that the conditions for $V$ in 
(\ref{mvec}) to be a $N=2$ primary of conformal weight zero are
\be  \label{cond1}
\nabla^{\a} \nabla_{\a} V_{\b\ddb} = 
\nabla^{\dda} \nabla_{\dda} V_{\b\ddb} = 
\nabla^{\a\dda} \nabla_{\a\dda} V_{\b\ddb} = 0
\ee
In addition, $V$ should be a primary field of weight one with respect
to the anti-holomorphic Virasoro algebra, yielding the additional
condition
\be \label{cond2}
\dif^{\b\ddb} V_{\b\ddb} = 0
\ee
These equations do not yet completely classify the inequivalent vertex
operators of the theory, because it is possible to perform certain 
gauge transformations that do not change the form of $V$. The integrated
vertex operator $V$ is invariant under
\be \label{ginv1}
\delta (V_{\a\dda} \bar{\Pi}^{\a\dda} ) = [\bar{L}_{-1},Z]   
\ee
and under
\be \label{ginv2}
\delta V_{\a\dda}  = 
\{G, e^{-i\rho} X_{\a\dda} \} + \{\bar{G}, e^{i\rho} \bar{X}_{\a\dda} \}
\ee
from which we find, up to some constants,
\be \label{ginv3}
\delta V_{\a\dda} = \nabla_{\a\dda} Z + \nabla^{\b} \nabla_{\b} X_{\a\dda}
+ \nabla^{\ddb} \nabla_{\ddb} \bar{X}_{\a\dda}.
\ee
One may verify that $V_{\a\dda}$, subject to (\ref{cond1}), (\ref{cond2})
and with the invariance (\ref{ginv3}) describe the on-shell content of
supergravity coupled to a tensor multiplet, 
with prepotential $V_{\a\dda}$ \cite{bersie}. To do this verification,
one chooses a Wess-Zumino gauge for $V_{\a\dda}$, after which one
can count the off-shell degrees of freedom contained in $V_{\a\dda}$.
There are 12 bosonic and 12 fermionic off-shell components, and by examining
their spins one finds that they indeed describe 
conformal supergravity coupled to
a tensor multiplet.

The structure of the linearized field equations and gauge fixing conditions
(\ref{cond1}), (\ref{cond2}), and the gauge invariances in (\ref{ginv3})
were all dictated by the particular form of the two-dimensional theory
we started with, and by the fact that the theory has an $N=2$ superconformal
symmetry. Different manifestly target space supersymmetric 
two-dimensional sigma models describing the
heterotic string might yield different off-shell descriptions 
of supergravity, but since the sigma model we discuss here is the only
known covariant description of the heterotic string, this is also the first
time we do actually see string theory selecting a particular off-shell
description.

Our next task is to write down the sigma model in an arbitrary curved
background. To do this, we should couple the sigma
model in flat space to the most general possible set of background
fields. In addition, we want to compare this with the vertex operators
(\ref{mvec}). To see what type of coupling to background fields the latter
predicts, we use the fact that for an arbitrary function $M$ of
$x,\theta,\bar{\theta}$ the following identity holds
\bea
[\bar{G},[G,M\}\} & = & 2 \Pi^{\a} \nabla_{\a} M -
\frac{1}{4} d^{\a}(\nabla^{\ddb} \nabla_{\ddb} 
\nabla_{\a} M) \nonu & & 
+i \Pi^{\a\dda} \nabla_{\dda} \nabla_{\a} M
-i d^{\dda} (\nabla^{\a} \nabla_{\a\dda} M) \nonu
& & -\a ' \dif \rho \nabla^{\a\dda} \nabla_{\dda}
\nabla_{\a} M
\eea
which can be rewritten as
\bea \label{vop}
[\bar{G},[G,M\}\} & = & (\Pi^{\a} \nabla_{\a} M - {\Pi}^{\dda}
\nabla_{\dda} M) + \dif M - \frac{1}{2i} \Pi^{\a\dda}
[\nabla_{\dda},\nabla_{\a} ] M \nonu
& & -\frac{1}{4} ( d^{\a} (\bar{\nabla}^2 \nabla_{\a} M)
- d^{\dda} (\nabla^2 \nabla_{\dda}  M ) ) 
-\frac{1}{4} d^{\dda} (\nabla_{\dda} \nabla^2) M
\nonu & & 
+\frac{i\alpha '}{8} (\dif \rho)(
\bar{\nabla}^2 \nabla^2 - \nabla^2 \bar{\nabla}^2 +
8  \nabla^{\a\dda} \nabla_{\a\dda} ) M,
\eea
where $\del^2 = \del^\a \del_\a$ and $\bar{\del}^2 = \del^{\dda} \del_{\dda}$.
Based on these observations, one is then led to propose\cite{bersie} 
for the sigma model in curved space exactly the form as given in 
(\ref{act2})
\be  
S=\frac{1}{\alpha '} \int d^2 z (
\frac{1}{2} \Pi^{\a\dda} \bar{\Pi}_{\a\dda}
+ d_{\a} \Pb^{\a}
+ d_{\dda} \bar{\Pi}^{\dda}
+\frac{1}{2} \bar{\Pi}^A \Pi^B B_{BA}
-\frac{\alpha'}{2} \difb \rho (\dif \rho + a_z))
\ee
with the definitions in (\ref{def:vielbein}), but now with
arbtitrary vielbeins and anti-symmetric tensor fields. One might
imagine also introducing terms like $\dif \rho \bar{\Pi}^A N_A$ and 
ghost dependend terms in the sigma model, but such terms are not
present in the vertex operators (the $\dif\rho$ term in (\ref{vop})
drops out when $M$ satisfies (\ref{cond1})), and are probably forbidden
by world-sheet superconformal invariance (for example, 
$\dif\rho$ terms would break world sheet $U(1)$ invariance).
Therefore we will not consider such couplings here. The generators 
$J,G,\bar{G}$ of the $N=2$ algebra are the same for the sigma model in
curved space and in flat space (see (\ref{generators1})), but
the stress energy tensor is in curved space replaced by
\be
T  =  \frac{1}{\alpha'} \left(
 -\frac{1}{2} \Pi^{\a\dda} \Pi_{\a\dda}  - d_{\a} \Pi^{\a}
- {d}_{\dda} \Pi^{\dda} 
+ \frac{\alpha'}{2} \dif \rho \dif \rho
\right).
\ee

As usual, the sigma model action contains potentials $B_{BA}$
and $E_M{}^A$ rather than a prepotentials like $V_{\a\dda}$. In
particular, the sigma model is manifestly space-time supersymmetric.
Of course, the anti-symmetric tensor field and the vielbeins contain
many more degrees of freedom than the prepotential $V_{\a\dda}$.
We can at this stage already see what we expect will happen when we
do a perturbative analysis of (\ref{act2}). The single poles in
(\ref{mopes}) arise from tree-level contractions, whereas the double
poles arise from double contractions, corresponding to one-loop
diagrams. Since it is the double poles that give rise to the linearized
field equations for the physical components of the vertex operator,
we expect that in our sigma model the equations of motion will
only appear at one-loop, and that at tree level we will find a set
of constraints that reduces the field content from the vielbein and
anti-symmetric tensor field to the field strenghts of 
conformal supergravity
coupled to a tensor multiplet. (As the sigma model is manifestly
supersymmetric, we expect to find the field strengths rather than the
prepotentials of supergravity.) Later, we will see that this is
indeed what happens (in a rather subtle way),
but before we do a perturbative computation
in (\ref{act2}), we first have to include a Fradkin-Tseytlin term
in the action to take the dilaton into account. Including the dilaton,
the field content becomes that of Poincar\'e supergravity coupled
to a tensor multiplet.

\subsection{The Fradkin-Tseytlin term}

The Fradkin-Tesytlin term in the action will be a direct generalization
of the dilaton coupling $\int d^2 z \sqrt{g} R \Phi$ of
the bosonic string \cite{fratse}. The dilaton is not part of the anti-symmetric
tensor field $B_{AB}$, because it should not couple classically
in the action.
%
%
As we will see later from the tree-level constraints, 
the tensor multiplet is expressed completely in terms of the torsions,
and in particular 
the constraints break target space conformal invariance. 
Therefore the dilaton must be part of the remaining superfield, which
is the conformal compensator. In a more conventional formulation
one would use the conformal invariance to gauge fix the conformal
compensator rather than the tensor multiplet, but in the sigma model 
it is the other way around. Something similar happens in
the bosonic string: the dilaton there is the compensator for conformal
transformations, a fact which is reflected in the wrong sign of the
kinetic term of the dilaton. This is very particular for the `string
gauge' for target space conformal transformations
that the sigma model selects. By going from the string gauge
to another gauge by means of a field redefinition, it is possible
to obtain a kinetic term for a scalar degree of freedom
which does have the right sign. However, this scalar degree
of freedom should no longer be called the dilaton, the dilaton
being by definition the scalar field which counts the number
of string loops. Similar
statements apply to heterotic superstring and will be discussed
in more detail in section~5.

To write down the explicit form of the dilaton term, we need to know
a little about $N=(2,0)$ world-sheet supergravity \cite{brmu} and
$N=(2,0)$ superspace. The latter contains two extra anti-commuting
coordinates which we will call $\k$ and $\bbk$. The two (holomorphic)
superderivatives are $D=\dif_{\k}-\bbk \dif$ and $\bar{D}=\dif_{\bbk}-
\k \dif$, whereas the supersymmetry generators are
$Q=\dif_{\k}+\bbk \dif$ and $\bar{Q}=\dif_{\bbk}+
\k \dif$. They satisfy $\{D,\bar{D}\}=-2\dif$ and $\{Q,\bar{Q}\}=2\dif$. 
If we want to identify the action of $Q$ and $\bar{Q}$ with those
of $G\equiv\oint \frac{dz}{2\pi i} G$ and $\bar{G}$,
we find that a $(2,0)$ superfield is expressed in terms of its
lowest components as follows
\be
\Phi=\phi + \k[G,\phi\} + \bbk[\bar{G},\phi\} + 
\frac{1}{2} \k\bbk ([G,[\bar{G},\phi\}\} - [\bar{G},[G,\phi \}\}).
\ee
In particular, if $\phi$ is target space chiral
\be \label{chir}
\Phi=\phi + \k[G,\phi\} - \k \bbk \dif \phi
\ee
and, correspondingly, if $\phi$ is target space anti-chiral
\be \label{antichir}
\bar{\Phi}=\bar{\phi} + \bbk[\bar{G},\bar{\phi}\} +
 \k \bbk \dif \bar{\phi}.
\ee
With these conventions, we find that the $(2,0)$ world-sheet
super stress-energy tensor $\Theta$ is given by the component expansion
\be
\Theta = J - \k G + \bbk \bar{G} + 2 \k \bbk T .
\ee
On a curved $(2,0)$ world sheet, the derivatives $D$, $\bar{D}$ and
$\dif$ get replaced by covariant derivatives $\nabla_{\k}$, $\del_{\bbk}$,
and $\del_z$. In the case of a $(0,0)$ world-sheet, the curvature arises
in the commutator $[\del_z,\del_{\bar{z}}]\sim RM$, where $M$ is the
generator of Lorentz transformations. In the same way, the curvatures
of a $(2,0)$ world-sheet can be extracted from the commutators
$[\del_{\bbk},\del_{\bar{z}}]\sim \Sigma (M+iY)$ and 
$[\del_{\k},\del_{\bar{z}}]\sim \bar{\Sigma} (M-iY)$, where $Y$
is the generator of $U(1)$ transformations, with respect to which
$\k$ and $\bbk$ have weights $\pm \frac{1}{2}$, and $\Sigma$ and
$\bar{\Sigma}$ are respectively chiral and anti-chiral world-sheet
supercurvatures. The components of $\Sigma$ look like $\Sigma\sim
\chi+\k(R+iF) - \k \bbk \dif \chi$, where $R$ and $F$ are the world-sheet
curvature and $U(1)$ field strength, and $\chi$ is the gravitino
field strength \cite{brmu}. If now $\Phi$ and $\bar{\Phi}$
are the chiral and anti-chiral world-sheet 
superfields associated to target
space chiral and anti-chiral superfields $\phi,\bar{\phi}$ as in
(\ref{chir}) and (\ref{antichir}), we can write down the 
Fradkin-Tseytlin term \cite{bersie}
\be \label{FTterm}
\int d^2 z d\k \e^{-1} \Phi \Sigma +
\int d^2 z d\bbk \bbe^{-1} \bar{\Phi} \bar{\Sigma}
\ee
where $\e$ is the $(2,0)$ world-sheet chiral density. The fields $\Phi$
and $\bar{\Phi}$ have to be chiral and anti-chiral, otherwise 
(\ref{FTterm}) would break local $(2,0)$ world-sheet supersymmetry.

The precise form of the Fradkin-Tseytlin
term will not be very important to us in the remainder, since
we will always work in superconformal gauge. In superconformal gauge,
$\Sigma \sim i \k \dbar a_z$ and 
the only remnant of the Fradkin-Tseytlin term is an
additional term in the action
\be \label{extraterm}
\Delta S = - \frac{i}{2} \int d^2 z \dbar(\phi-\bar{\phi}) a_z.
\ee
In addition, the Fradkin-Tseytlin term modifies the generators
of the $N=(2,0)$ superconformal algebra. This modification can 
easily be determined from (\ref{FTterm}), and reads
\be \label{nn1}
\Theta \rightarrow \Theta + \dif(\Phi-\bar{\Phi}).
\ee
However, there is an additional modification of the stress-energy
tensor induced by the extra term (\ref{extraterm}), which tells
us that it is no longer $\rho$ but rather $\rho+i(\phi-\bar{\phi})$
which is the chiral boson. Together with the components of (\ref{nn1})
we then find for the generators in the presence of a dilaton
\bea 
J & \rightarrow & J + \dif \phi - \dif \bar{\phi} \nonu
G & \rightarrow & G - \frac{1}{\sqrt{2\a'}i} \dif 
\left( e^{i\rho} d^{\gamma} \del_{\gamma} \phi \right)  \nonu
\bar{G} & \rightarrow & \bar{G} - \frac{1}{\sqrt{2\a'}i} \dif 
\left( e^{-i\rho} d^{\ddg} \del_{\ddg} \bar{\phi} \right) \nonu
T & \rightarrow & 
 \frac{1}{\alpha'} \left(
 -\frac{1}{2} \Pi^{\a\dda} \Pi_{\a\dda}  - d_{\a} \Pi^{\a}
- {d}_{\dda} \Pi^{\dda}  \right. \nonu
& & \left. 
+ \frac{\alpha'}{2} \dif (\rho+i(\phi-\bar{\phi}))  
   \dif (\rho+i(\phi-\bar{\phi}))
\right) - \frac{1}{2} \dif^2 (\phi+\bar{\phi}).
\label{generators2}
\eea

Previously we required that the expansion around a flat background
of the sigma model in curved background should to first order reproduce
the massless vertex operators of the theory. This is true for all
massless vertex operators except the dilaton (for a discussion 
see e.g. \cite{gsw1}). There is a corresponding statement for
the dilaton coupling, the soft dilaton theorem \cite{sofdil1,sofdil2}, 
which states that
the insertion of a dilaton vertex operator in a correlation function
measures the world-sheet curvature at that point, exactly as one
would expect from the Fradkin-Tseytlin term (for a more recent 
discussion of the dilaton, see \cite{sofdil3}). Thus one could
attempt to further justify the form (\ref{FTterm}) by showing that
in this formulation of the heterotic string there are two dilaton-type
vertex operators, that respectively compute the world-sheet curvature and 
$U(1)$ curvature. 

Instead of (\ref{FTterm}), there may be other possibilities to 
couple the dilaton. For example, in \cite{banese} the dilaton 
is coupled to the ghost current rather than to the world-sheet
curvature. The difference between the two couplings is 
given by a term quadratic in the dilaton which through a field
redefinition be absorbed in the target space metric. In either
case the part linear in the dilaton is uniquely given by the
Fradkin-Tseytlin term. Dilaton couplings with a different Fradkin-Tseytlin
type term have been suggested in \cite{gates3}, but it is not
clear how to implement these in the present context.

In our case we will find it convenient to 
redefine the $U(1)$ gauge field $a_z$ by
\be
a_z \rightarrow a_z +i \dif(\phi-\bar{\phi})
\ee
which does not change the theory but introduces a quadratic
dilaton term in the action, which now reads
\bea 
S & = & \frac{1}{\alpha '} \int d^2 z (
\frac{1}{2} \Pi^{\a\dda} \bar{\Pi}_{\a\dda}
+ d_{\a} \Pb^{\a}
+ d_{\dda} \bar{\Pi}^{\dda}
+\frac{1}{2} \bar{\Pi}^A \Pi^B B_{BA}
\nonu & & \qquad
-\frac{\alpha'}{2} (\difb \rho+i \difb(\phi-\bar{\phi}))
 (\dif \rho +i\dif(\phi-\bar{\phi})+ a_z))
\label{finalact}
\eea

After this redefinition, the action is invariant under the following
local spacetime $U(1)$ transformations
\be
\delta \phi = -\frac{1}{2} \Lambda, \qquad 
\delta \bar{\phi} = \frac{1}{2} \Lambda,
\ee\be
\delta \Pi^{\a} = -\frac{1}{2} \Lambda \Pi^{\a}, \qquad
\delta \Pi^{\dda} = \frac{1}{2} \Lambda \Pi^{\dda}, 
\ee\be
\delta d_{\a} = \frac{1}{2} \Lambda d_{\a}, \qquad
\delta d_{\dda} = -\frac{1}{2} \Lambda d_{\dda}, \qquad
\delta \rho =  i \Lambda 
\label{u1}
\ee
In addition, the action (\ref{finalact}) is invariant under 
local target space Lorentz transformation, and the covariant
derivatives $\del_A$ with a flat index $A$
contain therefore both a spin and a $U(1)$
connection (the conventions we use for these are given in appendix A); 
in particular the dilaton $\phi$ should be covariantly
chiral, a $U(1)$ invariant statement. Both the Lorentz and $U(1)$
symmetries need to be preserved after quantization of the sigma model,
in order to arrive at a non-anomalous target space theory, but
both can suffer from sigma-model anomalies. 
One does indeed expect such anomalies to be present, due to the
fields $d_{\alpha}$ and $d_{\dda}$,
which are like chiral fermions, although their
conformal weight is one. 
We postpone the discussion of the 
anomalies they cause to section~\ref{anom}, and we will for the time
being assume that the Lorentz and $U(1)$ symmetries are unbroken.
Notice that one necessary condition for this to hold is indeed satisfied,
namely the generators of the $N=2$ algebra in (\ref{generators1})
together with (\ref{generators2}) are $U(1)$ and Lorentz invariant.

Having described the sigma model and some of its properties,
we now turn to its perturbative analysis.

\section{The sigma model at tree level}
\setcounter{equation}{0}

\subsection{Perturbation theory in the sigma model}

Before setting up the perturbation theory in $\a'$
for (\ref{finalact}) using
a covariant background formalism, we notice that 
there is one immediate problem, and
that is the field $\rho$. Its kinetic term does not have an explicit
factor of $1/\a'$ in front, and therefore an arbitrary number
of $\rho$-loops
contributes to any given order in $\a'$. In addition, 
in (\ref{finalact})
$\rho$ couples
to the other fields of the theory through $\phi-\bar{\phi}$, and 
it is not obvious whether or not it is possible to find explicit
results for all diagrams with a fixed number of $\rho$-loops and 
to sum them up. 
Furthermore, in (\ref{finalact}) the world-sheet Lagrange multiplier
$a_z$ imposes the constraint $\difb (\rho + i(\phi-\bar{\phi}))=0$
which is difficult to handle. Luckily, these last two problems 
disappear if we make the field redefinition 
\be \label{fredef}
\rho  \rightarrow  \rho-i(\phi-\bar{\phi})
\ee
after which
$\rho$ becomes a chiral boson which can be quantized
exactly, and which does not interact with the other fields in the
theory. Our perturbation theory will therefore consist of a hybrid
combination of conventional perturbation theory for all fields except
$\rho$, and exact conformal field theory results for $\rho$. 
After this field redefinition, the generators of the $N=2$
algebra read

\bea 
J & = & -i \dif \rho \nonu
G & = &  \frac{1}{i \alpha' \sqrt{8\alpha'}} e^{i\rho}
 e^{\phi-\bar{\phi}}
 d^\a d_\a 
 - \frac{1}{\sqrt{2\a'}i} \dif 
\left( e^{i\rho} 
 e^{\phi-\bar{\phi}}
d^{\gamma}  \del_{\gamma} \phi \right)  \nonu
\bar{G} & = &  \frac{1}{i \alpha' \sqrt{8\alpha'}} e^{-i\rho}
 e^{\bar{\phi}-\phi}
 d^{\dda} d_{\dda} 
 - \frac{1}{\sqrt{2\a'}i} \dif 
\left( e^{-i\rho} 
 e^{\bar{\phi}-\phi}
d^{\ddg} \del_{\ddg} \bar{\phi} \right) \nonu
T & \rightarrow & 
 \frac{1}{\alpha'} \left(
 -\frac{1}{2} \Pi^{\a\dda} \Pi_{\a\dda}  - d_{\a}  \Pi^{\a}
- {d}_{\dda}  \Pi^{\dda} 
+ \frac{\alpha'}{2} \dif \rho
   \dif \rho
\right) - \frac{1}{2} \dif^2 (\phi+\bar{\phi})
\label{e-gen}
\eea


Next, we describe the kind of calculation we intend to do. Our goal
is to obtain the equations of motion of the low-energy effective
target space action that is obtained from (\ref{finalact})
by integrating out all world-sheet fields. In the case of the
bosonic string, the equations of motion are equivalent to
finiteness of the sigma model and can therefore be obtained
from the beta-functions for the background fields 
\cite{bosbeta1,bosbeta2,fratse}. It is, due to the presence
of the $\rho$-field, not obvious that a similar computation would
provide the equations of motion in our case. A second complication
is that finiteness of the sigma model only implies that the
sigma model is conformal, and does not guarantee the full
$N=(2,0)$ superconformal invariance. The latter would only follow
from a standard supersymmetric 
$\beta$-function calculation if the model could
be formulated in $N=(2,0)$ superspace on the world-sheet, which
does not seem possible. In view of these complications we have
chosen to employ an alternative approach, which we will compare
to $\beta$-function calculations in section~\ref{betafie}.

Our approach is an extension of the one that has been developed for the
bosonic string in \cite{banese}. In that paper it is shown
that the equations of motion for the bosonic string can also
be derived by requiring nilpotency of the BRST-operator of the
sigma model through one-loop. This is in turn equivalent to
demanding that the Virasoro algebra is preserved at one-loop.
The equivalent statement in our case would be that the
$N=2$ algebra in (\ref{n2algebra}) is preserved at one-loop,
and this is certainly something we can check. One-loop
here refers to our hybrid way of doing perturbation theory, i.e.
one-loop in all fields except $\rho$ and to all orders in the field
$\rho$. 

To verify (\ref{n2algebra}) at one-loop, we use a covariant
background field expansion in the action of the sigma model,
and so that every field gets a non-zero vacuum expectation value,
except $\rho$, which we treat exactly.
Next, we require that the singular parts of the one-loop expectation
values of each of the pairs of operators on the left-hand sides
in (\ref{n2algebra}) matches the one-loop expectation value 
of the right-hand side. For example, for the Virasoro algebra,
we want to verify the identity
\be \label{vircheck}
<T(z) T(w)> = \frac{c/2}{(z-w)^4} + \frac{2<T(w)>}{(z-w)^2} + 
\frac{<\dif T(w)>}{(z-w)} + {\rm regular}
\ee
through one-loop. This computation will be desribed 
in section \ref{computation},
but before that we first have to describe the background field
expansion of the sigma model and to analyze it at tree-level.

\subsection{Covariant background field expansion}

In the background field method all fields are decomposed in terms of
background
fields which satisfy the classical field equations\footnote{If one
is only interested in the 1PI effective action it is not
necessary to require that the background fields satisfy the
classical equations of motion; in our case we want to verify
identitites like (\ref{vircheck}) and then it is more convenient
to impose the equations of motion.} and quantum fields which
represent the quantum fluctuations around the background fields.
We want to develop an expansion which is covariant with respect to the
target space local symmetries. In our case, this means a local
Lorentz and $U(1)$ covariant expansion. The traditional way to
achieve such an expansion is to use Riemann normal coordinates\cite{normal}.
Riemann normal coordinates are defined as follows:
Let $Z_0$ be a point on a (super)-manifold and $y^M$ a tangent vector
at $Z_0$. Let $Z(t)$ be the geodesic that satisfies $Z(0)=Z_0$
and $\frac{d}{dt}Z(t)|_{t=0} = y^M$, then the $y^M$ define coordinates
in a neighborhood of $Z_0$ by associating to $y^M$ the point 
$Z(Z_0;y^M)\equiv Z(t)|_{t=1}$ of $\cm$. In Riemann normal coordinates,
one labels the point $Z(Z_0;y^M)$ by $Z_0^M+y^M$,
and in those coordinates the geodesics are are just straight lines
eminating from $Z_0^M$. In general the geodesic $Z(t)$ is given by
$Z(t)=Z(Z_0;ty^M)$.

The geodesic equation can be written as
\be \label{geo-eq}
\frac{dZ^M}{dt} \del_M \frac{dZ^N}{dt} = 0.
\ee
Now let us denote by $y^N(Z_0;y^M)$
the tangent vector at $Z(Z_0;y^M)$ obtained by
parallel transporting the tangent vector $y^N$ at $Z_0$ to
the point $Z(Z_0;y^M)$ along the geodesic $Z(Z_0;ty^M)$. In other 
words, $y^N(Z_0;y^M)$ satisfies $y^N(Z_0;0)=y^N$ and
\be \label{aux1}
\frac{dZ^P(Z_0;t y^M)}{dt} \del_P y^N(Z_0;ty^M) =0.
\ee
The geodesic equation implies that (\ref{aux1}) is solved 
in particular by 
\be
y^M(Z_0;ty^P) \equiv \frac{d}{dt} Z^M(Z_0;ty^P),
\ee
which allows us to rewrite (\ref{aux1}) as
\be \label{aux2}
y^M(Z_0;y^P) \del_M y^N(Z_0;y^P) =0.
\ee
In particular, in Riemann normal coordinates, the tangent space at
$Z(Z_0;y^M)=Z_0+y^M$ is naturally identified with the tangent space
at $Z_0$, and (\ref{aux2}) is solved by $y^N(Z_0;y^M)= y^N$.

The covariant background field expansion is achieved by performing
a power series expansion in
the Riemann normal coordinates and then covariantizing the result
\be
{\bf \tilde{T}}' \equiv {\bf T}(Z_0+y)
= \sum \frac{1}{n!} (\pa_{M_1} \cdots \pa_{M_n}{\bf T})(Z_0) \ \
y^{M_n} \cdots y^{M_1}, \label{old}
\ee
where ${\bf T}(Z)$ is any tensor. The coefficients of the
expansion are covariantized by writing the derivatives in (\ref{old})
as the difference of a covariant derivative and a connection
piece, and by subsequently using relations between non-covariant
objects, like derivatives of connections, and covariant 
tensors that are only valid in Riemann normal coordinates.
For example, in the purely bosonic case we have in normal
coordinates relations of the form
\be
\pa_k \G^l_{mn} = \frac{1}{3} (R^l_{\ mkn} + R^l_{\ nkm}).
\ee
If, however, one is only interested in the expansion of scalar quantities
then there is a better algorithm available\cite{mukhi,attick}.
Observe that (\ref{old}) can be also written as
\be
{\bf \tilde{T}}' = e^{y^M \pa_M} {\bf T}. \label{old1}
\ee
If ${\bf T}=S$ is a scalar we can immediately
covariantize this expression by just converting the derivative into
a covariant derivative.
\be
\tilde{S}'= e^{y^A \nabla_A} S. \label{new}
\ee
Notice that in the exponent we also replaced curved indices $M$ by 
the more convenient flat indices $A$, using a vielbein to
convert one index into the other:
$y^A=y^M E_M{}^A$ and $\nabla_M=E_M{}^A \nabla_A$. Whereas
$\nabla_M$ contained a Christoffel connection, $\nabla_A$ contains
a spin connection, and its explicit form is given in (\ref{del}).

If $S$ is composed of several tensors then (\ref{new}) implies (by the
chain rule) an ``induced expansion'' for them which is given by exactly
the same formula (\ref{new}),
\be
{\bf T}' = e^{y^A \del_A} {\bf T}. \label{new1}
\ee
Let us further define
\be
\D {\bf T} = [y^A \del_A, {\bf T}]
\ee
Clearly ${\bf T}'$ can be obtained by applying iteratively the 
operator $\D$.

Notice that in (\ref{new1}) we have used a different symbol than in 
(\ref{old}). Both (\ref{new1}) and (\ref{old}) are covariant, but they 
differ by a Lorentz rotation
\be \label{aux8}
{\bf \tilde{T}}' = \Omega(Z_0;y) {\bf T}'.
\ee
When composing tensors to construct a scalar, the dependence on the
Lorentz transformation in (\ref{aux8}) will drop out, and the result
will be the same regardless of whether we used (\ref{old}) or
(\ref{new1}). In practise, (\ref{new1}) is much easier to use. Equation
(\ref{new1}) has a simple geometrical interpretation: the left hand
side is the result obtained by parallel transporting ${\bf T}(Z(Z_0;y))$
back from $Z(Z_0;y)$ to $Z_0$. Again, when composing tensors to 
form a scalar, these parallel transport terms drop out, and a third,
equivalent way to construct a covariant background field formalism
for scalars would be to simple take ${\bf T}(Z(Z_0;y))$. This
has the disadvantage that the expansion of a tensor ${\bf T}$ will
contain explicit connection terms, because ${\bf T}(Z(Z_0;y))$ 
lives at $Z(Z_0;y)$ whereas it is expanded in terms of objects that
live at $Z_0$, but, again, in scalar quantities the connection terms
cancel. Although we will only use (\ref{new1}), the distinction between
the different approaches is important in order to understand the
at first awkward looking rules like (\ref{drule}).

We are interested in carrying out the background field expansion of the
action $I$ and the generators $T,G,\bar{G},J$ of the superconformal
algebra. All of them are both Lorentz and $U(1)$ scalars. Therefore,
we can use (\ref{new1}). However, they are
not composed directly of tensors but rather they are made up from vielbeins
and connections. Hence, we need to determine the expansion of
$E_A^{\ M}, \ome_{A\b}^{\ \ \ \g}$ and $\G_A$.
To this end, consider ${\bf T}=\del_A$. Using the definitions and
conventions in appendix~A in (\ref{del}) and (\ref{com}), we obtain
from $\D \del_A = [y^B \del_B,\del_A]$ that
\bea
\D E_A{}^{M} &=& - [(\del_A y^B) - y^C T_{CA}{}^B] E_B{}^{M}  \\
\D \ome_{A\b}{}^{\g} &=& -[(\del_A y^B) - y^C T_{CA}{}^{B}]
\ome_{B\b}{}^{\g} + y^B R_{BA\b}{}^{\g} \\
\D \G_A &=& -[(\del_A y^B) - y^C T_{CA}{}^{B}] \G_B + y^B F_{BA}
\eea

Furthermore, let ${\bf T}=y^A$, then using the geodesic equation we
immediately get $\D y^A=0$. Recalling that $\P^A = \pa Z^M E_M{}^{A}$,
$\del = \P^A \del_A$, and defining
\be
\del y^A = \dif y^A + \P^B [\ome_{B \b}{}^{\g} M_{\g}{}^{\b}
+ \ome_{B \ddb}{}^{\ddg} M_{\ddg}{}^{\ddb} + \G_B Y,y^A]
\ee
we finally obtain, using that $\dif Z^M$ evaluated at $Z^M(Z_0;y)$ and
parallel transported back to $Z_0$ is just $\dif Z_0^M$, 
\bea
\D \P^A &=& \del y^A - \P^B y^C T_{CB}{}^{A} \\
\D (\del y^A) &=& - y^D \P^B y^C R_{CBD}{}^{A}
- w(A) y^A \P^B y^C F_{CB}
\eea
where $w(A)$ the $U(1)$ charge associated with the index $A$, i.e.
$w(a) =0$, $w(\a) =1/2$ and $w(\dda) =-1/2$.

The action (\ref{finalact}) contains in addition to $(\P^A,\bar{\P}^A)$
the world-sheet fields $d_{\a}$ and $\rho$. 
Although these are world-sheet
fields, one sees from the action (\ref{finalact}) and the generators
(\ref{e-gen}) that they should be viewed as being located at
$Z(Z_0;y)$. Correspondingly, in order to do the
same covariant background field expansion for
them as for the other fields in the sigma model, they should also
be parallel transported to $Z_0$.  This can be
achieved using the same equation (\ref{new1}) as we use for the other fields
in the theory, where only the connection piece in $\del_A$ in (\ref{new1})
acts on $d_{\a}$ and $\rho$. On $\rho$, which is $U(1)$ neutral after
the field redefinition (\ref{fredef}) and also a Lorentz scalar,
the connection piece acts trivially, so that $\rho$ has a trivial
background field expansion, $\Delta \rho=0$. However, $d_{\a}$
transforms under Lorentz and $U(1)$ transformations, leading to the rule
\be
\D d_\b = y^A (\ome_{A\b}{}^{\g} + \frac{1}{2} \G_A \d_\b{}^{\g}) d_\g.
\label{drule}
\ee
This rule is only valid to first order in $y$. To obtain the higher 
order terms
in the expansion of $d_\a$, one should first work out
$e^{y^A \nabla_A}d_\a $ to the required order and then put 
$\dif_A d_\a=0$ in this expansion.
Although this leads to terms that explicitly depend on the connections in the
background field expansion, as one already sees 
(\ref{drule}), the theory is still Lorentz and $U(1)$ invariant,
because $d_{\a}$  transforms as if it were located at $Z(Z_0;y)$,
and the explicit expressions for the transformation rules
(in terms of objects that live at $Z_0$) will
contain connection pieces as well, rendering the total background field 
expansion Lorentz and $U(1)$ invariant. 

Although (\ref{drule}) seems to be the most natural background field
expansion for $d_{\a}$, one can always make a field redefinition of
$d$ to arrive at any other background field expansion with
\be \label{ddU}
\Delta d_{\a}= y^C U_{c\a}{}^{\b} d_{\b}. 
\ee
Such field redefinitions
do in principle have jacobians that can affect the one-loop 
properties of the sigma model. It is our point of view that one
could in principle take any background field expansion of $d$ and
take that as the definition of the sigma model. In a similar
way one has the freedom to make a field redefinition of the
quantum variable $y^A$.
With this in mind,
we will from now on also choose the trivial background field 
expansion for $d_{\a}$, $\Delta d_{\a}=0$. The advantage of this
choice is that Lorentz symmetry is completely manifest, with
$d_{\a}$ transforming naively. In addition, we will give a background
expectation value $D_\a$ to $d_{\a}$,
\be
d_\a \rightarrow d_\a + D_\a.
\ee


%

We are now ready to perform the covariant expansion of the action $S$.
For the purpose of our calculation we only need to go up to two background
fields in the part quadratic in the quantum fields (denoted by $S^{(2)}$),
and up to one background field in the part with three
quantum fields (denoted by $S^{(3)}$). 
This can be seen by either a general analysis of the
orders of $\a'$ in the theory, of by an inspection of the Feynman
diagrams that occur. Terms with $4$ quantum fields and no background
fields contribute only when two of the quantum fields are contracted
as in diagram 34 in figure~1, but since massless tadpoles vanish in
dimensional regularization, we have not explicitly given the results
for these terms. The results for $S^{(2)}$ read
\bea
S^{(2)} &=& \frac{1}{2} \del y^a \bar{\del} y^a
+ d_{\ta} \bar{\del} y^{\ta} \nonumber \\
&&+ \frac{1}{2} \bar{\del} y^a y^C (\P^B T_{BC}{}^{a})
+ \frac{1}{2} \del y^a y^C (\bar{\P}^B T_{BC}{}^{a})
- \frac{1}{4} \bar{\del} y^C y^B
(\P^a T_{BC}{}^{a} + 2 \P^A H_{ABC}) \nonumber \\
&&+ \frac{1}{2} \bar{\del} y^C y^B (T_{BC}{}^{\ta} D_{\ta})
+ d_{\ta} y^C (\bar{\P}^B T_{BC}{}^{\ta})
- \frac{1}{4} \del y^C y^B
(\bar{\P}^a T_{BC}{}^{a} - 2 \bar{\P}^A H_{ABC}) \nonumber \\
&&+ \frac{1}{4} y^B y^C [(\P^D \bar{\P}^a + \bar{\P}^D \P^a)
T_{DCB}{}^{a}
-2 \bar{\P}^D T_{DCB}{}^{\ta} D_{\ta} \nonumber \\
& & \hspace{1cm} + \bar{\P}^D
((-1)^{E(D+B)+CD +1} T_C^{\ Ea} T_{DB}{}^{a}
+ H_{DCB}{}^{E}) \P_E ] \label{sexp2}
\eea
whereas for $S^{(3)}$ we have
\bea
S^{(3)} &=& - \frac{1}{4} (\bar{\del} y^a \del y^B y^C
+ \del y^a \bar{\del} y^B y^C)T_{CB}{}^{a} \nonumber \\
&&+ \frac{1}{3} \bar{\del} y^A \del y^B y^C H_{CBA}
+ \frac{1}{2} d_{\ta} y^B \bar{\del} y^C T_{CB}{}^{\ta} \nonumber \\
&&+\frac{1}{4} (\bar{\del} y^a y^B y^C \P^D
+ \del y^a y^B y^C \bar{\P}^D)T_{DCB}{}^{a}
+ \frac{1}{2} d_{\ta} y^B y^C \bar{\P}^D T_{DCB}{}^{\ta} \nonumber \\
&&+ y^B y^C \del y^D [\frac{1}{12} \bar{\P}^a
(T_{DCB}{}^{a} + (-1)^{CD} \del_C T_{DB}{}^{a}) \nonumber \\
& & \hspace{1.7cm}+ (-\frac{1}{3} H_{DCB}{}^{A}
+\frac{1}{6} T_{DC}{}^{E} H_{EB}{}^{A}
- \frac{1}{4} T_{DC}{}^{e} T_B{}^{Ae})\bar{\P}_A] \nonumber \\
&& +  y^B y^C \bar{\del} y^D [\frac{1}{12} \bar{\P}^a
(T_{DCB}{}^{a} + (-1)^{CD} \del_C T_{DB}{}^{a}) \nonumber \\
& & \hspace{1.7cm}+ (\frac{1}{3} H_{DCB}{}^{A}
-\frac{1}{6} T_{DC}{}^{E} H_{EB}{}^{A}
- \frac{1}{4} T_{DC}{}^{e} T_B{}^{Ae})\P_A] \nonumber \\
& & -  \frac{1}{6} y^B y^C \bar{\del} y^D [T_{DCB}{}^{\ta}
+ (-1)^{CD} \del_C T_{DB}{}^{\ta}] D_{\ta},
\label{sexp3}
\eea
where
\bea
T_{DCB}{}^{A}&=& R_{DCB}{}^{A} + w(A) F_{DC} \d_B{}^{A}
+ T_{DC}{}^{E} T_{EB}{}^{A} + (-1)^{CD} \del_C T_{DB}{}^{A}, \\
H_{DCBA}&=&\del_C H_{DBA}(-1)^{CD}-T_{CA}{}^{E} H_{EDB}
(-1)^{A(B+D)+CD}
+T_{DC}{}^{E} H_{EBA}. 
\eea

Next, we give the expansions of the
leading parts of the 
generators of the superconformal algebra in (\ref{e-gen}).
Again to the order we are interested in
they are given by
\bea
T &=& - \del y^a \P^a - D_{\ta} \del y^{\ta} - d_{\ta} \P^{\ta}
-y^C \P^B (T_{BC}{}^{a} \P_a - T_{BC}{}^{\ta} D_{\ta}) \nonumber \\
&&- \frac{1}{2} \del y^a \del y^a - d_{\ta} \del y^{\ta} \nonumber \\
&&- \del y^a y^C \P^B T_{BC}{}^{a} +
\frac{1}{2} \del y^B y^C \P^a T_{BC}{}^{a}
- \frac{1}{2} \del y^B y^C T_{BC}{}^{\ta} D_{\ta} \nonu
&&- d_{\ta} y^B \P^C T_{CB}{}^{\ta} 
-\frac{1}{2} y^B y^C \P^D [ T_{DCB}{}^{a} \P^a
- T_{DCB}{}^{\ta} D_{\ta} \nonu
&&- (-1)^{E(D+B)+CD} T_C^{\ Ea} T_{DB}{}^{a} \P_E] \nonu
G & = & e^{i\rho} e^{\phi-\bar{\phi}} (d+D)^2
 (1 + y^A \del_A(\phi-\bar{\phi}) +
 \frac{1}{2} y^A y^B \del_B \del_A (\phi-\bar{\phi}) \nonu
& & \qquad \qquad \qquad + \frac{1}{2}(y^A \del_A(\phi-\bar{\phi}) )^2 )
\nonu
\bar{G} & = & e^{-i\rho} e^{\bar{\phi}-\phi} (\bar{d}+\bar{D})^2
 (1 + y^A \del_A(\bar{\phi}-\phi) +
 \frac{1}{2} y^A y^B \del_B \del_A (\bar{\phi}-\phi)  \nonu
& & \qquad \qquad + \frac{1}{2}(y^A \del_A(\bar{\phi}-{\phi}) )^2 )
\label{gen-exp}
\eea
Finally, the relevant expansions of the dilaton pieces of the
superconformal algebra in (\ref{e-gen}) read
\bea
G_{\rm dil} & = & -\frac{1}{\sqrt{2 \a'} i} \dif
(e^{i\rho} 
 e^{\phi-\bar{\phi}}
d^{\gamma} \del_{\gamma} \phi + 
 e^{i\rho} 
 e^{\phi-\bar{\phi}}
D^{\gamma} y^A \del_A \del_{\gamma} \phi
\nonu & & \quad +
 e^{i\rho} e^{\phi-\bar{\phi}}
 D^{\gamma} \del_{\gamma} \phi y^A \del_A(\phi-\bar{\phi}) ) \nonu
\bar{G}_{\rm dil} & = & -\frac{1}{\sqrt{2 \a'} i} \dif
(e^{-i\rho} 
 e^{\bar{\phi}-\phi}
d^{\ddg} \del_{\ddg} \bar{\phi} + 
 e^{-i\rho}
 e^{\bar{\phi}-\phi}
 D^{\ddg} y^A \del_A \del_{\ddg} \bar{\phi} \nonu & & \quad +
 e^{-i\rho} 
 e^{\bar{\phi}-\phi}
 D^{\ddg} \del_{\ddg} \bar{\phi}  y^A \del_A(\bar{\phi}-\phi)
) \nonu
T_{\rm dil} & = & -\frac{1}{2} \dif^2 (y^A \del_A (\phi+\bar{\phi}))
\label{dgen-exp}
\eea

Besides Lorentz and $U(1)$ invariance, the background field expansion
of (\ref{finalact}) has an additional set of symmetries,
which we denote by `shift symmetries'. These originate in the fact that
the original action (\ref{finalact}) depends only on the vielbeins,
not on connections. The relation between torsions, connections and vielbeins
is given by (\ref{torsion}). Adding a covariant tensor to any of the 
connections in (\ref{torsion}) and subtracting a corresponding covariant term
from the torsions will leave the vielbeins invariant. This `shift symmetry'
should therefore be an invariance of the action. In our case one can
indeed verify that the action and the $N=2$ generators
are invariant under the following 
non-linearly realized shift symmetry:
\bea
\delta \omega_{AB}{}^C & = &  Y_{AB}{}^C \nonu
\delta \Gamma_A & = &  X_A \nonu
\delta T_{AB}{}^C & = &  Y_{[AB\} }{}^C + 
 w(C) X_{[A} \delta_{B ) }{}^C \nonu
\delta y^A & = & -\frac{1}{2} y^B y^C (Y_{CB}{}^R + w(A) X_C \delta_B{}^A)
 + {\cal O}(y^3) \nonu
\delta d_{\a} & = & (y^M Y_{M\a}{}^{\b} + \frac{1}{2} y^A X_A 
  \delta_{\a}{}^{\b} ) (d_{\b} + D_{\b}) + {\cal O}(y^2) 
\label{ssym}
\eea
where $X_A$ and $Y_{AB}{}^C$ are arbitrary covariant tensors,
such that $Y_{AB}{}^C$ has the same symmetries as the spin connection and
preserves the fact that the Lorentz group acts reducibly in target
space (see (\ref{ir1})-(\ref{ir4})).  In the next section we
will gauge this `shift symmetry' by putting certain torsions equal to
zero. These gauge fixing conditions are usually called `conventional
constraints'.

Apart from these observations, we will completely ignore all non-covariant
terms in the background field expansion, since in the absence of anomalies
the final results of our computation should be Lorentz and $U(1)$ invariant
(we will come back to this in section~\ref{anom};
in the bosonic case explicit calculations show all explicit 
dependence on the spin connection indeed drops out \cite{dbha}). 

\subsection{Tree-level constraints}

We are now ready to examine whether or not the sigma model has an
$N=2$ algebra at tree level. 
By examining the orders of $\a'$ in (\ref{n2algebra}), and
using exact results for $\rho$ we find that at tree level we only
need to verify
\bea \label{treecheck}
T(z) T(w) & = & \frac{2T(w)}{(z-w)^2} +\frac{ \dif T(w)}{(z-w)} + 
 {\cal O}((z-w)^0) \nonu
T(z) \eff
d^2(w) & = & \frac{2\eff d^2(w)}{(z-w)^2} + 
\frac{\dif(\eff d^2)(w)}{(z-w)} + 
 {\cal O}((z-w)^0) \nonu
T(z) \effb \bar{d}^2(w) & = & 
\frac{2 \effb \bar{d}^2(w)}{(z-w)^2} 
+ \frac{\dif (\effb \bar{d}^2)(w)}{(z-w)} + 
 {\cal O}((z-w)^0) \nonu
\eff d^2(z) \eff d^2(w) & = & {\cal O}((z-w)^1) \nonu
\effb \bar{d}^2(z) \effb \bar{d}^2(w) & = & {\cal O}((z-w)^1) \nonu
\eff d^2(z) \effb \bar{d}^2(w) & = & {\cal O}((z-w)^{-1}),
\eea
where  $T$ denotes here the $\rho$-indepent piece of $T$.
Notice the peculiar orders of $z-w$ to which order these identities
should hold; these follow from the OPE's of the exponentials of
$\rho$ appearing in $G$ and $\bar{G}$.

In addition to (\ref{treecheck}),
the generators of the $N=2$ superconformal algebra
should be holomorphic, implying $\dbar T=\dbar(\eff d^2) = \dbar(\effb 
\bar{d}^2)=0$.
Clearly, the presence of terms proportional to for example 
$(\bar{z}-\bar{w})/(z-w)^3$ in the right hand side of
(\ref{treecheck}) would imply that
the fields on the left hand side are not holomorphic, but the 
converse need not be true,
namely that the absence of nonholomorphic pieces in the OPE's implies
that the currents themselves are holomorphic.

The stress-energy tensor $T$ is the Noether current associated to the
symmetry
\be \Pi^A \rightarrow \Pi^A \dif \epsilon + \dif \Pi^A \epsilon,
\qquad \bar{\Pi}^A \rightarrow \Pi^A \dbar \epsilon + 
\epsilon \dif \bar{\Pi}^A  \label{aux9} \ee
\be d^{\a} \rightarrow d^{\a} \dif \epsilon + \dif d^{\a} \epsilon,
\qquad 
 d^{\dda} \rightarrow d^{\dda} \dif \epsilon 
+ \dif d^{\dda} \epsilon
  \label{aux10}
\ee
and this immediately implies (i) that $\dbar T=0$ and (ii) that the
first three OPE's in (\ref{treecheck}) are satisfied, as can be seen
by comparing the transformations of $T$, $\eff d^2$ and $\effb \bar{d}^2$ that
follow from (\ref{aux9}) and (\ref{aux10}) to those that are generated
by $\oint \frac{dz}{2\pi i} \epsilon(z) T(z)$ according to the OPE's
(\ref{treecheck}). It therefore remains to analyze $\eff d^2$ and 
$\effb \bar{d}^2$. 

Since it is easier to work out the conditions 
$\dbar (\eff d^2) = \dbar(\effb \bar{d}^2)=0$
than to compute the tree diagrams contributing to (\ref{treecheck}),
we start with the former. For this we need the equations of motion of
(\ref{finalact}). 
These can be found in a very easy way by applying the
operator $\D$ which generates the covariant background field expansion
(see the previous section) once to (\ref{finalact}). 
This yields
the following field equations
\bea
0&=& \bar{\P}^{\ta}  \label{deq} \\
0&=&
\del \bar{\P}_a + \bar{\del} \P_a - \P^C T_{Ca}{}^{b} \bar{\P}_b
- \bar{\P}^c T_{ca}{}^{b} \P_b
+ 2 \bar{\P}^c T_{ca}{}^{\tb} d_{\tb}
+ 2 \P^C \bar{\P}^b H_{bCa}  \hspace{1cm} \label{zaeq} \\
0&=&\bar{\del} d_\a + \frac{1}{2}(\P^C T_{C\a}{}^{b} \bar{\P}_b
+ \bar{\P}^c T_{c\a}{}^{b} \P_b)
- \bar{\P}^c T_{c\a}{}^{\tb} d_{\tb}
- \P^C \bar{\P}^b H_{bC\a}  \label{zaleq}
\eea
These can all be used to express $\delbar(\mbox{\rm background field})$
in terms bilinear in the background fields, using the identity
\be \del \bar{\Pi}^A - \delbar \Pi^A = -\bar{\Pi}^B \Pi^C T_{CB}{}^A.
\ee

Using the equations of motion (\ref{zaleq}), we find that
$\dbar(\eff d^2) = \dbar(\effb \bar{d}^2) = 0 $ if and only if
\bea
T_{\ta \tb a} - 2 H_{\ta \tb a} &=& 0 \label{tree1} \\
T_{a \tb c} + T_{c \tb a} = H_{a \tb c} &=& 0 \label{tree2} \\
T_{a \b}{}^{\ddb} = T_{a \ddb}{}^{\b} &=& 0 \label{tree3} \\
T_{a \b}{}^{\b} + \del_a(\phi-\bar{\phi})
 &=& 0 \ \ \label{tree4} \\
T_{a \ddb}{}^{\ddb} + \del_a(\bar{\phi}-\phi) &=& 0 \ \ \label{tree5}
\eea
Notice the explicit dilaton terms in (\ref{tree4}) and (\ref{tree5}).
They are crucial in order that the constraints are invariant under
the `shift symmetry' mentioned in the previous section. Although it
is not obvious at this stage, imposing (\ref{tree4}) and (\ref{tree5}) 
puts the theory partially on-shell. It is, however, not necessary
to impose (\ref{tree4}) and (\ref{tree5}) in order to have a consistent
tree-level theory. If (\ref{tree4}) and (\ref{tree5}) are not satisfied,
$G$ and $\bar{G}$ satisfy equations of the form $\dbar G= UG$,
$\dbar \bar{G} = \bar{U} \bar{G}$ for some $U$. With $U$ nonzero,
it is still possible to couple the theory to world-sheet supergravity,
with a modified transformation rule for the gravitino under
superconformal transformations. At the linearized level, this is
particularly easy to see. Consider the world-sheet action given by
\be
S=S_{\mbox{\rm sigma model}} + \int d^2 z (\nu G + \bar{\nu} \bar{G} )
\ee
where $\nu,\bar{\nu}$ are gauge fields for the superconformal
transformations generated by $G$ and $\bar{G}$. Under such
a transformation with parameters $\epsilon,\bar{\epsilon}$, the
action transforms, up to irrelevant numerical factors, into
\be
\delta S = \int d^2 z ( \epsilon(\dbar G +UG)+ \bar{\epsilon}( \dbar
\bar{G} 
+ \bar{U} \bar{G} ) + \delta_{\epsilon} \nu G + 
\delta_{\bar{\epsilon}} \bar{\nu}
\bar{G})
\ee
Therefore, if we take $\delta_{\epsilon} \nu = \dbar \epsilon - U
\epsilon$ and similarly for $\delta_{\bar{\epsilon}}\bar{\nu}$,
the action is invariant. 

Violating any of the other constraints (\ref{tree1})-(\ref{tree3}) 
would imply that $\dbar G$ would no longer be proportional to $G$,
and this is not acceptable. Therefore, (\ref{tree1})-(\ref{tree3})
have to be imposed. 

We continue by analyzing the tree-level OPE of $d^2$ with $\bar{d}^2$.
There is only one diagram that can possibly contribute, which is
diagram (49) in figure~1, which contains a table of all diagrams
that are relevant for this paper. Diagrams with more than two external
background field lines are always of order ${\cal O}((z-w)^{-1})$,
and are therefore not relevant for (\ref{treecheck}). As one sees from the
the kinetic term in $S^{(2)}$ in (\ref{sexp2}), there is no
propagator from $d_{\a}$ to $d_{\ddb}$, which is what would be
needed to obtain a nonzero contribution from diagram (49). We conclude
that diagram (49) vanishes and that the OPE of $d^2$ and $\bar{d}^2$
has the required form.

It remains to analyze the OPE of $d^2$ with itself, the analysis for
$\bar{d}^2$ with itself is similar. Again, diagram (49) yields no
contribution. However, we now have to include all diagrams with
up to four background fields, as we need the OPE up to order
${\cal O}((z-w)^{1})$. The diagrams with three or four 
background fields that contribute are diagrams (50), (51), (52),
(55) and (59). Diagrams (53), (54) and (56) vanish as they
involve the product of at least three $D$'s. 

The relevant diagrams can be worked out in a straightforward fashion in
either coordinate or momentum space, and although they are completely
finite there is an ambiguity in the results\footnote{A more
detailed discussion of momentum and coordinate space methods will
be given in section~4.}, which we discuss in
a moment. Diagrams (50), (51) and (52) contribute,
up to an overall factor,
\bea
< \eff d^2(z) \eff d^2(w) > & \sim & 
\frac{(\bar{z}-\bar{w})^2}{(z-w)^2}
(\dbar D^{\a} D^\g \bar{\Theta}_{\g\a}) \nonu
& & +
\frac{\bar{z}-\bar{w}}{z-w}
(\dif D^{\a} D^\g \bar{\Theta}_{\g\a}+ 
\dbar D^{\a} D^\g {\Theta}_{\g\a}) \nonu
& & +
(\dif D^{\a} D^\g {\Theta}_{\g\a}) 
\label{nn2}
\eea
where we defined
\bea
\Theta_{\rho\b} & = & \frac{1}{4} T_{\rho\b}{}^{a} \Pi^a
 - \frac{1}{2} T_{\rho\b}{}^{\ta} D_{\ta} + \frac{1}{2} \Pi^B
 H_{B\rho\beta} 
 - \frac{1}{2} D_{\rho} \del_{\beta}(\phi-\bar{\phi} ) 
\nonu
\bar{\Theta}_{\rho\b} & = & \frac{1}{4} T_{\rho\b}{}^{a} \bar{\Pi}^a
 - \frac{1}{2} \bar{\Pi}^B H_{B\rho\beta}.
\eea
Since the background fields satisfy the field equations, we
could in principle replace $\dbar D^{\a}$ in (\ref{nn2}) by a bilinear in the
background fields, yielding terms with four background fields, that
might potentially cancel against contributions from diagram (55).
However, something similar is not possible for the terms containing
$\dif D^{\a}$, and these terms should vanish by themselves, from which
we deduce that $\dif D^{\a} D^\g {\Theta}_{\g\a}$ and 
$\dif D^{\a} D^\g \bar{\Theta}_{\g\a}$ should vanish completely. 
This leads to the constraints
\be \label{tree6}
T_{\alpha\beta}{}^{c} = T_{\alpha\beta}{}^{\ddg} = H_{\alpha\beta c} = 
H_{\alpha\beta\ddg} = H_{\alpha\beta\gamma} = 0
\ee
and 
\be \label{tree7}
T_{\alpha\b}{}^{\b} + 2 \del_{\alpha} (\phi-\bar{\phi}) = 0.
\ee
The constraints (\ref{tree6}) include the representation preserving
constraints needed to be able to define chiral superfields in
target space. On the other hand, (\ref{tree7}) looks very awkward.
On can easily verify that it is not invariant under the 
shift symmetry (\ref{ssym}). Apparently, we have broken the shift
symmetry in this tree level calculation. One may wonder how this
is possible, since one could also have done a Poisson bracket
calculation in the sigma model without using any background field
expansion, in which case the shift symmetry is manifest. However,
a Poisson bracket calculation for the $\rho$-independent part
of the sigma model action would provide no information about the
regular part of the OPE of $\eff d^2$ with itself, which is
precisely where the problematic term (\ref{tree7}) came from.

One explanation of the ambiguities that underly (\ref{tree7}) is as
follows. Imagine adding a term to the action of the form
\be \label{totder}
\int d^2 z (\delbar D_{\gamma} + \ldots ) (y^{\alpha} y^{\beta}
 S_{\beta\alpha}{}^{\gamma} )
\ee
where $(\delbar D_{\gamma} + \ldots)$ is the field equation (\ref{zaleq})
with $d$ replaced by $D$. Since $D$ satisfies the field equation,
the extra term in (\ref{totder}) is zero. On the other hand, 
if one partially integrates the $\delbar$ in $\delbar D_{\gamma}$,
one gets a series of vertices with two background fields, and
precisely one vertex with one background fields, namely
\be \label{totder2}
\int d^2 z (-2 D_{\gamma} \delbar y^{\alpha} y^{\beta}
 S_{\beta\alpha}{}^{\gamma} ).
\ee
If we include this vertex in the calculation, we find an
extra contribution in (\ref{tree7}) proportional to $S_{\alpha\b}{}^{\b}$.
This represents an inherent ambiguity in our calculation, and since
we unfortunately do not know of any manifest shift symmetric scheme
to compute things, we will simply choose $S_{\alpha\b}{}^{\gamma}$ 
so as to cancel (\ref{tree7}) completely. Of course, to be 
self-consistent, we have to use the same vertex (\ref{totder2})
also in our one-loop calculation.

Another way to look at this ambiguity is to observe that on the
one hand $\delbar D$ consists of terms bilinear in the background
fields, but if one in a diagram encounters a propagator 
$(\dbar)^{-1}$ and integrates this against $\delbar D$, one
can obtain contributions containing a single $D$. Clearly, 
a better understanding of these issues would be desirable.

We are left with diagrams (55) and (59) to analyze. 
Before doing so, we will first 
try to simplify these remaining calculations as much as possible,
by supplementing the constraints obtained so far
((\ref{tree1}), (\ref{tree2}), (\ref{tree3}) and (\ref{tree6}))
with a maximal set of 
conventional constraints, and by subsequently solving the Bianchi 
identities for the connections $\del_A$, $\{\del_A,\{\del_B,\del_C\}\}
+\mbox{\rm cycl.}=0$. This will further constrain the torsions and curvatures
and provide us with the largest possible set of constraints
implied by 
(\ref{tree1}), (\ref{tree2}), (\ref{tree3}) and (\ref{tree6}).



Quite remarkably, it turns out that 
(\ref{tree1}), (\ref{tree2}), (\ref{tree3}) and (\ref{tree6})
plus
a maximal set of conventional constraints is already sufficient
to completely solve the Bianchi identities and reduce the field
content of the theory to that of 
conformal supergravity coupled to a linear
multiplet. This in complete accordance with the analysis of the
vertex operators in section~(\ref{subsect13}), see in particular
the discussion at the end of section~(\ref{subsect13}), where
we argued that all contraints should appear at tree level and the
field equations should appear at one-loop.

Let us now determine a 
maximal set of conventional constraints.  All but one of the 
conventional constraints can be viewed as a gauge fixing of the shift
symmetry (\ref{ssym}) discussed previously. 

Recall that the parameters of the shift symmetry are
covariant tensors
$X_A$ and $Y_{AB}{}^C$,
where $Y_{AB}{}^C$ has the same symmetries as the spin connection and
preserves the fact that the Lorentz group acts reducibly in target
space (see (\ref{ir1})-(\ref{ir4})).
%
%
We can not use this to put the spin connection and $U(1)$ connection
equal to zero, since zero is not a globally well-defined connection,
but we can use it to put some torsions equal to zero. To determine
how many torsion coefficients can be removed  we analyze the spin
content of the covariant tensors $X$ and $Y$.
The tensor $Y_{\a \b}{}^{\g}$ contains a spin-3/2 and a spin-1 piece
and can be used to remove the torsion $T_{\a \b}{}^{\g}$ completely.
The tensor $Y_{\dda \b}{}^{\g}$ removes
the symmetric part $T_{\dda (\b}{}^{\g)}$ of the torsion
whereas  $X_{\dda}$ removes the remaining antisymmetric part
$T_{\dda \b}{}^{\b}$. Similarly, $Y_{a\b}{}^{\g}$ is
used
to put $T_{a(\b}{}^{\g)}$ equal to zero, and we use $X_a$ to
put $T_{a\b}{}^{\b} - T_{a\ddb}{}^{\ddb}$ equal to zero.
To summarize, we impose
the following conventional contraints:
\be
T_{\a \b}{}^{\g} =
T_{\dda \b}{}^{\g} =
T_{\a \ddb}{}^{\ddg} =
T_{\dda \ddb}{}^{\ddg} =
T_{a (\b}{}^{\g)} = 
T_{a (\ddb}{}^{\ddg)} = 
T_{a\b}{}^{\b} - T_{a\ddb}{}^{\ddb}=0.
\ee

There is still one more conventional constraint, of a somewhat different
origin, that we can impose.
One can always
redefine the vielbeins by performing a local Lorentz tranformation,
since it is not the vielbein itself but only the metric
$G_{MN}=E_M{}^{A} E_{NA}$ that appears in the action. The 
Lorentz group acts reducible in target space, in other words independently
on the indices $\a$, $\dda$ and $a$. Using a local Lorentz rotation
that acts only on the vector indices $a$, we can always achieve that
\be
T_{\a \ddb}{}^{c}= c \d_{\a}{}^{\g} \d_{\ddb}{}^{\ddg}, \label{adota}
\ee
where $c$ is a constant. After choosing this gauge for the vector piece
of the Lorentz gauge transformations, Lorentz transformations acting
on the spinor indices must be accompagnied by one on the vector indices
in order to preserve (\ref{adota}). This leads to the identifications
(\ref{ir1}) and (\ref{ir3}).
Comparing with the exact result for flat
superspace in (\ref{flattor}) we find that $c=-2i$, and this
will be confirmed by our one-loop calculations in the remaining sections.
Equation (\ref{tree1}) now implies
\be
H_{\a \dda b} = - i C_{\a \b} C_{\dda \ddb} \label{hada}
\ee
in accordance with (\ref{flath}).

We are now ready to solve the Bianchi identities that will reduce
the field content to that of 
conformal
supergravity coupled to a tensor multiplet.
The involved algebra is rather tedious and here we just present the
final result, postponing the details to appendix B
\bea
\{\del_{\a}, \del_{\b}\} &=& 0,
\label{bian1} \\
\{\del_{\a}, \del_{\ddb} \}
&=& -2i \del_{\a \ddb} -4i H_{\ddb \g} M_{\a}{}^{\g}
+4i H_{\ddg \a} M_{\ddb}{}^{\ddg} +4i H_{\ddb \a} Y, 
\label{bian2} \\
\left[\del_{\a}, \del_{b} \right] &=&
-2 \del_{\b} H_{\ddb \g} M_{\a}{}^{\g}
\nonumber \\
&\ &+[-2i C_{\a \b} \bar{W}_{\ddb \ddg}{}^{\ddd}
+ C_{\ddb \ddg} (\del_{(\a} H^{\ddd}{}_{\b)} -
\frac{1}{3} C_{\a \b} \del^{\e} H^{\ddd}{}_{\e})] M_{\ddd}{}^{\ddg}
\nonumber \\
&\ &+ 2 \del_{\b} H_{\ddb \a} Y  \label{bian3} \\
\left[\del_{a}, \del_{b} \right] &=&
\{-2 H_{\dda \b}  \del_{\a \ddb} \nonumber \\
&\ &+[\frac{i}{2} C_{\a \b} \del_{(\dda} H_{\ddb)}{}^{\g}
+C_{\dda \ddb}(-\frac{i}{6} C^{\g}{}_{(\a|} \del^{\dde} H_{\dde| \b)}
+ W_{\a \b}{}^{\g})] \del_{\g} \nonumber \\
&\ &+\left[ C_{\dda \ddb} \left(\frac{1}{24} \del_{(\a} W_{\b \g}{}^{\d)}
+\frac{1}{4}
(C^{\d}{}_{\a} \del_{(\b|\dde} H^{\dde}{}_{|\g)} + \a \leftrightarrow \b)
\right. \right.
\nonumber \\ 
&\ &
\left. \left.
+\frac{i}{6} C_{\a \g} C^{\d}{}_{\b} \del^{\dde} \del^{\e} H_{\dde \e}
\right)
+\frac{i}{2} C_{\a \b} \del_{\g} \del_{(\dda} H_{\ddb)}{}^{\d}\right]
M_{\d}{}^{\g} \nonumber \\
&\ &-\frac{i}{2} C_{\a \b} \del^{\d} \del_{(\dda} H_{\ddb)\d} Y + {\rm c.c.} \}
\label{bian4}
\eea
where `c.c.' denotes our definition of complex conjugation, see appendix A,
$W_{\a \b \g}$ is completely symmetric chiral superfield,
\be
\del_{\ddd} W_{\a \b \g} = 0,
\ee
and $H_{\dda \b}$ is
defined as follows (see also appendix B)
\be
H_{a b c} = C_{\g \a} C_{\ddg \ddb} H_{\dda \b} -
C_{\g \b} C_{\ddg \dda} H_{\ddb \a}. \label{habc}
\ee
$W_{\a \b \g}$ and $H_{\dda \b}$ satisfy the following differential
relations
\be
\del_a H^a = 0, \
\del^{\g} W_{\g \a \b} = \frac{i}{6} \del_{(\a|} \del^{\ddg} H_{\ddg| \b)}
+\frac{i}{2} \del^{\ddg} \del_{(\a|}  H_{\ddg| \b)}, \
\del^{\b} \del_{\b} H_a = 0.
\ee
Furthermore, all the components of the field strength $H_{ABC}$ vanish
except the ones that are given in (\ref{hada}) and (\ref{habc}).
Compactly, we have
\be
T_{ABc} + (-1)^{AB} 2 H_{ABc} = 0.
\ee
A similar supergravity algebra has been obtained in \cite{gates}.

From (\ref{bian1})-(\ref{bian4}) we can read off how all torsions
and curvatures can be expressed in terms of $H_{\dda\b}$ and
$W_{\a\b\gamma}$. In particular, the background field expansion for
the action will then contain only these two fields, and this is the
form of the action with the fewest number of vertices and therefore
the most suitable one to use in calculations. 

One can verify that with maximally simplified form of the action
the remaining tree diagrams (55) and (59) yield no contribution,
which concludes the tree-level analysis of the theory. We 
next turn our attention to one-loop calculations.

\section{The sigma model at one-loop} 
\setcounter{equation}{0}
\label{computation}

\subsection{Coordinate space and momentum space techniques}

The starting point for our calculations will be the covariant
background field expansion for the action, with all the constraints
and explicit expressions obtained from the tree-level constraints
and the Bianchi identities inserted in it. In addition, as we
explained before, we will for the time being drop all explicit
connection terms. The part of the action quadratic in the quantum
fields given in (\ref{sexp2}) contains a piece 
\be
S_{\rm kin} = \frac{1}{\a'} \int d^2 z 
(\frac{1}{2} \dif y^a \dbar y^a + d_{\ta} \dbar y^{\ta} ) 
\label{skin}
\ee
which we take as the kinetic term, since $S_{\rm kin}$ describes
free fields for which we know the propagators explicitly. The
remainder of (\ref{sexp2}) will together with (\ref{sexp3}) 
provide the vertices of the theory. 

The propagators obtained from (\ref{skin}) are in coordinate space
given by
\be
<y^a(z) y^b(w)> = \a' \delta^{ab}
 G(z,w) \equiv -\a' \delta^{ab}
\log |z-w|^2 
\label{prop1}
\ee\be
<d_{\a}(z) y^{\b}(w) > = \frac{\a' \delta_{\a}{}^{\b} }{(z-w)} = 
- \a' \delta_{\a}{}^{\b} \dif_z G(z,w)   \label{prop2}
\ee
It is now straightforward to write down coordinate space expressions
for the Feynman diagrams given in figure~1.
We are only going to compute
the contributions to $<T(z)T(w)>$, $<T(z)G(w)>$ etc. which involve
at most two background fields (or more precisely, where the sum of
the conformal weights of the background fields is at most two).
For $<G(z)\bar{G}(w)>$ this is all
we need (cf. (\ref{treecheck})), as it is for $<T(z)T(w)>$ by
virtue of its symmetry under $z\leftrightarrow w$. Thus, the only
contributions we are missing are those to $<T(z)G(w)>$ and
$<T(z)\bar{G}(w)>$ with three background fields and those to
$<G(z)G(w)>$ and $<\bar{G}(z) \bar{G}(w)>$ with four background
fields. It is a very tedious calculation to determine these contributions,
and in addition we believe that they will not lead to any new equations,
although we have no general proof of this.

In diagrams with two background fields, we can assume that the background
fields are independent of the coordinates, or in momentum space,
that they have zero momentum. If we would make a Taylor
expansion of the background fields around a fixed point on the 
world-sheet, all derivatives of the background fields would have
higher conformal weight and can be ignored. For diagrams with one
or zero background fields one has to be more careful and expand
everything up to the relevant order. In addition, there may be ambiguities
in this similar to those discussed below (\ref{tree7}), but these
particular ambiguities don't play a role in the part of the
calculation we present below.

In coordinate space, we then encounter various integrals not containing
background fields, like for example
\be \label{exam1}
\int d^2 u \frac{1}{z-u} \frac{1}{u-w} 
\ee
Such integrals can then be manipulated using the fact that they are 
translationally invariant, using partial integrations inside the integral
and using identities such as
\be \dif_{\bar{z}} \frac{1}{z-w} = \delta^{(2)}(z-w).
\label{fund}
\ee
Here, $\delta^{(2)}$ is the delta function with respect to our measure
$d^2z$ which was defined as $dxdy/\pi$ with $z=x+iy$. For our example
(\ref{exam1}) one then finds
\be 
\int d^2 u \frac{1}{z-u} \frac{1}{u-w}  = \frac{\bar{z}-\bar{w}}{z-w}
\ee
by writing $1/(u-w)$ as $\dbar_{\bar{u}} ((\bar{u}-\bar{w})/(u-w))$,
using a partial integration and (\ref{fund}). One has to be
very careful using such naive manipulations, since one-loop
diagrams can have divergences, and this can sometimes lead
to ambiguous answers. For example, diagram (9) can lead to integrals
like
\be
\int d^2 u d^2 v \frac{1}{z-u} \frac{1}{u-v} \delta^{(2)}(v-w) 
\delta^{(2)}(u-w) 
\ee
If we first do the $u$ integral, we get
\be
\int d^2 v \frac{1}{z-w} \frac{1}{w-v} \delta^{(2)}(v-w) = 
- \frac{1}{2} \frac{1}{z-w}
\int d^2 v \dbar_{\bar{v}} \frac{1}{(v-w)^2} = 0.
\ee
whereas first integrating over $v$ leads to
\be
\int d^2 u \frac{1}{z-u} \frac{1}{2} \dbar_{\bar{u}} \frac{1}{(u-w)^2} = 
\int d^2 u \delta^{(2)}(u-z) \frac{1}{2} \frac{1}{(u-w)^2} = 
 \frac{1}{2} \frac{1}{(z-w)^2}.
\ee
This ambiguity reflects the fact that we should regularize the 
(relatively convergent) integrals. Although there are proposals 
for a consistent regularization in coordinate space \cite{fried},
these are not particularly useful in our case. In our case,
we have found indications that it might be possible to
formulate a set of rules in coordinate space that allow one to
express every integral in terms of unambiguous integrals and
only one particular integral whose evaluation in coordinate
space seems to be very difficult,
\be
D_2=\int d^2u \d^2v \frac{1}{z-u} \left[ \delta^{(2)}(u-v) \right]^2
\frac{1}{v-w}.
\label{ambint}
\ee
The precise value of this integral then depends on the choice of
regularization prescription (in dimensional regularization
it is $\frac{1}{6(z-w)^2}$), but its precise value
is not important because one can always choose finite local counterterms
in the generators of the $N=2$ algebra so that in the final result
$D_2$ does not appear. Unfortunately, we have not been able
to make these statements more precise, and therefore
we have chosen to use dimensional regularization in momentum space
instead. 

In dimensional regularization, we immediately run into two problems.
First, there are terms in the action coming from the antisymmetric
tensor field in (\ref{finalact}), and these contain two-dimensional
anti-symmetric tensors $\epsilon_{ij}$ when written in a world-sheet
covariant way. Second, there are `chiral' world-sheet
fields like $d_{\a}$ and $y^{\a}$. We have chosen the following version
of dimensional regularization to deal with these problems. First
of all, we keep all interactions in exactly two dimensions, 
and in particular the $\e$-tensor will remain in two dimensions.
Secondly, we view $d_{\a}$ as a vector field on the world-sheet with
only a $z$-component, and we will in other dimensions
still view it as a vector field with only a component in the $z$-direction.
Besides this, the kinetic terms will be taken in $2-2\e$ dimensions.
With such a type of regularization, one has to be careful when going
beyond one
loop, in which case evanescent counterterms have to be taken into account
(see \cite{wit} for a detailed discussion of the two-loop
renormalization of a two-dimensional sigma model, including problems
with the $\e$ tensor in two dimensions)\footnote{There
exist several other approaches to regularize the integrals
that appear in chiral and non-local two-dimensional field theories,
such as analytic regularization \cite{anreg}, exponential regularization
\cite{expreg} and Pauli-Villars \cite{pvreg}. The situation beyond one-loop
is unclear, but at one-loop
we expect each of these to yield equivalent answers.}. 
We will not further discuss these
issues here, since we will only be doing a one-loop calculation.

In the same way as we defined the measure $d^2 z$ as $dxdy/\pi$, we
will define $d^2k$ also as $dk_x dk_y /\pi$, and it turns out that
with this choice we never see any explicit 
factors of $\pi$ in our calculation. Then $G(z,w)$ has the 
following representation
\be
G(z,w) = \int d^2 k \frac{e^{ik(z-w) + i\kb(\zb-\wb)}}{|k|^2}
\ee
and the propagators in momentum space read
\bea
<y^a(k) y^b(l)> & = & \a' \delta^{ab}
 \delta^{(2)}(k+l) \frac{1}{|k|^2} \nonu
<d_{\alpha}(k) y^{\b} (l) > & = & \a'
\delta_{\a}{}^{\b} \delta^{(2)}(k+l) \frac{-ik}{|k|^2} 
\label{propk}
\eea
where $k$ is shorthand notation for $k_z$, the $z$-component of $k$,
and in dimensional regularization we keep $k$ in one dimension, and
only continue the denominators in (\ref{propk}) outside two dimensions.
To work out the Feynman diagrams we write down the corresponding
expression in momentum space, continue to $d=2-2\e$ dimensions, and
include a factor 
$\G(1-\e) (4\pi)^{-\e}(2\pi)^{2\e}$ for each loop, which removes a lot
of unwanted constants like the Euler constant, see \cite{wit}. The
$d$-dimensional measure $d^d k $ is equal to the standard $d$-dimensional
measure divided by $\pi$. 
The relevant integrals come out as follows
\bea
&&\!\!\!\!\!\!\!\! \G(1-\e) (4\pi)^{-\e} (2\pi)^{2\e}
\int d^d p \frac{p^a \pb^b}{[|p|^2]^{\a} [|p-k|^2]^{\b} }
=  \nonumber \\ & &\!\!\!\!\!\!\!\!= 
k^{a+1-\a-\b} \kb^{b+1-\a-\b} |k^2/\m^2|^{-\e}
\frac{\G[1-\e] \G[1-\e-\a+b]}{\G[\a] \G[\b]
\G[2-\a-\b+b-\e]}   \nonumber   \\ 
&&\!\!\!\!\!\!\!\! \times \sum_{r=0}^{a} 
\left(
\begin{array}{c}  a   \\ r \end{array} 
\right)
\frac{\G[2-\a-\b+b+r-\e] \G[\a+\b-1-r+\e] \G[1-\e-\b+r]}{
\G[2-2\e-\a-\b+r+b]}
\label{dimreg}
\eea
as one proves by differentiating both sides with respect to $k$
and $\kb$, starting from the known result for $a=b=0$.
The parameter $\mu$ is the usual dimensional regularization
mass parameter.

We have worked out all our diagrams in momentum space using (\ref{dimreg}),
and then afterwards reexpressed them in coordinate space using the
following translation table
\bea
-\frac{k^3}{3 \bar{k}^3} & \leftrightarrow & 
\frac{(\bar{z}-\bar{w})^2}{(z-w)^4} \nonu
\frac{k^2}{2 \bar{k}^2} & \leftrightarrow & 
\frac{\bar{z}-\bar{w}}{(z-w)^3} \nonu
-\frac{k}{\bar{k}} & \leftrightarrow & 
\frac{1}{(z-w)^2} \nonu
\frac{k}{\epsilon \bar{k}} + \frac{k}{\bar{k}} (1-\log(|k|^2/\mu^2))
& \leftrightarrow & \frac{G(z,w)}{(z-w)^2} 
\label{transla}
\eea
The first three follow by Fourier transformation, the last one,
which contains an $1/\e$ divergence, was chosen by comparing the
result for $<y(z)y(w)><\dif y(z) \dif y(w)>$ in coordinate space
and momentum space. In the right hand side 
one could in principle have chosen a different
finite term proportional to $k/\bar{k}$, but since all divergences
will cancel in our calculation, this would not make any
difference for the final answers.

There is one more important thing we have to discuss before
we present the results of our calculation. Namely, we have 
dropped in the expansion of $G$ and $\bar{G}$ in (\ref{gen-exp})
all terms that came from the background field expansion of
$\eff$ and $\effb$. Including these terms leads to a 
number of additional contributions in all OPE's except that of
$T$ with $T$, each of which contain the dilaton. However, we
claim that one can systematically remove all these terms. 
For this one needs (i) one-loop counterterms, (ii) ambiguities
similar to the one discussed below (\ref{tree7}) and (iii) the
possibility to choose different background field expansions for
$d$. The precise details are rather cumbersome and will be 
reported elsewhere \cite{dbsk}. We suspect there may be a better
explanation in which the shift symmetry in (\ref{ssym})
plays a crucial role, but for the time being all this remains 
rather mysterious. 

Having described how we performed them, we now turn to the results of
our calculations.

\subsection{One-loop results}

\begin{sloppypar}
Below we tabulate the contributions of diagrams (1)-(19) to
$<~G(z) \bar{G}(w)>$, $<G(z)G(w)>$,
$<T(z)G(w)>$ and $<T(z) T(w)>$. We postpone
the discussion of the other diagrams for the time being; most
of them do not give any contribution, except for the ones involving
an insertion of the dilaton terms in
(\ref{dgen-exp}), and these will be tabulated separately.
\end{sloppypar}

Denoting by $I_d$ the contribution of diagram number $d$,
the non-zero diagrams for $<G(z)\bar{G}(w)>$ come out as follows
(note that we have also taken the $\rho$-contributions into
 account, whose only effect is to shift the order of the poles by
 one, and
$\lambda$ is the coefficient in front of $G$, $\lambda=1/(i\a' 
\sqrt{8\a'} )$)
\bea
I_4 & = & \frac{1}{(z-w)} ( 8  \lambda^2 \a'^2 \Pi^a \Pi^a)
\nonu
I_{10} & = & \frac{1}{(z-w)} (+16 \lambda^2 \a'^2 
D_{\b} \Pi^{\b} - \lambda^2 \a'^2 D^{\a} R^{\ddg}{}_{\a\ddg}{}^{\ddb}
D_{\ddb} - \frac{1}{2} \lambda^2 \a'^2 D^{\a} D^{\ddg} F_{\ddg \a})
\nonu
I_{15} & = & \frac{1}{(z-w)} (16 \lambda^2 \a'^2 
D_{\ddb} \Pi^{\ddb} - \lambda^2 \a'^2 D^{\dda} R^{\g}{}_{\dda\g}{}^{\b}
D_{\b} - \frac{1}{2} \lambda^2 \a'^2 D^{\a} D^{\ddg} F_{\ddg \a})
\nonumber
\eea
We have also computed all one-loop contributions to $<G(z)\bar{G}(w)>$,
without imposing any constraints, except 
(\ref{tree1})-(\ref{tree5}), and found many more terms contributing
to $<G(z)\bar{G}(w)>$, but the result remained finite, i.e. the
terms proportional to $G(z,w)$ canceled each other. This is a 
reflection of the fact that when $G$ and $\bar{G}$ are conserved 
quantities, their OPE should be finite.
Adding up the contributions we get 
\be
<G(z)\bar{G}(w)>={1 \over (z-w)} (2 T(w) + \frac{2i}{\a'}
D^\a D^{\ddb} H_{\ddb \a})
\label{ggbi}
\ee
The first term is precisely the term we expect (see (\ref{n2algebra})).
The second term will be removed by choosing a different background
field expansion for $d$.
We defer a furhter discussion of the background field expansion for $d$ and
the one-loop counterterms till after we have discussed all OPE's.

\begin{sloppypar}
None of the diagrams (1)-(19) contributes to $<G(z)G(w)>$ or
$<~\bar{G}(z)\bar{G}(w)>$, in agreement with (\ref{n2algebra}). 
\end{sloppypar}

Next we turn to $<T(z)G(w)>$. Below are the results for this
computation, and we have extracted a factor of $\lambda \a'e^{i\r} \eff$
To obtain the contributions to $<T(z)G(w)>$, one should still
multiply everything by this factor.
\bea
I_5 & = & \frac{(\zb-\wb)}{(z-w)^3}
(-2 \Pb^d R_d{}^{\b}{}_{\b}{}^{\a} D_{\a} - 
\Pb^d F_d{}^{\g} D_{\g} ) \nonu
I_7 & = & \frac{1}{(z-w)^2}(
\Pi^d R_d{}^{\b}{}_{\b}{}^{\a} D_{\a} + \Pi^{\ddd} 
R_{\ddd}{}^{\b}{}_{\b}{}^{\a} D_{\a} + \frac{1}{2} \Pi^d F_d{}^{\g}
D_{\g} \nonu
&& \qquad + \frac{1}{2} \Pi^{\ddd} F_{\ddd}{}^{\g} D_{\g} 
- 8\Pi_{\dda} \Pi^{\dda}) \nonu
I_{10} & = & \frac{(\zb-\wb)}{(z-w)^3} (
(+2 \Pb^d R_d{}^{\b}{}_{\b}{}^{\a} D_{\a}  +
\Pb^d F_d{}^{\g} D_{\g} ) \nonu
&+&\frac{1}{(z-w)^2} (
-\frac{1}{2}\Pi^d R_d{}^{\b}{}_{\b}{}^{\a} D_{\a} 
-\frac{1}{2} \Pi^{\ddd} 
R_{\ddd}{}^{\b}{}_{\b}{}^{\a} D_{\a} - \frac{1}{4} \Pi^d F_d{}^{\g}
D_{\g} \nonu
&& \hspace{2cm}- \frac{1}{4} \Pi^{\ddd} F_{\ddd}{}^{\g} D_{\g} + 8
\Pi_{\dda} \Pi^{\dda}) \nonu
I_{14} & = & 
\frac{(\zb-\wb)}{(z-w)^3} (-2i \Pb^c T_{c,}{}^{\g}{}_{\ddb,}{}^{\ddb}
D_{\g} ) 
\nonu
I_{15} & = & 
\frac{(\zb-\wb)}{(z-w)^3} (
\Pb^a (\frac{1}{2} R^{\g b}{}_{ba} + 3i T_{a,}{}^{\g}{}_{\ddb,}{}^{\ddb}
+ R_a{}^{\b\g}{}_{\b} - \frac{1}{2} F_a{}^{\g} )D_{\g} ) \nonu
&+&\frac{1}{(z-w)^2} (
\Pi^a (-\frac{1}{4} R^{\g b}{}_{ba} + \frac{i}{2} 
T_{a,}{}^{\g}{}_{\ddb,}{}^{\ddb} ) D_{\g} ) \nonu
I_{18} & = & \frac{1}{(z-w)^2} (3i \Pi^c T_{c,}{}^{\g}{}_{\ddb,}{}^{\ddb}
D_{\g} ) \nonu
I_{19} & = & \frac{1}{(z-w)^2} (-\frac{13i}{6}
\Pi^c T_{c,}{}^{\g}{}_{\ddb,}{}^{\ddb}
D_{\g} )
\eea
Adding up all diagrams we obtain
\bea
<T(z)G(w)>&=&\frac{\zb-\wb}{(z-w)^3} \Pb^a(R_a{}^\b{}_\b{}^\g
-\half F_a{}^\g + \half R^{\g b}{}_{ba} + iT_{a,}{}^\g{}_{\ddb,}{}^{\ddb})
D_\g \nonu
&+&\frac{1}{(z-w)^2}[\P^a(\half R_a{}^\b{}_\b{}^\g -{1 \over 4}
R^{\g b}{}_{ba}+{1 \over 4}F_a{}^\g \nonu
&& \qquad \!\!\!+{4 \over 3} 
iT_{a,}{}^\g{}_{\ddb,}{}^{\ddb}) D_\g  
+\P^{\dda}(\half R_{\dda}{}^\b{}_\b{}^\g+{1 \over 4}F_{\dda}{}^{\g})D_\g]
\eea
Inserting the expicit expressions of the curvatures and torsions
from (\ref{table3}) we get
\bea
<T(z)G(w)>&=&
\frac{\zb-\wb}{(z-w)^3}[-4\Pb^a \del_\a H_{\dda}{}^\g D_\g]
\nonu
&+&
\frac{1}{(z-w)^2}[-{17 \over 6} \P^a \del_\a H_{\dda}{}^\g D_\g  
+ 4i \P^{\dda} H_{\dda}{}^\g D_\g] \label{tgfinal}
\eea

Finally, we list the results for $<T(z)T(w)>$. Due to the symmetry
of this expectation value under $z \leftrightarrow w$, a diagram
and its mirror image (like diagrams (2) and (7)) give identical
contributions, and this is reflected in the following table
\bea
I_1 & = & 
\frac{G(z,w)}{(z-w)^2} (- (\Pi^c H_{cmn} -
\frac{1}{2} T_{mn}{}^{\ta} D_{\ta} )^2 ) \nonu
& & + \frac{1}{(z-w)^2} 
 ( (\Pi^c H_{cmn} -
\frac{1}{2} T_{mn}{}^{\ta} D_{\ta} )^2 
\nonu & & \qquad
-4i (\Pi^{\dda} \Pi^c T_{c,\b\dda,}{}^{\b} +
 (\Pi^{\a} \Pi^c T_{c,\a\ddb,}{}^{\ddb} ) )
\nonu
I_2 & = & 
\frac{G(z,w)}{(z-w)^2} (2 (\Pi^c H_{cmn} -
\frac{1}{2} T_{mn}{}^{\ta} D_{\ta} )^2 ) \nonu
& & + \frac{1}{(z-w)^2} 
 (-4 (\Pi^c H_{cmn} -
\frac{1}{2} T_{mn}{}^{\ta} D_{\ta} )^2 ) \nonu
I_3 & = & I_2 \nonu
I_4 & = & 
 \frac{1}{(z-w)^2} 
 (2 (\Pi^c H_{cmn} -
\frac{1}{2} T_{mn}{}^{\ta} D_{\ta} )^2 ) \nonu
I_5 & = & \frac{(\zb-\wb)}{(z-w)^3} 
(- \Pi^d \Pb^a ( 
\frac{1}{2} R_d{}^m{}_{ma} +
\frac{1}{2} R_a{}^m{}_{md} + \del^m H_{amd} ) 
-\frac{1}{2} \Pi^{\td} \Pb^a R_{\td}{}^m{}_m{}^a 
\nonu
& & \qquad +
i\Pb^d T_{d,}{}^{\b\ddb}{}_{,(\b} \Pi_{\ddb)} 
+2 \Pb^d (\frac{1}{2} \del^m T_{dm}{}^{\ta} - 
H_{dme} T_{em}{}^{\ta} ) D_{\ta}  )
\nonu
I_6 & = & 
\frac{G(z,w)}{(z-w)^2} (-4 (\Pi^c H_{cmn} -
\frac{1}{2} T_{mn}{}^{\ta} D_{\ta} )^2 ) \nonu
& & + \frac{1}{(z-w)^2} 
 (6 (\Pi^c H_{cmn} -
\frac{1}{2} T_{mn}{}^{\ta} D_{\ta} )^2 ) \nonu
I_7 & = & 
\frac{1}{(z-w)^2} (\Pi^d \Pi^a (\frac{1}{2} R_{dcca} 
-4 H_{dec} H_{cea} ) \nonu
& & \qquad + \frac{1}{2} \Pi^{\td} \Pi^a R_{\td c c a} 
+ \Pi^d( H_{dec} T_{ce}{}^{\ta} + \frac{1}{2}
\del_c T_{cd}{}^{\ta} ) D_{\ta} ) \nonu
I_8 & = & I_7 \nonu
I_{9} & = & 
\frac{(\zb-\wb)}{(z-w)^3} 
(-i \Pb^c (\Pi_{\n} T_{c}{}^n{}_{\ddn} 
+\Pi_{\ddn} T_{c}{}^n{}_{\n} ) ) \nonu
&+&\frac{G(z,w)}{(z-w)^2} (+2 (\Pi^c H_{cmn} -
\frac{1}{2} T_{mn}{}^{\ta} D_{\ta} )^2 ) \nonu
&+& 
\frac{1}{(z-w)^2} ( -
(-\Pi^c H_{cmn} +\frac{1}{2}
 T_{mn}{}^{\ta} D_{\ta})
(-3 \Pi^c H_{cmn} +
\frac{1}{2} T_{mn}{}^{\ta} D_{\ta}) )
\nonu
I_{10} &  =&  \frac{(\zb-\wb)}{(z-w)^3} (
\Pi_m \Pb_a (\frac{1}{4} R_{mnna} + \frac{3}{4}R_{annm}
+\frac{1}{2} \del_n H_{nma} -\frac{1}{2} \del^{\tg} T_{am\tg} ) 
\nonu 
& & \qquad
+ \Pi^{\b} \Pb^a (
-\frac{1}{4} R_{m\b ma} + \frac{3i}{2} T_{am}{}^{\ddm}
\delta_{\b}{}^{\m} 
+\frac{1}{2} R_a{}^{\g}{}_{\b\g} -\frac{1}{4} F_{a\b} ) 
\nonu 
& & \qquad
+ \Pi^{\ddb} \Pb^a (
-\frac{1}{4} R_{m\ddb ma} + \frac{3i}{2} T_{am}{}^{\m}
\delta_{\ddb}{}^{\ddm} 
+\frac{1}{2} R_a{}^{\ddg}{}_{\ddb\ddg} +\frac{1}{4} F_{a\ddb} ) 
\nonu 
& & \qquad
+D^{\tg} \Pb^a (\del_m T_{ma\tg} + 2 H_{aem} T_{me\tg} ) ) 
\nonu &+&\frac{1}{(z-w)^2}
(\Pi_m \Pi_a (\frac{1}{8} R_{nmna} +\frac{3}{8} R_{anmn}
+ H_{nem} H_{nae} ) \nonu
& & \qquad
+\Pi_m \Pi^{\a} (\frac{3}{8} R_{\a nmn} + \frac{1}{8}
R_{n \a nm})  \nonu
& & \qquad
+\Pi_m \Pi^{\dda} (\frac{3}{8} R_{\dda nmn} + \frac{1}{8}
R_{n \dda nm}) 
\nonu 
& & \qquad
+ \Pi_m D^{\ta} \frac{1}{2} (H_{nem} T_{ne\ta} + \del_n T_{mn\ta} ) )
\nonu
I_{13} & = & 
\frac{G(z,w)}{(z-w)^2} (- (\Pi^c H_{cmn} -
\frac{1}{2} T_{mn}{}^{\ta} D_{\ta} )^2 ) \nonu
&+&\frac{1}{(z-w)^2}
 ( (\Pi^c H_{cmn} -
\frac{1}{2} T_{mn}{}^{\ta} D_{\ta} )^2 
+ 3i \Pi_{(\n|} \Pb^c T_c{}^n{}_{|\ddn)} )
\nonu
I_{14} & = & I_{9} \nonu
I_{15} & = & I_{10} \nonu
I_{18} & = & I_{13} \nonu
I_{19} & = &
\frac{G(z,w)}{(z-w)^2} ( -(\Pi^c H_{cmn} -
\frac{1}{2} T_{mn}{}^{\ta} D_{\ta} )^2 ) \nonu
&+&\frac{1}{(z-w)^2}
 ( (\Pi^c H_{cmn} -
\frac{1}{2} T_{mn}{}^{\ta} D_{\ta} )^2 \nonu
&& \qquad -\frac{13i}{6} (\Pi_{\ddn} T_m{}^n{}_{\n}
 + \Pi_{\n} T_m{}^n{}_{\ddn} ) \Pi_m )
\eea
Adding up all contributions we get
\bea
<T(z)T(w)>&=&\frac{\zb-\wb}{(z-w)^3}[\P^a \Pb^b(R_{bcca} 
+ \del_{\tg} T_{ba}{}^{\tg}) \nonu
&&\qquad +\P^\b \Pb^a(R_a{}^\g{}_{\b \g} -\half F_{a\b}) \nonu
&&\qquad +\P^{\ddb} \Pb^a(R_a{}^{\ddg}{}_{\ddb \ddg} +\half F_{a\ddb}) \nonu
&&\qquad +D^{\tg} \Pb^b(\del_a T_{ab\tg} -2H_{bde} T_{de\tg})] \nonu
&+&\frac{1}{(z-w)^2}[-{i \over 6} 
(\P^{\dda} \P^c T_{c,\b \dda}{}^\b +\P^{\a} \P^c T_{c,\a \ddb}{}^{\ddb})
\nonu
&& \qquad +4 \P^a \P^b H_{acd} H_{bcd}]
\eea
After some simplifications we obtain
\bea 
<T(z)T(w)>&=&\frac{\zb-\wb}{(z-w)^3}[\P^a \Pb^b R_{bcca} \nonu
&&\qquad+ \P^\b \Pb^a(-2\del_\a H_{\dda \b}) 
+ \P^{\ddb} \Pb^a (2\del_{\dda} H_{\ddb \a}) \nonu
&&\qquad +D^{\tg} \Pb^b(\del_a T_{ab\tg} -2H_{bde} T_{de\tg})] \nonu
&+&\frac{1}{(z-w)^2}[{1 \over 6} \P^\a \P^b \del_\b H_{\ddb \a}
-{1 \over 6} \P^{\dda} \P^b \del_{\ddb} H_{\dda \b}
\nonu
&& \qquad +4 \P^a \P^b H_{acd} H_{bcd}]
\label{ttfinal}
\eea

We now move to the dilaton sector. Since the dilaton corrections
(\ref{dgen-exp})
to the superconformal generators enter
with an extra $\a'$ one only has to compute tree graphs.
To $<G(z)G(w)>$ only diagrams (60) and (61) contribute, and
we find
\be
I_{60} = I_{61} = \frac{1}{(z-w)^3} (4\lambda^2 \a'^2 
e^{2i\rho} \eff D^2 \del^2 \eff).
\label{dilfieldeq}
\ee
To cancel this term, one can add a one-loop counterterm to the
action of the form $\Delta S \sim \int d^2 z \del^2 \eff
\Pi^{\a} \bar{\Pi}_{\a}$, which does not interfere with any
of the other one-loop calculations. If one, however, takes the
point of view that it should not be allowed to add one-loop
counterterms containing the dilaton to the theory, then one 
is forced to require $\del^2 \eff=0$.
Interestingly enough, we will see
later that the low-energy effective action does have
$\del^2 \eff=0$ as one of its equations of motion, so that either
point of view is consistent with our further results.

It is rather easy to check that there is no dilaton contribution to 
$<G(z) \bar{G}(w)>$. In the tree graph the pole due to the contraction
between $d_\a$ and $y^\a$ is cancelled by the $\r$ contribution
and the result is regular.

The dilaton contributions to $<T(z) G(w)>$ are as follows
(again a factor of $\l \a'$ has been extracted).
\bea
I_{61} &=& {1 \over (z-w)^2} (-2e^{i\r} D^\a \pa(\del_\a \eff)) \nonu
I_{62} &=& {1 \over (z-w)^3} 2 e^{i\r}\eff D^\a \del_\a \f \nonu
&+&{\zb-\wb \over (z-w)^3} 2e^{i\r}\eff D^\a \bar{\pa}(\del_\a \f) \nonu
I_{63} &=& {1 \over (z-w)^3}(-4e^{i\r} D^\a \del_\a \eff) \nonu
&&+{\zb-\wb \over (z-w)^3} (-4 e^{i\r} \bar{\pa}(D^\a) \del_\a \eff) \nonu
&&+{1 \over (z-w)^2}(-2)(\pa(e^{i\r}) D^\a \del_\a \eff +
e^{i\r} D^\a \pa (\del_\a \eff)
+ 2 e^{i\rho} (\dif D^{\a}) \del_\a \eff) \nonu
I_{65} &=& -I_{67}=4i \P^{\dda} \P_{\a \dda} \pa_w({1 \over z-w}
e^{i\r}\del^\a \eff) \nonu
I_{69} &=& {1 \over (z-w)^3}(2e^{i\r} D^\a \del_\a \eff) \nonu
&&+{1 \over (z-w)^2}(\pa (e^{i\r}D^\a\del_\a \eff)
 -2 \pa(e^{i\r})D^\a\del_\a \eff)
\eea
Here, diagram (69) refers to a contraction of the term $\frac{1}{2}\dif\r
\dif \r$ in $T$ with the dilaton term in $G$. 
Adding up these contributions we get that 
the cubic poles cancel,
$G_{\rm dil}$ has conformal weight $3/2$ and the remaining non-holomorphic
part is equal to 
\be
{\zb-\wb \over (z-w)^3} 2e^{i\r}\eff D^\a \Pb^b \del_b \del_\a \f
\label{tgdil}
\ee

We are now ready to write down the first field equation. 
As in the bosonic string the field equations are determined from the 
non-holomorphic part. The holomorphic part determines the counterterms
to the currents. From (\ref{tgfinal}) and (\ref{tgdil}) we obtain
the following field equation
\be
2 \del_\a H_{\dda \b} = \del_\b \del_a (\f + \fb). \label{eqtg}
\ee
 
The dilaton contribution to $<T(z) T(w)>$ is tabulated below
\bea
I_{62}+I_{63}&=&{1 \over (z-w)^2}2(-\half \pa^2 (\f+\fb)) \nonu
&&+ {\zb-\wb \over (z-w)^3}(2 \P^A \Pb^b \del_b \del_A (\f+\fb)
-\bar{\pa} \pa (\f+\fb)) \nonu
I_{64}&=&I_{65}={1 \over (z-w)^2} (-\half \P^b T_{bc}{}^{\ta} D_{\ta}
\del_c(\f+\fb)) \nonu
I_{66}&=&I_{67}= {1 \over (z-w)^2} (\half \P^b T_{bc}{}^{\ta} D_{\ta}
\del_c(\f+\fb)) \nonu 
&&\hspace{1.1cm}
+{\zb-\wb \over (z-w)^3}(-D_{\ta}\Pb^bT_{cb}{}^{\ta} \del_c(\f+\fb)
\nonu
&&\hspace{2.5cm}
+\P^a \Pb^b (T_{ab}{}^\g \del_\g \f + T_{ab}{}^{\ddg} \del_{\ddg} \fb))
\eea
Adding up these contributions we obtain that $T_{\rm dil}$ has conformal
weight $2$ plus the following non-holomorphic contribution
\be
{\zb-\wb \over (z-w)^3}(\P^A \Pb^b \del_b \del_A (\f+\fb)
+\P^a \Pb^b T_{ab}{}^{\tg} \del_{\tg} (\f+\fb) +
D^{\ta}\Pb^bT_{cb\ta} \del_c(\f+\fb)). \label{ttdil}
\ee

Thus, from (\ref{ttfinal}) and (\ref{ttdil}) follows that
the following three equations should hold
\bea
&&2 \del_\a H_{\dda \b} = \del_\b \del_a (\f + \fb), \label{fieldeq1}\\
&&R_{abbc} = -\del_a \del_c (\f + \fb) + T_{ac}{}^{\td} 
\del_{\td} (\f + \fb), \label{fieldeq2}\\
&&\del_a T_{ab}{}^{\tg} -2H_{bde} T_{de}{}^{\tg} = T_{ba}{}^{\tg} 
\del_a (\f + \fb). \label{fieldeq3}
\eea
Observe that (\ref{fieldeq1}) is precisely equation (\ref{eqtg})!
Even more, the last two equations follow from the first
as we now show!

Let us first introduce some notation.
Let $T_A = \del_A (\f + \fb)$.
We also define
\bea
&& \b_{\b a} = 2 \del_\a H_{\dda \b}; \ \ \ \ 
\b_{a c} = R_{abbc};  \nonu
&& \b^D_{b \tg} = \del_a T_{ab\tg} -2H_{bde} T_{de\tg}
\eea
In this notation equations (\ref{fieldeq1}) read
\bea  
&&\b_{\b a} = \del_a T_\b \label{eq1} \\
&&\b_{a c} = -\del_a T_c + T_{ac}{}^{\td} T_{\td} \label{eq2} \\
&&\b^D_{b \tg} =  T_{ba \tg} T_a \label{eq3}
\eea  

We shall show that (\ref{eq2}) follows from (\ref{eq1})
by differentiating once and (\ref{eq3}) by differentiating twice.
In the process we will need a few identities that we 
tabulate below.
\bea
&&[\del_{\dda}, \del_b] V_\a = -2i T_{ab}{}^\g V_\g, \label{id1} \\
&&[\del_{\a}, \del_b] V_\g = 2 C_{\g \a} \del_\b H_{\ddb}{}^\d V_\d, \\
&& \{ \del^{\a}, \del^{\dda} \} T_{\gamma}  =  -2i 
 (\del^{\a\dda} T_{\gamma} - 2 H^{\dda\b} T_\b C^{\a}{}_{\gamma}) \\
&&T_{\a\dda, \b\ddb, \g} - T_{\g\dda, \b \ddb, \a} = 
i C_{\a \g} \del_{\ddb} H_{\dda \b}, \\
&& \del^\a T_{\a\dda, \b\ddb, \g} = 2 \del_b H_{\dda \g} 
-i \del_{\dda} \del_{\b} H_{\ddb \g} \\
&& R_{abbc} = {1 \over 2}[i C_{\a \g} C_{\dda \ddg} 
\del^{\dde} \del^{\e} H_{\dde \e} 
+C_{\dda \ddg} \del_{(\a|\dde} H^{\dde}{}_{|\g)} 
-C_{\a \g} \del_{\e(\dda} H_{\ddg)}{}^\e \nonu 
&& \hspace{2cm}
+{i \over 2} ([\del_\a, \del_{\dda}] H_{\ddg \g} + \a \leftrightarrow \g +
\dda \leftrightarrow \ddg)], \label{curva}\\
&&\del_aT_{ac\ddd} = {1 \over 2} \del^{\a}[\{\del_\g, \del_{\ddg}\}H_{\ddd \a}
-\del_\a \del_{\ddg} H_{\ddd \g}] \nonu
&&\hspace{2cm}+4 H^{\dda \a} T_{\g\dda,\a\ddg,\ddd} 
-2i H_{\ddd}{}^\a \del_{\g} H_{\ddg \a} \label{delt} \\
&&\del^\a \del_{\ddb} \del^\g H_{\ddd \a} = 
\del^\a \del_{\ddd} \del^\g H_{\ddb \a}. \label{ident}
\eea
In these identities $V_{\ta}$ is assumed to have the $U(1)$ charge of its 
index, namely $1/2$ if the index is undotted and $-1/2$ if the index is 
dotted.
One may use the first three identities to derive how 
(anti-)commutators act on tensors.
However, one has to be careful when the $U(1)$ charge of the tensor is 
different from the one its indices indicate. 
Such an example is the 
torsion $T_{ab\g}$ that has $U(1)$ charge $-1/2$ and not $1/2$.
In such cases it is preferable to work out the commutators
directly. However,
when the tensor involved has the same $U(1)$ charge as its indices
the identities may be used safely. 

We are now ready to derive equations (\ref{eq2}) and (\ref{eq3})
from (\ref{eq1}).
Equation (\ref{eq2}) is derived by differentiating (\ref{eq1}) by
$\del_\a$ and then adding to the resulting equation the complex conjugate 
one. Then the result follows from (\ref{curva}).
To get equation (\ref{eq3}) we start by defining the following object
\bea
\O_{\a \b \ddb}&=&{i \over 2} [ \del_\b \del_{\ddb} T_\a
+ \del_{\ddb} \del_\b T_\a - \del_\a \del_\b T_{\ddb}
- \del_\a \del_{\ddb} T_\b] \nonu
&&+ 2 C_{\a \b} H_{\ddb}{}^\d T_\d 
\eea
Using the definition of $T_\a$ one may prove that $\O_{\a \b \ddb}$ is 
equal to zero. Nevertheless, we momentarily ignore this fact and treat
$T_a$ as a tensor that only satisfies (\ref{eq1}).
We compute the quantity $C^{\a \b} \del_c \O_{\a \b \ddb}$. The idea
is to commute $\del_c$ all the way to the right and then to use (\ref{eq1}).
After some rather tedious algebra that involves use of the identities
(\ref{id1})-(\ref{ident}) we get
\bea
C^{\a \b} \del_c \O_{\a \b \ddb}&=&
{i \over 2} C^{\a\b} (\del_\b \del_{\ddb} \b_{\a c}
+ \del_{\ddb} \del_\b  \b_{\a c} -
\del_\a \del_\b \b_{\ddb c} -
\del_\a \del_{\ddb} \b_{\b c}) \nonu
&& 4 H_{\ddb}{}^\d \b_{\d c} -2 T^{\a \dda}{}_{c,\ddb} \del_{(\dda} T_{\a)}
+ T_{\a \ddb,c}{}^\d \del^{(\a} T_{\d)}. 
\eea
Using now the definition of $T_A$ we see that the last term vanishes,
whereas the one but the last is equal to $4i T_{ac\ddb} \del_a (\f +\fb)$
which is just (up to the factor of $4i$) the right hand side of (\ref{eq3}).
To complete the proof we have to show that 
\be
\b^D_{b \tg} =  -{1 \over 8} C^{\a\b} (\del_\b \del_{\ddb} \b_{\a c}
+ \del_{\ddb} \del_\b  \b_{\a c} -
\del_\a \del_\b \b_{\ddb c} -
\del_\a \del_{\ddb} \b_{\b c}) + i H_{\ddb}{}^\d \b_{\d c}
\ee
This follows upon using (\ref{delt}) and (\ref{ident}).

Finally, let us rewrite the equation (\ref{fieldeq1}) in a form that 
it will appear in the next section. Observe that the left hand side of
(\ref{fieldeq1}) can be rewitten as $[\del_a, \del_\b](\f - \fb)$.
It follows that the field equation is equal to 
\be
\del_\g \del_a (\f - \fb)=0. \label{eqfinal}
\ee

Having discussed the field equations we now turn to the remaining 
holomorphic pieces in the OPE's. We already discussed the $<G(z)G(w)>$
previously. In the $<T(z)T(w)>$ in (\ref{ttfinal}) there are
several holomorphic terms, each of which can be easily
cancelled by adding the following one-loop counterterms to $T$
\be
T \to T - {1 \over 12}  \P^\a \P^b \del_{\b} H_{\ddb \a}
+{1 \over 12}   \P^{\dda} \P^b \del_{\ddb} H_{\dda \b} 
  -2  \P^a \P^b H_{acd} H_{bcd}.
\ee
This leaves an unwanted holomorphic term in $<G(z) \bar{G}(w)>$ of
the form (see (\ref{ggbi})
\be
{1 \over (z-w)} (\frac{2i}{\a'}
D^\a D^{\ddb} H_{\ddb \a})
\label{aux01}
\ee
and in $<T(z)G(w)>$ that looks like
\be
\frac{1}{(z-w)^2}[-3\P^a \del_\a H_{\dda}{}^\g D_\g  
+ 4i \P^{\dda} H_{\dda}{}^\g D_\g]
\label{aux02}
\ee
It seems impossible to cancel (\ref{aux02}) by means of a counterterm. 
The remaining possibility is then to try to modify the background
field expansion of $d$ in such a way as to cancel (\ref{aux02}).
Suppose we modify the background field expansion of $d$ as follows
\be
(d_{\a}+D_{\a}) \rightarrow (d_{\a} + D_{\a})(1+y^A y^B N_{BA}).
\label{propos}
\ee
A straightforward calculation reveals that by taking 
$N_{a\g}=N_{\g a} = \frac{3}{2} \del_{\a} H_{\dda \g }$ and
$N_{\a\dda}= - N_{\dda\a} = 2i H_{\dda a}$ one can indeed remove 
(\ref{aux02}) completely. A pleasant side effect of this new
background field expansion for $d$ is that it also at the same
time completely removes (\ref{aux01}), while it does not affect
any of the other results. While everything at one-loop now
seems to work, it would certainly be nice to have a further,
maybe even geometrical, understanding of the right background
field expansion of $d$. 

\subsection{Lorentz and $U(1)$ anomalies}
\label{anom}

In this section, we discuss the anomalies of the world-sheet
sigma model. The two symmetries of the theory, local Lorentz and
local $U(1)$ invariance, can be discussed at the same time.
We certainly expect the local Lorentz and $U(1)$ gauge transformations to be
anomalous, due to the presence of the `chiral' fields $d_{\tta},y^{\tta}$
which, from the world-sheet point of view, closely resemble
chiral fermions. Chiral fermions are the usual source of sigma
model anomalies \cite{MoNe}. In the RNS formalism of the heterotic
string, the local Lorentz and gauge anomalies of the low-energy
effective action are closely related to the world-sheet sigma model
anomalies and the latter can be cancelled by a Green-Schwarz
mechanism \cite{GrSc,HuWi,HuTo}. Since our sigma model is 
supposedly a reformulation of the RNS string, a similar anomaly
cancellation should be possible in our case as well, as we will
now demonstrate.

The local Lorenz anomaly comes only from diagrams with $d_{\tta},y^{\tta}$
loops and for the sake of a one-loop analysis it is sufficient
to look only at the kinetic term of $d_{\tta},y^{\tta}$ which we
write as 
\be \int d^2 z d_{\tta} (\delta_{\ttg}{}^{\tta} \dbar +
\Omega_{\ttg}{}^{\tta} ) y^{\ttg}
\ee
where $\Omega\equiv -\bar{\Pi}^B (\omega_{B \ttg}{}^{\tta}+
 w(\ttg) \Gamma_B \delta_{\ttg}{}^{\tta} ) $ 
contains the spin and $U(1)$ connection and 
depends only on the background fields. 
Local Lorentz and $U(1)$
transformations act on $d_{\tta},y^{\tta},\Omega_{\ttg}{}^{\tta}$
 as follows
\bea
\delta y^{\ttb} & = & \Lambda^{\ttb}{}_{\tta} y^{\tta} \nonu
\delta d_{\tta} & = & -d_{\ttb} \Lambda^{\ttb}{}_{\tta} \nonu
\delta \Omega_{\ttg}{}^{\tta} & = & \Lambda^{\tta}{}_{\ttb} 
\Omega_{\ttg}{}^{\ttb} - \Omega_{\ttb}{}^{\tta} \Lambda^{\ttb}{}_{\ttg}
 - \dbar \Lambda^{\tta}{}_{\ttg}
\eea
where $\Lambda$ is the parameter of the local Lorentz 
and $U(1)$ transformation
that depends only on the background fields. The full Lorentz 
and $U(1)$ transformation
rule for $d_{\tta}$ may involve further $y^A$-dependent terms, depending
on the choice of background expansion, but these do not contribute at
one loop. 

If we define an effective action by
\be 
e^{-\frac{1}{\a '}\Gamma[\Omega]} = \int Dd Dy 
e^{-\frac{1}{\a '} \int d^2 z d_{\tta} (\delta_{\ttg}{}^{\tta} \dbar +
\Omega_{\ttg}{}^{\tta} ) y^{\tta}  }
\ee
one obtains for its anomalous variation
\be \label{anomvar}
\delta \Gamma[\Omega] = - \alpha' \int d^2 z \dif \Lambda^{\tta}{}_{\ttg}
\Omega_{\tta}{}^{\ttg}.
\ee
To obtain this result one can e.g. use a point splitting regularization
as in \cite{Pol}. The full result for $\Gamma[\Omega]$ is 
proportional to  $\Gamma_{WZW}[g]$, the Wess-Zumino-Witten action,
where the group valued field $g$ is related to $\Omega$ through
$\Omega= g^{-1} \dbar g$, and is therefore 
a non-local functional of $\Omega$. One easily checks this correctly
reproduces (\ref{anomvar}). 

Our next task is to examine whether or not we can cancel (\ref{anomvar})
by a suitable transformation of the vielbein and the antisymmetric tensor.
The relevant equation reads
\be \label{centeq}
\delta\left( - \int d^2 z (
\frac{1}{2} \Pi^a \bar{\Pi}^a + D_{\tta} \bar{\Pi}^{\tta} + 
\frac{1}{2} \bar{\Pi}^A \Pi^B B_{BA} ) \right) 
 + \alpha' \int d^2 z \dif \Lambda^{\tta}{}_{\ttg}
\Omega_{\tta}{}^{\ttg} =0.
\ee
We will assume that the background fields are on-shell, in particular
that $\bar{\Pi}^{\tta}=0$. Furthermore we use
$\Omega_{\tta}{}^{\ttg} = -\bar{\Pi}^A \Omega_{A \tta}{}^{\ttg} $ and
will abbreviate $\dif_B \Lambda^{\tta}{}_{\ttg} 
(\omega_{A \tta}{}^{\ttg} + w(\ttg) \Gamma_A \delta_{\tta}{}^{\ttg})$
 by $\trr(\dif_B \Lambda \hat{\omega}_A)$.
Then we find from (\ref{centeq}) that
\be \delta \Pi^A = \Pi^B X_B{}^A, \qquad \delta\bar{\Pi}^A = 
\bar{\Pi}^B X_B{}^A \ee
with
\bea X_{ab} & = & -\frac{\a'}{2} \trr ( \dif_b \Lambda \hat{\omega}_a + 
 \dif_a \Lambda \Omega_b ) \nonu
 X_{\tta b} & = & -\a'  \trr ( \dif_{\tta} \Lambda \hat{\omega}_b + 
 \dif_b \Lambda \Omega_{\tta} ).
\eea
In order that $\delta \bar{\Pi}^{\tta}=0$, one has to choose $X_{a \ttb}=0$,
but $X_{\tta \ttb}$ can still be chosen arbitrarily. For the variation
of the antisymmetric tensor we find
\be
\delta B_{BA} = -\a' \trr (\dif_B \hat{\omega}_A - (-1)^{AB} \dif_A \Lambda
 \hat{\omega}_B) - X_B{}^S B_{SA} + (-1)^{AB} X_A{}^S B_{SB},
\ee
and with these variations (\ref{centeq}) is indeed
satisfied. 

What is the interpretation of these results? From the variations of 
the vielbein, we find that the metric $G_{MN}$ varies in exactly the
same way as $-\frac{\a'}{2} \trr ( \hat{\omega}_M \hat{\omega}_N)$. So instead
of introducing new variations of the vielbein, we could also have 
redefined the metric as $G_{MN} \rightarrow G_{MN} - 
\frac{\a'}{2} \trr (\hat{\omega}_M \hat{\omega}_N)$, 
which is the same as adding a local
counterterm to the world-sheet action. We have done an explicit calculation
of the one-loop OPE's without neglecting the explicit $U(1)$ connections,
and we found $U(1)$ noninvariant terms. All $U(1)$ noninvariant
terms could be cancelled by a counterterm in the action of the form
$\int d^2 z \Gamma^A \Gamma^B \Pi_B \bar{\Pi}_A$, in perfect agreement
with the general anomaly analysis presented above. 

The variation of the antisymmetric tensor field cannot be canceled by a
local counterterm. However, in the background field expansion all terms
containing at least two quantum fields contain only the field strength $H$,
and there is a unique local term one can add to $H$ in order to make
it into a covariant quantity under the modified transformation rule
of $B$,
\be \label{newH}
H_{MNP} \rightarrow H_{MNP} - \frac{\a'}{2} \trr
 (\hat{\omega}_{[M} \dif_N \hat{\omega}_{P\} } + \frac{2}{3} 
 \hat{\omega}_{[M} \hat{\omega}_N \hat{\omega}_{P\} }). 
\ee
In other words, one should add a 
super Chern-Simons three-form \cite{gn} to $H$ in order
to make it covariant. We know that with the modified transformation rules
discussed above, the Lorentz and $U(1)$
anomalies cancel, and therefore the result
obtained from a perturbative calculation that takes all the spin and
$U(1)$
connections properly into account, can also be obtained in a simpler way
by first ignoring all the spin and $U(1)$
connections and by subsequently performing
the substitution (\ref{newH}) in the final answer. In our computation 
we have ignored the spin and $U(1)$
connections, but we have only done a one-loop
calculation and the shift in (\ref{newH}) would therefore only
affect two-loop results. Altogether this shows that if we would have
taken the spin and $U(1)$
connections into account, we would have been able to 
get rid of all non-covariant terms by means of suitable local counterterms,
and the final field equations would still come out the same, and we
verified this explicitly for the $U(1)$ connections, as mentioned above.

Similarly, the inclusion of gauge field from the internal 
six dimensional space or the heterotic fermions would give rise to
a Chern-Simons term for the gauge fields in $H$, and these
Chern-Simons terms play a crucial role in the anomaly cancelation
through the Green-Schwarz mechanism.

\subsection{Standard $\beta$-function calculations}
\label{betafie}

In this section we briefly examine the relation between the diagrammatic 
results and a `standard' beta-function calculation. In a standard
beta-function calculation (see e.g. \cite{gsw1}, \cite{tseyt}, and
for the Green-Schwarz string \cite{beta10,beta2}), 
one couples the
sigma model to a world-sheet metric and then continues the
theory to $2+2\epsilon$ dimensions. If the $2+2\epsilon$-dimensional
metric is purely conformal, $g_{\a\b} \sim e^{\sigma} \delta_{\a\b}$,
then the action picks up an overall factor $e^{\epsilon\sigma}$.
In addition, the Fradkin-Tseytlin term
 $\int d^2 z \sqrt{g} R (\phi+\bar{\phi})$ becomes proportional to
$\int d^2 z (\phi+\bar{\phi}) \dif \dbar \sigma$, and it 
explicitly breaks conformal invariance. This is not a problem,
because the Fradkin-Tseytlin term is of higher order in $\a'$,
so that classically the sigma model is still conformally invariant.
If we now vary the action with respect to $\sigma$, we find that
$\delta S/\delta \sigma \sim \epsilon {\cal L} + \alpha' c \dif \dbar (\phi+
\bar{\phi})$, where ${\cal L}$ is the Lagrangian density, $S=\int d^2 z 
{\cal L}$, and $c$ is some constant. 
If the theory is UV divergent at one-loop, it should be 
renormalized with a counterterm $\frac{\a'}{\epsilon} {\cal L}_{\rm div}$.
Including such a counterterm, the condition for the theory to be
conformally invariant at one-loop becomes ${\cal L}_{\rm div} \sim
\dif \dbar(\phi+\bar{\phi})$. The only one-loop diagrams that
are UV divergent are those of the type of diagram (41), with the cross
indicating a vertex from the action. If we denote the relevant
vertices in (\ref{sexp2}) by $\int d^2 z y^A y^B C^{(2)}_{BA}$, we find
a UV divergent contribution $\frac{1}{\epsilon} C^{(2)}_{aa}$. Thus,
conformal invariance of the sigma model requires
$C^{(2)}_{aa} \sim \dif \dbar({\phi}+\bar{\phi})$. Using the equations
of motion of the background fields and the explicit form of the
vertices $C^{(2)}_{BA}$ in (\ref{sexp2}), we find that this
reproduces exactly the equations (\ref{fieldeq1})--(\ref{fieldeq3}).
Although this derivation is much simpler than the diagrammatic
one in the previous sections,
it is not clear to us whether these beta-function calculations
would also be sufficient to guarantee the full $N=2$ superconformal
invariance of the theory. As this example illustrates, the standard
beta-function calculation reproduces precisely the equations of
motion that are implied by requiring the presence of a Virasoro algebra
only, and it may be that at higher loop this is a weaker condition
than requiring a full $N=2$ invariance. Nevertheless, it would
certainly be a useful exercise to compute the two-loop beta-functions
of the sigma model, as these do provide necessary conditions for
$N=2$ superconformal invariance. If the number of two-loop equations
thus obtained would match the number of one-loop field equations,
this would be a strong indication that one did in fact
find the complete set of equations up to two-loop. These could then
be used to find the $\alpha'$ corrections to the supergravity action,
which is discussed in the next section. 

An alternative approach would be to first formulate the sigma model
as a sigma model for a $(2,0)$ super world-sheet, and to then do
a manifestly supersymmetric beta-function calculation (see e.g.
\cite{susybeta}). This would automatically guarantee the full
$N=(2,0)$ superconformal invariance. Unfortunately, no such
super world-sheet formulation if the sigma model is known to us.

\section{The supergravity action}
\setcounter{equation}{0}

In this section we derive the low energy effective action.
The fields that appear 
in the theory are the vierbein $E_M{}^A$ and the dilaton
complex chiral superfield $\f$. 
The anti-symmetric tensor $B_{AB}$ is not an indepent field,
as it is expressed purely in terms of the vielbein by means of
the tree-level constraints. We will 
find it convenient to define 
$\F=\exp \f$ and $\Fb=\exp \fb$, so that $\F$ and $\Fb$ have 
multiplicative $U(1)$ charge.

We have already seen that the field content of the low-energy theory
is that of `16-16 supergravity' \cite{16-16}. This supergravity 
can be expressed as a non-minimal
supergravity with an additional constraint (such that the 
number of off-shell
components of the non-minimal supergravity is reduced from 
40 to 32). Solving for this constraint one sees that `16-16 supergravity'
may be either old-minimal supergravity coupled to a tensor multiplet
or a formulation of
new-minimal supergravity coupled to a chiral multiplet \cite{wa1}.
(Now the counting of the number of off-shell components yields 24+8;
24 being the number of supergravity components and 8 the 
number of 
tensor or chiral
multiplet components).
The action that describes this non-minimal supergravity
theory is given by \cite{wa1} 
\be
S = \int E^{-1} G^{3n+1} (\F \Fb)^{-2n} \label{waction}
\ee  
where $G$ is a real linear multiplet and $\F$ is a chiral multiplet
(the $\F$ in (\ref{waction}) is equal to $(\F')^{3/2}$,
where $\F'$ is the chiral multiplet that appears in \cite{wa1};
this follows from the $U(1)$ charge of our $\F$ or equivalently from 
its conformal weight (which is $3/2$)).
Now in (\ref{waction}) either $G$ or $\F$ is a compensator field.
Which one of the two is the compensator 
depends on the sign of the kinetic term of the 
scalar in the linear and chiral multiplet.
 If $G$ is a compensator we are dealing with 
a new-minimal theory, whereas in case $\F$ is the compensator
the theory is old minimal. To determine the sign let us go to the 
gauge $\F=1$ and then shift $G \to G+1$. The factor $1$ gives the 
usual supergravity action. The term quadratic in $G$ has coefficient
$3n(3n+1)$. If this coefficient is positive then the 
kinetic term for the scalar in the 
tensor multiplet has
the correct sign and $\F$ is the compensator.
 
The constraints that the supergravity algebra satisfies imply that
we are in the ``string gauge'' $G=1$. In this gauge the field content of 
the tensor multiplet has been moved to the supergravity fields.
The antisymmetric tensor that couples in the sigma model is sitting in the 
supergravity multiplet and not in the tensor multiplet.
This follows from the sigma-model constraints
\be 
T_{ABc}+(-1)^{AB} 2 H_{ABc}=0.
\ee
In addition, this relation implies
\be 
H_{\a \ddb c} = -i C_{\a \g} C_{\ddb \ddg}
\ee
In rigid superspace the Bianchi identities for the tensor mupliplet 
give
\be 
H_{\a \ddb c} = -i G C_{\a \g} C_{\ddb \ddg}.
\ee
So we indeed get $G=1$.   

The action (\ref{waction}) in the string gauge takes the form
\be
S=\int d^4x d^4\theta E^{-1} (\F \Fb)^{-2n}. \label{sugra}
\ee
We emphasize that the fact that one can gauge away $G$ does not mean that 
$G$ is a compensating multiplet. It is only the signs of the kinetic terms 
that determine which multiplet is the physical one, and which is the
compensator.
 
To derive the field equations we will use a method developed by Wess and 
Zumino \cite{wezu}.
We define $\D_A{}^B=\d E_A{}^M E_M{}^B$. 
Then using
\be
\d E^{-1} = - E^{-1} (-1)^A \D_A{}^A.
\ee
the variation of the action is given by
\be
\d S= \int E^{-1}[-(-1)^A \D_A{}^A -2n (\d \f + \d \fb)](\F \Fb)^{-2n}.
\ee
However, $\D_A{}^A, \d \f$ and $\d \fb$ are not unconstrained superfields.
The variations of the superfields
should be along the constrained surface defined by the original constraints.
This imposes constraints on the variations.
In other words, given $\del_A$ that satisfies (\ref{bian1})-(\ref{bian4}),
$\del_A + \d \del_A$, where
\be
\d \del_A = \D_A{}^B \del_B + \d \o_{A\tb}{}^{\tg} M_{\tg}{}^{\tb}
+ \d \G_A Y,
\ee
should satisfy the same supergravity algebra to first order in the
variations.
We proceed to solve the constraints by
expressing the variations in terms of prepotentials. 

For the purpose of obtaining the field equations one need not solve
for all the variations. All that we need is to express 
$\D_a{}^a, \D_{\ta}{}^C, \G_{\ta}$ in terms of uncontrained
superfields. The need for $\D_{\dda}{}^C, \G_{\dda}$ follows
from the chirality of $\F$.  The variation of $\F$ should respect its
chirality
\be
0 = \d(\del_{\dda} \F) = \D_{\dda}{}^C \del_C \F - 
{1 \over 2} \d \G_{\dda} \F + \del_{\dda}(\d \F), \label{fvar}
\ee
and similarly for  $\D_{\a}{}^C, \G_{\a}$. 

We start by solving (\ref{bian1}) (from now on ``solving'' means
``solving at the linearized level''). It is rather easy to solve
all equation that arise. The results are 
\bea 
&&\D_\a{}^\b = \del_\a \j^\b - \j_\a{}^\b - {1 \over 2} C_{\a}{}^{\b} \j
\nonu
&& \D_\a{}^{\ddb} = \del_{\a} \bar{M}^{\ddb} \nonu
&& \D_\a{}^b = 2i C_{\a}{}^{\b} \bar{M}^{\ddb} + \del_\a L^b \nonu
&& \d \G_\a = \del_\a \j -4i H_{\ddb \a} \bar{M}^{\ddb} 
+ 2 \del_\b H_{\ddb \a} L^b \nonu
&& \d \o_{\a\g}{}^\d = \del_\a \j_\g{}^\d 
+ 2i \bar{M}^{\ddg} H_{\ddg(\g|}C_{\a|}{}^{\d)} 
- L^e \del_{\e}H_{\dde(\g|}C_{\a|}{}^{\d)} \nonu
&& \d \o_{\a\ddg}{}^{\ddd} = \del_\a \fb_{\ddg}{}^{\ddd}
-2i\bar{M}^{(\ddd}H_{\ddg)\a} -2i L_{\a\dde}\bar{W}^{\dde}{}_{\ddg}{}^{\ddd}
\nonu
&& \hspace{1.5cm} +{1 \over 6} L_{\a(\ddg|} \del^{\g|} H^{\ddd)}{}_\g
+{1 \over 2}(L^\e{}_{\ddg} \del_{(\a|}H^{\ddd}{}_{|\e)} 
+ \ddg \leftrightarrow \ddd) \label{eqbian1}
\eea
where $\j^\a, \j_\a{}^\b, \j, \bar{M}^{\dda}, L^a,  \fb_{\ddg}{}^{\ddd}$
are Lorentz irreducible superfields (e.g. $\psi_{\a}{}^{\a}=0$). 
At this point they are unconstrained but they
will be constrained when we will impose the rest of the supergravity
algebra. Notice that we have also solved for the spin connections.
We did this because they will be needed in the subsequent analysis.
The complex conjugate  equations 
are obtained from the above by just interchanging dotted with undotted
indices. Note, however, that since $\d \G_{\a} \to -\d \G_{\dda}$ 
(which follows from $Y^\dagger = -Y$, see Appendix A) we have
$\j \to -\bar{\j}$.

Next, the torsion part of (\ref{bian2}) implies 
\bea
\D_a{}^\b&=&{i \over 2}[\del_{(\a} \D_{\dda)}{}^\b + \d \o_{\dda \a}{}^\b
+{1 \over 2} \d \G_{\dda} C_\a{}^\b], \nonu
\D_{a}{}^b&=&{i \over 2}[\del_{(\a} \D_{\dda)}{}^b 
-2i \D_\a{}^\b C_{\dda}{}^{\ddb}
-2i \D_{\dda}{}^{\ddb} C_{\a}{}^{\b}], \label{eqbian2}
\eea
From the curvature and $U(1)$ 
parts of (\ref{bian2}) one can determine
$\d H_{\dda \a}, \d \o_{a \tb}{}^{\tg}, \d \G_a$. However, since 
we do not need these variations for the derivation of the field equations we 
will not solve the corresponding equations.

The commutator in (\ref{bian3}) does not have any torsion part.
This imposes certain constraints on the variations. Explicitly, 
we get that the following combinations should vanish
\bea
&&
-\del_b \D_\a{}^\b + \D_\a{}^c T_{cb}{}^\g + \del_\a \D_b{}^\g
-\d \o_{b\a}{}^\g - {1 \over 2} \d \G_b C_\a{}^\g = 0 \\
&& 
-\del_b \D_\a{}^{\ddg} + \D_\a{}^c T_{cb}{}^{\ddg}
+ \del_\a \D_b{}^{\ddg}=0 \label{biancon1}\\
&&
-\del_b \D_\a{}^{c}+\D_\a{}^d T_{db}{}^c
+\d \o_{\a \b}{}^\g C_{\ddb}{}^{\ddg}
+\d \o_{\a \ddb}{}^{\ddg} C_{\b}{}^{\g} 
+\del_\a \D_b{}^c+2i\D_b{}^{\ddg}C_{\b}{}^{\g} = 0 \phantom{xxxxxx}
 \label{biancon2}
\eea
The first equation will constrain the prepotentials for 
$\d \o_{a \tb}{}^{\tg}, \d \G_a$. Since we are not interested in them 
we will not discuss further this equation. Equations (\ref{biancon1})
and (\ref{biancon2}) constrain the prepotentials 
$\j^\a, \j_\a{}^\b, \j, \bar{M}^{\dda}, L^a,  \fb_{\ddg}{}^{\ddd}$
as we now show.

Inserting into (\ref{biancon1}) the results from (\ref{eqbian1}) and
(\ref{eqbian2}) and after some rather tedious algebra we get that
certain combinations of the prepotentials are linear superfields.
We solve these constraints to get 
\bea
&&\del^{\dda}(\bar{\j}_{\dda} - \bar{M}_{\dda}) + \j - \bar{\j} = 
\del^\a \L_\a \nonu
&& \del_{(\dda}(\bar{\j}_{\ddb)}- \bar{M}_{\ddb)}) 
+ 2 (\fb_{\dda \ddb} - \jb_{\dda \ddb}) = \del^\a K_{\a \dda \ddb}
\label{linear1}
\eea
where $\L_\a, K_{\a \dda \ddb}$ are (at this point) unconstrained
superfields. The calculation involves a rather delicate cancellation
among all terms that are not linear in the fields or in the variations. 
One may anticipate such a cancellation since
the solution of the full non-linear Bianchi's involves $H_{\dda \a}$
and $W_{\a \b \g}$ only linearly. 

Procceding in a similar way we get from (\ref{biancon2})
(again after tedious algebra and many cancellations)
\bea 
&&{i \over 2} \del_{\dda}(L^a-\Lb^a) - \L^\a +2(M^\a - \j^\a)
=\del^\b \q_{\b}{}^\a + \del^\a \q \\
&&K^\a{}_{\ddb}{}^{\ddg} + {i \over 2} \del_{(\ddb|} 
(L^{\a|\ddg)}- \Lb^{\a|\ddg)}) = \del^\b \S_\b{}^\a{}_{\ddb}{}^{\ddg}
+ \del^\a  \S_{\ddb}{}^{\ddg}
\label{linear2}
\eea 
where $\q_{\b}{}^\a, \S_\b{}^\a{}_{\ddb}{}^{\ddg}, \S_{\ddb}{}^{\ddg}$
are Lorentz irreducible.

We see that these equations only depend on certain differences of 
superfields. We define
\be
\hj=\j - \jb; \ \ 
\hf_{\a \b} = \f_{\a \b} - \j_{\a \b}; \ \ 
\hL^{a}=L^a - \Lb^a; \ \ 
\hM^\a=M^\a - \j^\a.
\ee
In terms of these fields the equations (\ref{linear1}), (\ref{linear2})
and their complex conjugate read
\bea
&&\del_{\dda} \hMb^{\dda} + \hj = \del^\a \L_\a \label{psi1} \\
&&\del_{\a} \hM^{\a} + \hj = \del^{\dda} \Lab_{\dda} \label{psi2} \\  
&&{i \over 2} \del_{\dda}\hL^{a} - \L^\a + 2 \hM^{\a}=\del^\b \q_\b{}^\a 
+ \del^\a \q \label{la1} \\
&&-{i \over 2} \del_{\a}\hL^{a} - \Lab^{\dda} + 2 \hMb^{\dda}=
\del^{\ddb} \qb_{\ddb}{}^{\dda} 
+ \del^{\dda} \qb \label{la2} \\
&&-\del_{(\dda} \hMb_{\ddb)} + 2 \hfb_{\dda \ddb} 
= \del^\a K_{\a \dda}{}^{\ddb}
\\
&&-\del_{(\a} \hM_{\b)} + 2 \hf_{\a \b} = \del^{\dda} \Kb_{\dda \a}{}^{\b}
\\
&&K^\a{}_{\ddb}{}^{\dda} 
+ {i \over 2} \del_{(\ddb|} \hL^{\a|\dda)} = 
\del^\b \S_\b{}^\a{}_{\ddb}{}^{\ddg}
+ \del^\a  \S_{\ddb}{}^{\ddg} 
\\
&&\Kb^{\dda}{}_{\b}{}^{\a} 
- {i \over 2} \del_{(\b} \hL^{\a)\dda} = 
\del^{\ddb} \Sib_{\ddb}{}^{\dda}{}_{\b}{}^{\g}
+ \del^{\dda} \Sib_{\b}{}^{\g}.
\eea 
The last four equations are irrelevant for the derivation of the 
field equations and, therefore, we will not deal with them any further.  
The first four equations constrain the superfields $\hj, \hM, \L^\a,
\q_{\a \b}, \q$. We now solve them in terms of new prepotentials
(prepotentials-for-prepotentials). 
Equations (\ref{la1}) and (\ref{la2}) can be used to solve for 
$\L^\a$ and $\L^{\dda}$. Eliminating $\hj$ from (\ref{psi1}) and
(\ref{psi2}) we get
\be
-\del^\a \hM_\a + \del^{\dda} \hMb_{\dda} =
-\del^\a({i \over 2} \del^{\dda} \hL_a  + \del_\a \q)
+\del^{\dda}(-{i \over 2} \del^{\a} \hL_a  
+ \del_{\dda} \qb)
\ee
The general solution of this equation is a particular solution $\hM^1_\a$ plus 
the most general solution $\hM^0_\a$ of the homogeneous equation.
It is very easy to find a particular solution
\be
\hM^1_\a = {i \over 2} \del^{\dda} L_a + \del_\a \q
\ee
The general solution of the homogeneous equation is given by
\be
\hM^0_\a = \bar{\del}^2 \del_\a V + \del^\b \c_{\b \a} + \del^2 \c_\a.
\label{homosol}
\ee
where $V$ is a real unconstrained superfield and $\c_{\a \b}, \c$
are Lorentz irreducible unconstrained complex supefields.
To prove the last statement we rewrite the equation as
\be 
\del^\a \hM_\a =  \del^{\dda} \hMb_{\dda}
\label{homo}
\ee
The general solution of this equation is a particular solution
plus a general element in the kernel of $\del_\a$. 
The last two terms in (\ref{homosol}) are clearly the most general
elements in the kernel of $\del^\a$.
It remains to show that the first term in the right hand side of 
(\ref{homosol}) is a particular solution of  (\ref{homo}).
This follows from the identity
\be
\del^\a \bar{\del}^2 \del_\a V = \del^{\dda} \del^2 \del_{\dda} V,
\ee
which  can be proven by direct computation. 
We have, thus, shown that
\be
\hM_\a = {i \over 2} \del^{\dda} L_a + \del_\a \q + 
\bar{\del}^2 \del_\a V + \del^\b \c_{\b \a} + \del^2 \c_\a
\label{msol}
\ee
Furthermore, 
\bea
\L^\a &=& -{i \over 2} \del_{\dda} \hL^a + 2 \bar{\del}^2 \del^\a V
+ \del^\a \q + \del^\b(2 \c_{\b}{}^\a - \q_{\b}{}^\a) +2 \del^2 \c^\a,
\nonu
\hj &=& {i \over 2} [\del_\a, \del_{\dda}]L^a +3 \del^\a \bar{\del}^2 \del_\a V
+\del^2 \q + \bar{\del}^2 \bar{\q}. \label{psisol}
\eea

Next, we are going to express the variation of the dilaton superfield in terms 
of unconstrained superfields. We need  to solve equation (\ref{fvar}).
This is rather straightforward. The result is
\be
\d \F = \del^2 U - M^\a \del_\a \F - \Lb^b \del_b \F 
+{1 \over 2} \bar{\j} \F,
\ee
where $U$ is a complex unconstrained superfield.

We are now almost ready to derive the field equations.
We first need one more fact, namely that our supergravity 
algebra allows naive partial integrations.
This follows from the identity
\be
E^{-1} \stackrel{\leftarrow}{\del}_A = - E^{-1} (-1)^B T_{AB}{}^B
\ee
The proof of this identity can be found in 
\cite{book} p. 282, and it will not be repeated here.
Since in our supergravity algebra $(-1)^B T_{AB}{}^B = 0$ 
naive partial integrations are allowed.

The easiest equation to derive is the dilaton field equation. Integrating
by part the derivatives from $U$ and using the fact that $\Fb$ is 
anti-chiral we get 
\be
\del^2 \F = 0, \label{fieq1}
\ee
or in terms of $\f$
\be
\del^2 \f + \del^\a \f \del_\a \f= 0. \label{fieq1'}
\ee

We now move on to derive the remaining equations.
Direct computation gives
\bea
-(-1)^A \D_A{}^A &=& 2 [\del_\a (M^\a-\j^\a) 
+ \del_{\dda} (\Mb^{\dda}-\jb^{\dda})] + \j - \jb \nonu
&&-{i \over 2}(\del_\a \del_{\dda}\Lb^a+\del_{\dda} \del_\a L^a) 
+\del_\a \j^\a + \del_{\dda} \jb^{\dda} 
\label{det}
\eea
We see that some terms appear in the combinations of differences
that we have already seen but not all of them. However, 
an interesting conspiracy happens when we take into account the dilaton terms.
Integrating by parts the $M$-term in $\d \f$ the resulting $\del_\a M^\a$
combines with the bare $\del_\a \j^\a$ in (\ref{det}) to $\del_\a \hM^\a$.
Similarly, by integrating by parts the $L$ terms in 
$\d \f$ and $\d \fb$ we find that the $L$ terms also organize themselves
into $\hL^a$ (more precisely, we integrate by parts
the term $-\frac{1}{2}\Lb^b \del_b \F$, 
the other half we keep as it is; similarly
for the $L$ term in $\d \fb$). At this point we have
\be 
\d S = \int E^{-1}[\del_\a\hM^a+\del_{\dda} \hMb^{\dda}+
(1+n)\hj+{i \over 4} [\del_\a, \del_{\dda}]L^a
-n L^a \del_a(\f-\fb)](\F \F)^{-2n}
\ee
Inserting now (\ref{msol}), (\ref{psisol}) we get
\bea
\d S&=&\int E^{-1}[ {i \over 4} (2n+1)[\del_\a, \del_{\dda}]\hL^a
-n \hL^a \del_a(\f-\fb) +\nonu
&&
+n (\del^2 \q +\bar{\del}^2 \qb)    
+(1+3n) \del^\a \bar{\del}^2 \del_\a V](\F \Fb)^{-2n}
\phantom{xxxx} \label{varfinal}
\eea
Clearly, $\q$ and $\qb$ do not give any new equations.
For $n=-1/2$ the first term in (\ref{varfinal}) drops out and we get the 
following
field equation
\be
\del_a (\f-\fb) = 0, \label{fieq2}  
\ee
Clearly, this equation implies (\ref{eqfinal}). 
Equation (\ref{eqfinal}), however, is a slightly weaker condition than
that given by (\ref{fieq2}). We think that it is very unlkely that
it is possible to write down a different action
that gives directly (\ref{eqfinal}) as its field equation instead 
of just (\ref{fieq2}). It is clear though that in the class of theories
described by an action of the form (\ref{sugra}) the $n=-1/2$ theory
is the only one that can be the low energy effective action
of the heterotic superstring.
Furthermore, we expect that a two-loop
calculation will confirm the value $n=-1/2$. 

For $n=-1/2$  (which is the  $n=0, \tilde{n}=-1/2$ case in the 
language of  \cite{book}) $3n(3n+1)$ is positive, and therefore, the
low energy theory is old-minimal supergravity 
coupled to a tensor multiplet. The fact that the low energy effective 
action of the heterotic string is old-minimal supergravity 
was already predicted in \cite{wa2} using string field theory
arguments, and that it is in particular $n=-1/2$ 
`16-16 supergravity' in \cite{wa3}. Properties of low energy
effective actions similar to ours have been studied in \cite{action},
although the dilaton was incorrectly assumed to be part of $G$ rather
than $\Phi$.

It remains to determine the field equation that follows from $V$.
It turns out that this equation follows from the other two as 
we now describe.
Integrating by parts the derivatives from $V$ and using that $\F$ is 
chiral and linear (see (\ref{fieq1})) we get
\be
(\del_\a \del_{\ddb} \del^{\ddb} \del^\a \F) \Fb +
2 (\del_\a \del^{\ddb} \del^\a \F) (\del_{\ddb} \Fb) 
+2 (\del^{\ddb} \del^\a \F) (\del_\a \del_{\ddb} \Fb)
=0
\ee 
Adding to this equation its complex conjugate we get
\be
\del^a \f \del_a \fb - H^a H_a + {i \over 4}[\del_\a, \del_{\dda}]H^{\dda \a}
=0
\label{fieq3}
\ee
To get this result one has to use the following two identities
\bea
&& \del_\a \del^{\dda} \del^\a \f = 0 \nonu
&&{1 \over 2}(\del_\a H^{\dda \a} \del_{\dda} \fb -
\del_{\dda} H^{\dda \a} \del_{\a} \f) 
- H^{\dda \a} \del_{\dda} \fb \del_\a \f = 0. 
\eea
that both follow from (\ref{fieq2}).

It is now relatively straightforward to prove that (\ref{fieq3}) follows
from (\ref{fieq1}) and (\ref{fieq2}).
Equation (\ref{fieq2}) can be rewritten as
\be
H_{\dda \a} = - {i \over 4} [\del_\a, \del_{\dda}](\f + \fb) \label{eq2new}
\ee
Acting with $[\del_\a, \del_{\dda}]$ one gets
\be 
[\del_\a, \del_{\dda}]H^{\dda \a} = - i \del^a \del_a (\f + \fb) \label{eq2d}
\ee
Similarly by acting with $\del^{\dda} \del^\a$ on the dilaton field equation
we get
\be
\del^a \del_a (\f + \fb) + 2 \del^a \f \del_a \f - 2 H^a H_a - 
{i \over 2} [\del_\a, \del_{\dda}]H^{\dda \a} = 0.
\label{eq1d}
\ee
From  (\ref{eq2d}) and (\ref{eq1d}) one immediately gets (\ref{fieq3}).

\begin{figure}
\epsfig{figure=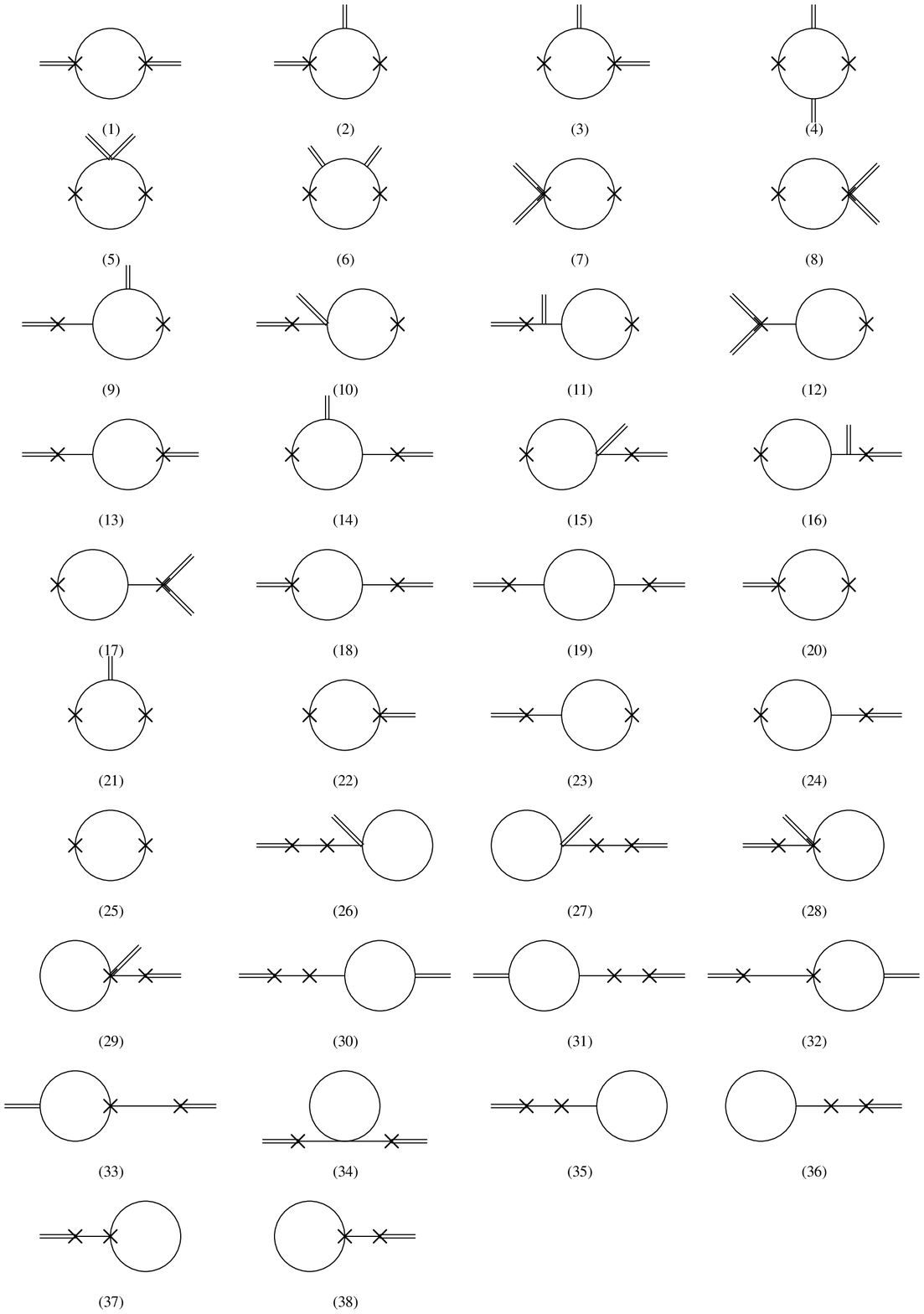,height=9in, width=6.25 in}
\end{figure}

\begin{figure} \label{figure}
\epsfig{figure=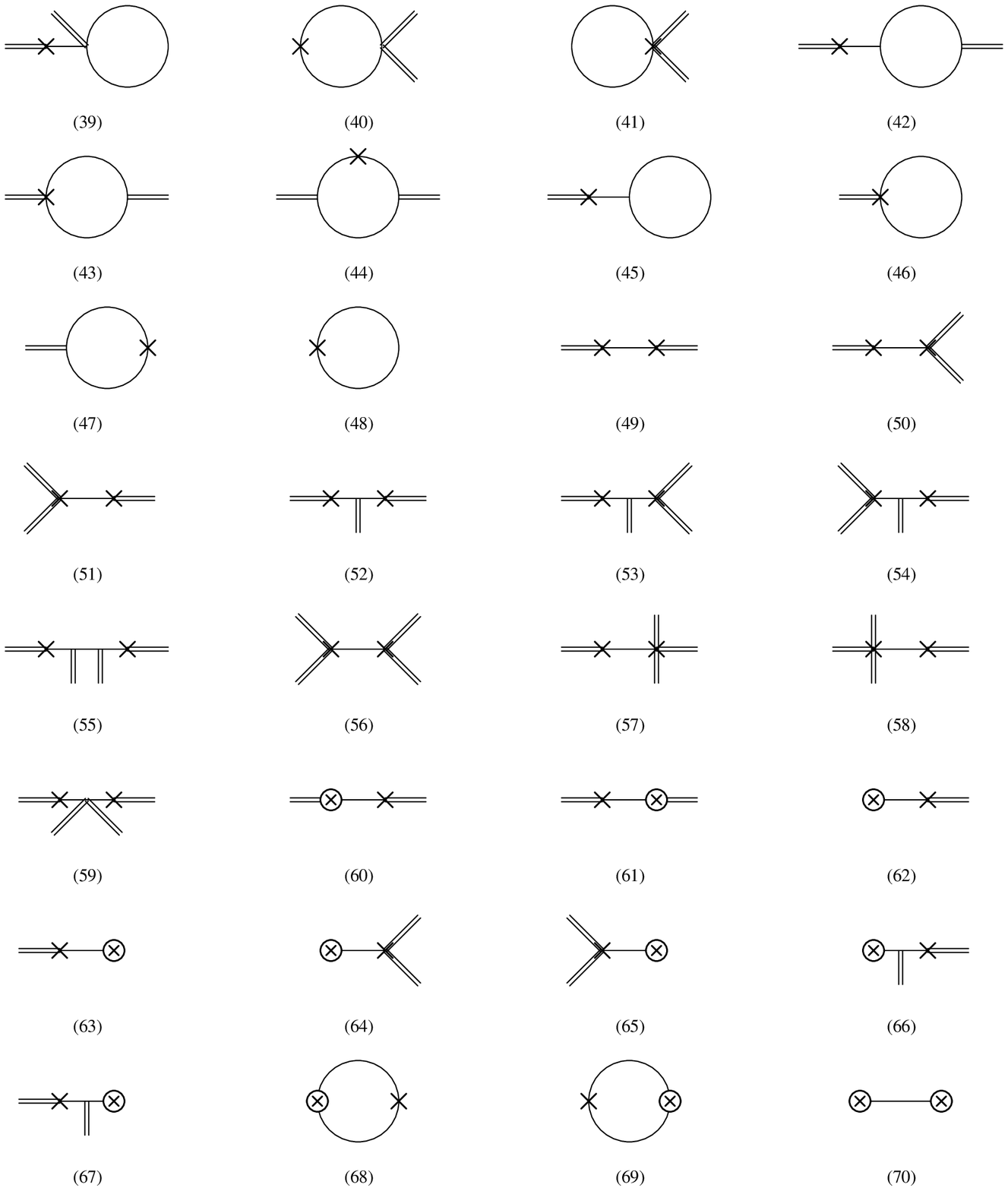,height=7.75in, width=6.25 in}
\caption{A table of Feynman diagrams; crosses ($\times$)
indicate vertices from
the non-dilaton parts of $T$,$G$, and $\bar{G}$; encircled crosses ($\otimes$)
indicate vertices coming from their dilaton parts; double lines
indicate background fields and single lines indicate quantum fields.
Whenever the diagram is a contribution to  $<A(z)B(w)>$, the leftmost
cross will always indicate a vertex from $A$, the rightmost cross a
vertex from $B$.}
\end{figure}

\section{Conclusions}
\setcounter{equation}{0}

We have performed a one-loop computation to determine
the low energy effective action of the heterotic superstring.
This was done by first coupling the Berkovits string to a curved
background and then requiring that it has an $N=2$ superconformal invariance
at the one-loop level. The tree-level analysis gave a set of constraints
that together with a maximal set of conventional constraints
determined completely the supergravity algebra through the Bianchi
identities.

Computing the OPE's of the superconformal algebra at the one-loop level
led to extra terms in the right hand side of the OPE's.
The extra holomorphic terms could be cancelled by a particular
and ill-understood choice of counterterms and background field
expansions. The non-holomorphic terms led to the field equations.
We found field equations only in the OPE's of $T$ with $G$ and
$T$ with $T$. This is accordance with the strange $\a'$ factors that
appears in the definition of $G$ and $\bar{G}$.
The one-loop calculation of $<G(z) \bar{G}(w)>$ yields only the
tree-level part of $T$,
and should therefore not contain any field equations
(similarly for $<G(z) G(w)>$, $<~\bar{G}(z) \bar{G}(w)>$).
The calculation of $<T(z) T(w)>$ led to three equations. However,
only one of those is an independent equation, namely
\be
\del_\g \del_a (\f - \fb)=0. \label{eqcon}
\ee
The other two
can be obtained by differentiating (\ref{eqcon}).
Vanishing of the
non-holomorphic part of $<T(z) G(w)>$ requires that precisely 
the same equation holds.

We then studied the problem of finding the supergravity action that
has (\ref{eqcon}) as its field equation. Starting with a general
form of the supergravity action (\ref{sugra}) we obtained that {\em only}
the $n=1/2$ case
\be
S=\int d^4x d^4\theta E^{-1} \F \Fb \label{sugra1}
\ee
is compatible with the string calculation.
Hence, we conlude that this is the low energy effective action.
This action describes old-minimal supergravity coupled to a tensor
multiplet.
Notice, however, that instead of equation (\ref{eqcon})
we find a slightly stronger field equation, namely
\be
\del_a (\f-\fb) = 0.
\label{abcde}
\ee
Another way to check that $n=1/2$ is consistent
is to look at the precise coefficient
of the dilaton term in the sigma model action, and to use that
one the sphere $\frac{1}{4}\int d^2 z R \sqrt{g} = -2$. 

The action (\ref{sugra1} leads to two more field equations
\bea
\del^2 \f + \del^\a \f \del_\a \f & = &  0 \\
\del^a \f \del_a \fb - H^a H_a + {i \over 4}[\del_\a \del_{\dda}]H^{\dda \a}
& = & 0
\eea
We saw a glimpse of the first one in the computation of
$<G(z)G(w)>$, and the second is a consequence of the first one and 
(\ref{abcde}). Both will probably follow from a two-loop calculation.

There are many different extensions and applications of our results.
One extension is to include
the compactification-dependent and the Yang-Mills background fields.
This extension should be straightforward since all our methods are
directly applicable.
One may also try to extend our results to two loops. 
This is, however, a much more difficult task since the number of graphs
one would need to compute grows enormously and, furthermore, one
runs into problems of renormalization. It would be desirable
to have a better understanding of the counterterms and different
background field expansions for $d$ before starting such a two-loop
calculation. At this moment, the easiest way to obtain two-loop
information seems to be to compute the conventional beta-functions,
as discussed in section~4.4, at two loops. 
Further generalizations include the case of the type II string. 
Since the sigma model treats Neveu-Schwarz fields on the same
footing as Ramond fields, the sigma model might provide interesting
information about the geometry of
the still, especially in the case of the type IIB
string, mysterious Ramond fields. It should also be possible
to generalize the covariant sigma model description of the heterotic
string to the case of $N=(1,2)$ heterotic strings. These string
theories have recently attracted some attention because they may
be of direct relevance to $M$-theory \cite{mart}. Another interesting
possibility is to consider $D$-branes in an open type II
version of the sigma model. Due to the simple nature of the Ramond
fields, the Ramond charges carried by a $D$-brane should be clearly
visible. In a somewhat 
different direction our results might shed light in the relation
between target space supersymmetry and $T$-duality, and lead
to new off-shell formulations for certain supergravity theories.

All in all, we believe that this world-sheet point of view
of string theory should provide a fruitful arena for further investigations.

\section*{Acknowledgements}

We would like to thank E. Sezgin, K. Stelle, 
and P. van Nieuwenhuizen for collaboration at the early
stages of this work, and M. Ro\v{c}ek and especially 
N. Berkovits and W. Siegel for very many helpful discussions
during the various stages of this project. We would further like
to thank G. Papadopoulos for a discussion on background field
expansions, and M.B. Halpern for discussions on the Feynman rules
we used.  This work was supported in part by
the National Science Foundation under grant PHY93-09888.

\appendix

\section{Conventions}
\renewcommand{\theequation}{A.\arabic{equation}}
\setcounter{equation}{0}

In this appendix we present the conventions used in this paper.
We use a two component notation. Our conventions are the ones of 
\cite{book}. For completeness we summarize the relevant points here.
Indices from the first part of the alphabet (Greek or Roman) are 
target space (or flat)  indices and from the middle part of the alphabet
are curved indices. Greek letters are reserved for spinor two components 
indices 
(for example, $\psi^\a$ is a two component spinor with $\a=+, -$) 
and Roman ones for vector indices. Each vector index is equivalent to
one undotted and one dotted index ($\phi^a = \phi^{\a \dda}$).
Capital indices denote both a spinor (dotted and undotted) and vector index, 
whereas a tilde on a Greek index denotes both a
dotted and undotted spinor index.
Indices are raised and lowered using 
an $sl_2$ invariant antisymmetric two dimensional matrix
$C_{\a \b}$. Since $C_{\a\b}$ is antisymmetric, we have to specify
how exactly we use it to raise and lower indices, and our convention
is the so-called
`down-hill'
rule from left to right for both
the undotted and the dotted sector. 

To illustrate the above we give couple of examples:
\bea
&\ & \j^A \c_A = \j^a \c_a + \j^{\ta} \c_{\ta} = 
\j^a \c_a + \j^{\a} \c_{\a} + \j^{\dda} \c_{\dda}  \\
&\ & \j^\a C_{\a \b} = \j_\b; \ \ C^{\a \b} \j_\b = \j^\a
\eea
In addition, we have the following identity
\be
C_{\a \b} C^{\g \d} = \d_{[\a}{}^{\g} \d_{\b]}{}^{\d},  \label{C-C}
\ee
where the square brackets indicate antisymmetrization. Similarly, parenthesis
indicate symmetrization, and square bracket from the left and parenthesis
from the right indicate graded antisymmetrization.
Our convention for (anti)-symmetrization is without
any additional factors, i.e.
\be
A_{(\a} B_{\b)} = A_\a B_\b + A_\b B_\a. 
\ee
Furthermore, indices between straight lines do not participate in 
(anti)-symmetrization. 

From (\ref{C-C}) we get
\bea
&\ & C^{\a \b} C_{\a \b} = \d_\a{}^{\a} = 2 \\
&\ & C_{\a \b} \c_\g - C_{\a \g} \c_\b  = - C_{\b \g} \c_\a \\
&\ & \c_\a \c_\b = - \frac{1}{2} C_{\a \b} \c^\g \c_\g
\eea
The last identity is an example of a Fierz identity. 
Finally, although we are not going to need it, we quote the explicit 
form of the $C$-matrices
\be 
C_{\a \b} = C^{\ddb \dda} = C_{\dda \ddb} = \s_2
=\left( \begin{array}{cc} 0 & -i \\ i & 0 \end{array} \right),
\ee
where $\s_2$ is the Pauli matrix.

Our conventions for covariant derivatives, torsion, curvature, etc. 
are as follows
\begin{equation}
\del_A = E_A{}^{M} \pa_M + \ome_{A \b}{}^{\g} M_{\g}{}^{\b}
+ \ome_{A \ddb}{}^{\ddg} M_{\ddg}{}^{\ddb} + \G_A Y, \label{del} 
\end{equation}
\be
[ \del_A, \del_B \} = T_{AB}{}^{C}\del_C 
+ R_{AB\g}{}^{\d} M_{\d}{}^{\g} 
+ R_{AB\ddg}{}^{\ddd} M_{\ddd}{}^{\ddg} 
+ F_{AB} Y, \label{com}
\ee
where $E_A{}^{M}$, $\ome_{A \tb}{}^{\tg}$ and $\G_A$ are the vielbein,
the spin connection and the $U(1)$ connection, respectively. 
$T_{AB}{}^{C}$,  $R_{AB\tg}{}^{\td}$ and $F_{AB}$ are the torsion,
the curvature tensor and the $U(1)$ curvature, repectively. 
$M_{\td}^{\ \tg}$ are the 
generators of the Lorentz group. They are symmetric in $\a, \b$ and they 
act as follows
\be 
[ \l_{\a}{}^{\b} M_{\b}{}^{\a}, s_\g ] = \l_{\g}{}^{\a} s_\a, 
\ee
Finally, $Y$ is the antihermitian generator of $U(1)$. The $U(1)$ charge of 
the various fields is given in the main text. In addition, we
sometimes use $w(A)$ to indicate the $U(1)$ weight of the index $A$,
which is defined by $w(\a)=+\frac{1}{2}$, $w(\dda)=-\frac{1}{2}$, and
$w(a)=0$. 
Let us also 
give the explicit dependence of the torsion on the vierbein and the 
connections
\be
T_{AB}{}^{C} = E_{[A}{}^{M} \pa_M E_{B)}{}^{N} E_N{}^{C}
+ \ome_{[AB)}{}^{C} + w(C) \G_{[A} \d_{B)}{}^{C}. \label{torsion}
\ee
 
Both the spin connection and the curvature are reducible with 
respect to the Lorentz group, in order to preserve (\ref{adota}).
In particular, one has
\bea
\ome_{A b}{}^{c} &=& \ome_{A \b}{}^{\g} \d_{\ddb}{}^{\ddg} + 
\ome_{A \ddb}{}^{\ddg} \d_{\b}{}^{\g} \label{ir1} \\
\ome_{A \b}{}^{c} &=& \ome_{A \b}{}^{\ddg} = 0 \label{ir2} \\
R_{A B c}{}^{d} &=& R_{A B \g}{}^{\d} \d_{\ddg}{}^{\ddd} + 
R_{A B \ddg}{}^{\ddd} \d_{\g}{}^{\d} \label{ir3} \\
R_{A B c}{}^{\d} &=& R_{A B \g}{}^{\g} 
= R_{A B \g}{}^{\d} = 0. \label{ir4}
\eea 

Finally, we use a version of hermitian conjugation, which we denote
by $c.c.$, and which acts as follows on the various objects
\be
(\del_{\a})^{\dagger}  =  \del_{\dda} ;\quad \quad
(C_{\a\b})^{\dagger}  =  C_{\dda\ddb} ;\quad \quad
(M_{\gamma}{}^{\delta})^{\dagger}  =  M_{\ddg}{}^{\ddd}  ;\quad \quad
(Y)^{\dagger}  =  -Y  \label{conjug1} ;
\ee\be
(\del_{a})^{\dagger}  =  -\del_{a}  ;\quad \quad
(H_{\a\ddb} )^{\dagger}  =  H_{\b\dda}  ;\quad \quad
(W_{\a\b\g})^{\dagger}  =  \bar{W}_{\dda\ddb\ddg}. 
\label{conjug2}
\ee

\section{Solving the Bianchi identities}
\renewcommand{\theequation}{B.\arabic{equation}}
\setcounter{equation}{0}

In this appendix we give some details of the solution of the Bianchi
identities. 
There are two different Bianchi identities that we will solve:
the Bianchi identity
associated with the antisymmetric tensor $B_{MN}$ and the 
Bianchi's associated with the covariant derivatives.
The first one just expresses the fact that $H_{LMN}$ is a closed 
3-form (notice that the indices in $H_{LMN}$ are curved indices).
The second ones follow from the Jacobi identities for the covariant 
derivatives.

Let us start from the latter. The equation we need to solve is the 
following \cite{book}
\be
\D_{ABC}{}^{D} = R_{[ABC)}{}^{D} 
+ w(D) F_{[AB} \d_{C)}{}^{D},
\label{bianchi}
\ee
where
\be
\D_{ABC}{}^{D} = \del_{[A} T_{BC)}{}^{D} - 
T_{[AB|}{}^{E} T_{[E|C)}{}^{D}.
\ee
There are two other equations that follow from the Jacobi identities.
One of them involves the covariant derivative of the curvatures 
$R_{ABC}{}^{D}$
and the other the covariant derivative of the $U(1)$ curvature.
They do not yield any further information, though, since they become 
true identities once (\ref{bianchi}) is satisfied 
(Dragon's theorem \cite{dragon}).

The second set of Bianchi identities is associated with the antisymmetric
tensor. The field strength $H_{LMN}$ is a closed 3-form. Therefore,
\be 
d H_{LMN} = 0. \label{bianchiH1}
\ee
One now writes (\ref{bianchiH1}) in terms of covariant derivatives and 
with all indices flat. The result is a sum of two terms. The first 
ones vanish if the supergravity Bianchi's are satisfied. So, we are
left with the following identity to solve
\be
\del_{[A} H_{BCD)} - T_{[AB|}{}^{E} H_{E|CD)} = 0.
\label{bianchiH2}
\ee
This equation has the same form as in flat superspace.

To solve (\ref{bianchi}) and (\ref{bianchiH2}) we decompose all 
tensors in irreducible 
representations of the Lorentz group and substitute these in (\ref{bianchi})
and (\ref{bianchiH2}).
By further decomposing the left and right hand side of  
(\ref{bianchi}) and (\ref{bianchiH2})
into Lorentz irreducible pieces we get relations among the different 
Lorentz components of the curvatures and the torsions. 
Eventually all tensors are expressed in terms of a few superfields that 
satisfy certain differential relations.

The Lorentz decomposition is achieved by just symmetrizing and 
antisymmetrizing all similar indices.
For example, the curvature $R_{\dda \b \g \d}$ contains
an (3/2,~1/2) and an (1/2, 1/2) irreducible  component\footnote{
Our notation $(a, b)$ refers to the
representation$(a, b)$ of the Lorentz group $SL(2, C)$, where $a$ is the 
spin carried by the undotted indices and $b$ is the spin carried by the 
dotted indices.}.
Explicitely we have
\be
R_{\dda \b \g \d} = r_{\dda, \b \g \d} + C_{\b (\g} r_{\d) \dda} 
\ee
where $r_{\dda, \b \g \d}$ is symmetric in $\b, \g, \d$ 
and represents the (3/2,~1/2) component and 
$r_{\d \dda}$ represents the (1/2, 1/2) piece.
To give a second example, consider the torsion component $T_{\a b c}$.
It contains the 6 different Lorentz components, namely 
$(3/2,0) \oplus (1/2,0) \oplus (1/2,0) 
\oplus (3/2,1) \oplus (1/2,1) \oplus (1/2,1)$. The tree level constraint
(\ref{tree2}), however, eliminates half of them, and we get the following
decomposition 
\be
T_{\a b c} = C_{\ddb \ddg} [K_{\a \b \g} + C_{\a (\b} K_{\g)}]
+ C_{\b \g} K_{\a \ddb \ddg} \label{albc}
\ee 
The Lorentz decomposition of the rest of the tensors is tabulated below
\bea
R_{\a \b \g \d} &=& r_{\a \b \g \d} 
+(C_{\d(\a} r_{\b)\g} + \g \leftrightarrow \d)
+C_{\a(\g} C_{\d)\b}  r \nonumber \\
R_{\dda b \g \d} &=& r_{\dda \ddb, \b \g \d} 
+ C_{\dda \ddb} r_{\b \g \d}+
C_{\b(\g} r_{\d) \dda \ddb}+
C_{\dda \ddb} C_{\b(\g} r_{\d)} \nonumber \\
R_{\a b \g \d} &=& r_{\ddb, \a \b \g \d}
+(C_{\a \b} r_{\g \d \ddb} + {\rm cyclic \ in} \ \b, \g, \d) 
-C_{\b (\g} r'_{\d) \a \ddb} + C_{\a(\g} C_{\d)\b} r_{\ddb} \nonumber \\
R_{a b \g \d} &=& C_{\dda \ddb} [ W_{\a \b \g \d} 
+(C_{\d(\a} W_{\b)\g} + \g \leftrightarrow \d) 
+C_{\a(\g} C_{\d)\b} W] 
+ C_{\a \b} W_{\dda \ddb, \g \d} \nonumber \\
F_{\a b} &=& C_{\a \b} f_{\ddb} + f_{\a \b, \ddb} \nonumber \\
F_{a b} &=& C_{\a \b} f_{\dda \ddb} + C_{\dda \ddb} f_{\a \b} \nonumber \\
T_{a b \g} &=& C_{\a \b} W_{\g,\dda \ddb} + C_{\dda \ddb}
(C_{\g(\a} W_{\b)} + W_{\a \b \g}) \nonumber \\
H_{a b c} &=& C_{\g \a} C_{\ddg \ddb} H_{\dda \b} -
C_{\g \b} C_{\ddg \dda} H_{\ddb \a} 
= \frac{1}{2} (C_{\dda \ddb} C_{\g(\a|} H_{\ddg|\b)} -
C_{\a \b} C_{\ddg (\dda} H_{\ddb) \g})\nonumber
\eea
The Lorentz decomposition of the remaining components can be 
obtained  from the ones listed above by hermitian conjugation,
see (\ref{conjug1}),(\ref{conjug2}).
For example,
\be
R_{\a \ddb \ddg \ddd} = \bar{r}_{\a, \ddb \ddg \ddd} + C_{\ddb (\ddg} 
\bar{r}_{\ddd) \a}. 
\ee

To organize the calculation we start with the Bianchi's of lowest 
dimension (vector indices have dimension 1 and spinor indices dimension 1/2;
then ``dimension = lower indices - upper indices'') 
and work our way up.
The first non-trivial equation appears at dimension 1/2 and it involves 
only torsions. As mentioned in Appendix A the Lorentz group acts 
reducibly in target space. This implies that in the equation
$(\dda, \b, \g, d)$\footnote{By $(\dda, \b, \g, d)$ 
we mean equation (\ref{bianchi})
with $A=\dda, B=\b, C=\g, D=d$.} the right hand side vanishes identically.
Then the equation yields 
\be
T_{(\a, \b)\ddb}^{\ \ \ \ \ \ c} = 0.
\ee
Using (\ref{albc}) we get that the torsion $T_{\a b}^{\ \ \ c}$ vanishes.

Because the curvatures $R_{ABcd}$ are determined from the curvatures 
$R_{AB\g \d}$ and $R_{AB\ddg \ddd}$ it is, in general, advisable to
analyze first the identities $(A, B, \tg, \td)$. We start, however, by
partly analyzing the equation $(\a, \ddb, c, d)$.
The right hand side is equal to $R_{\a \ddb c d}$.
The left hand side is given by
\be
\D_{\a \ddb c d} = 2i T_{\a\ddb cd} + 
i C_{\a \d} C_{\ddb \ddd} (T_{c \e}{}^{\e} +  T_{c \dde}{}^{\dde})
\label{b1}
\ee
To proceed further we need a result from the Bianchi's related to the 
antisymmetric tensor. From the equation $(a, \ddb, \g, \d)$ 
(where now by $(a, \ddb, \g, \d$) we mean
(\ref{bianchiH2}) with $A=a, B=\ddb, C=\g, D=\d$) we get 
\be
T_{\a \ddb, b c} = - 2 H_{\a \ddb, b c} 
+ {1 \over 2} C_{\g \d} C_{\ddg \ddd}
(T_{\a \ddb, \e}{}^{\e} +  T_{\a \ddb, \dde}{}^{\dde})
- {1 \over 2} C_{\a \d} C_{\ddb \ddd} (T_{c \e}{}^{\e} +  T_{c \dde}{}^{\dde})
\label{b2}
\ee
Inserting this into (\ref{b1}) we obtain
\be
\D_{\a \ddb c d} = - 4i H_{\a \ddb, b c}  + i C_{\g \d} C_{\ddg \ddd}
(T_{\a \ddb, \e}{}^{\e} +  T_{\a \ddb, \dde}{}^{\dde})
\ee
However, $\D_{\a \ddb c d}$ is antisymmetric in the indices $c$ and $d$ since
it is equal to $R_{\a, \ddb, c d}$.
Hence,
\be
T_{\a \ddb, \e}{}^{\e} +  T_{\a \ddb, \dde}{}^{\dde}=0,
\ee
which combined with the conventional constraint
\be
T_{\a \ddb, \e}{}^{\e} -  T_{\a \ddb, \dde}{}^{\dde}=0
\ee
yields
\be
T_{\a \ddb, \e}{}^{\e} = T_{\a \ddb, \dde}{}^{\dde}=0.
\ee
So, equations (\ref{b1}) and (\ref{b2}) become 
\bea
&&R_{\a \ddb c d}=- 4i H_{\a \ddb, b c} \nonumber \\
&&T_{a b c} + 2 H_{a b c} = 0. \label{TH}
\eea

We now proceed with the supergravity Bianchi's. We will return to 
the antisymmetric tensor Bianchi's after we solve completely the gravity ones.
In dimension 1
the left hand side of (\ref{bianchi}) vanishes identically (due to
constraints). 
There are 6 (non-trivial) equations at this level. 
Three of those, namely the equations 
$(\dda, \ddb, \g, \d), (\dda, \b, \g, \d),$ and  
$(\a, \b, \g, \d)$, yield 
\bea
&\ & R_{\dda \ddb \g \d} = r_{\dda, \b \g \d} = r_{\a \b \g \d} = 0, 
\nonumber \\
&\ & F_{\dda \ddb} = 0, \ \ F_{\dda \b} = - 2 r_{\b \dda}, \ \ 
F_{\a \b} = 4 r_{\a \b}. \label{lvl1}
\eea
The other three equations are the hermitian conjugates of these ones.
The results are the hermitian conjugates of the equations (\ref{lvl1}).
(One should remember that our $U(1)$ generators are anti-hermitian; this 
yields an extra minus sign in the $U(1)$ sector).

At dimension 3/2 we have 4 equations.
The equation $(\dda, b, \g, \d)$ yields
\bea
&\ & r_{\dda \ddb, \b \g \d} = 0, \ \ r_{\a \ddb \ddg} = i W_{\a \ddb \ddg},
\ \ r_{\a} = i W_{\a}, \ \ r_{\a \b \g} = -2i W_{\a \b \g}, \nonumber \\
&\ & \bar{f}_{\a} = 6i W_{\a}, \ \ 
\bar{f}_{\dda \ddb, \g} = -2i W_{\g \dda \ddb}
\eea
Next we consider the equation $(\a, b, \g, \d)$. It yields
\be
r_{\ddb, \a \b \g \d} = 0, \ \  
r'_{\a \b \ddg} = - \frac{1}{2} r_{\a \b \ddg} 
= \frac{1}{6} f_{\a \b, \ddg}, \ \
f_{\ddb} = - 2 r_{\ddb} 
\ee
Combining these results with their hermitian conjugates we get
\be
r_{\dda} = 3i \bar{W}_{\dda}, \ \ 
r'_{\a \b \ddg}= \frac{i}{3} \bar{W}_{\ddg \a \b}.
\ee
 
At the dimension 2, from the equation $(a, b, \g, \d)$ we get
\bea
&\ &W_{\a \b \g \d} = \frac{1}{24} \del_{(\a} W_{\b \g \d)}, \ 
W_{\a \b} = \frac{1}{4} \del_{(\a} W_{\b)} 
- \frac{1}{8} \del^{\d} W_{\d \a \b}, \nonumber \\
 &\ &W=\frac{1}{2} \del^{\d} W_{\d}, \ \
W_{\dda \ddb, \g \d} = \frac{1}{2} \del_{(\g} W_{\d), \dda \ddb}, \nonumber \\
&\ &f_{\a \b} = - \del^{\d} W_{\d \a \b} - \del_{(\a} W_{\b)}, \ \
f_{\dda \ddb} = - \del^{\d} W_{\d \dda \ddb}. \label{dim2}
\eea
The hermitian conjugate of this equations implies that
$f_{\a \b} = - \overline{(f_{\dda \ddb})}$, where the bar indicates that dotted
and undotted indices should be interchanged. Explicitely, we get
\be
\del^{\ddd} \bar{W}_{\ddd \a \b} + \del^{\d} W_{\d \a \b} 
+ \del_{(\a} W_{\b)}=0. \label{rel}
\ee

We now move to the case where the equations involve curvatures with 
two bosonic indices as the last two indices. There is no $U(1)$ contribution
in this sector because vector indices do not carry $U(1)$ charge.
The results are tabulated below 
\be
\begin{array}{ll}
(\a, \b, c, d): & r_{\a \b} = r = 0, \\
(\a, \ddb, c, d): & r_{\a \ddb} = - \bar{r}_{\ddb \a} = -2i H_{\ddb \a} \\
(\a, b, c, d): & \bar{W}_{\dda} = \frac{i}{6} \del^{\b} H_{\dda \b},
\ \bar{W}_{\dda \b \g} = - \frac{i}{2} \del_{(\b|} H_{\dda|\g)} \\
(a ,b, c, d): & 
\bar{W}_{\dda \ddb} = - \frac{1}{8} \del_{\g(\dda} H_{\ddb)}^{\ \ \ \g}, \  
W_{\a \b} = \frac{1}{8} \del_{(\a| \ddg} H^{\ddg}_{\ \ |\b)}, \
\bar{W} - W = - \frac{3}{2} \del_a H^a, \\ 
\ &\bar{W}_{\a \b, \ddg \ddd} - W_{\ddg \ddd, \a \b} = 
- \frac{1}{2} (\del_{\a(\ddg} H_{\ddd)\b} + \a \leftrightarrow \b)
\end{array}
\label{tableII}
\ee
In deriving the results in (\ref{tableII}) we have used (\ref{TH}).

Next we need to check the consistency between the results in (\ref{dim2})
and the ones in (\ref{tableII}). This yields
\be
W = \bar{W} = \frac{i}{12} \del^{\dda} \del^{\a} H_{\dda \a}, \      
\del_a H^a = 0.
\ee

All curvatures and torsions can now be expressed in terms of the superfields
$H_{\dda \b}$, $W_{\a \b \g}$ and $\bar{W}_{\dda \ddb \ddg}$.
The results are gathered below.
\bea
R_{\dda \b \g \d} &=& 2i H_{\dda (\g} C_{\d) \b} \nonumber \\
R_{\dda b \g \d} &=& -2i C_{\dda \ddb} W_{\b \g \d} 
+ [C_{\b \g} (-\frac{1}{2} \del_{(\dda} H_{\ddb) \d} + 
\frac{1}{6} C_{\dda \ddb} \del^{\dde} H_{\dde \d}) +  
\g \leftrightarrow \d] \nonumber \\
R_{\a b \g \d} &=& 
-\frac{1}{3} C_{\a \b} \del_{(\g|} H_{\ddb| \d)} \nonumber \\ 
&\ &-[C_{\a \g}(\frac{1}{3} \del_{(\d|} H_{\ddb|\b)} +  
\frac{1}{2} C_{\d \b} \del^{\e} H_{\ddb \e})
+ \frac{1}{6} C_{\b \g} \del_{(\d|} H_{\ddb|\a)})
+\g \leftrightarrow \d] \nonumber \\
&=& \del_{\b} H_{\ddb (\g} C_{\d) \a} \nonumber \\
R_{a b \g \d} &=&  \frac{1}{24} C_{\dda \ddb} \del_{(\a} W_{\b \g \d)}
\nonumber \\
&\ &+ [C_{\dda \ddb} (\frac{1}{8}
(C_{\d \a} \del_{(\b|\dde} H^{\dde}_{\ |\g)} + \a \leftrightarrow \b)
+ \frac{i}{12} C_{\a \g} C_{\d \b} \del^{\dde} \del^{\e} H_{\dde \e})
\nonumber \\
&\ &
+\frac{i}{4} C_{\a \b} \del_{\g} \del_{(\dda} H_{\ddb)\d} 
+ \g \leftrightarrow \d]
\nonumber \\
F_{\a b} &=& C_{\a \b} \del^{\g} H_{\ddb \g} + \del_{(\a|} H_{\ddb| \b)}
= 2 \del_{\b} H_{\ddb \a}
\nonumber \\
F_{a b} &=& -\frac{i}{2} [C_{\a \b} \del^{\d} \del_{(\dda} H_{\ddb)\d}
+C_{\dda \ddb} \del^{\ddd} \del_{(\a|} H_{\ddd| \b)}] \nonumber \\
F_{\a \ddb} &=& 4i H_{\ddb \a},  \nonumber \\
T_{a b \g} &=& \frac{i}{2} C_{\a \b} \del_{(\dda} H_{\ddb)\g}
+C_{\dda \ddb}[-\frac{i}{6} 
C_{\g (\a|} \del^{\dde} H_{\dde| \b)} + W_{\a \b \g}]
\label{table3}
\eea  
Furthermore, equation (\ref{rel}) becomes
\be 
\del^{\g} W_{\g \a \b} = \frac{i}{6} \del_{(\a|} \del^{\ddg} H_{\ddg| \b)}
+\frac{i}{2} \del^{\ddg} \del_{(\a|}  H_{\ddg| \b)}.
\ee

The last non-trivial equation is the equation $(a, b, \g, \ddd)$. It yields
\be
\del_{\ddd} T_{a b \g} = 0.
\ee
This, in turn, implies that $W_{\a \b \g}$ is a chiral superfield  and
$H_a$ is a linear superfield
\be
\del_{\ddd} W_{\a \b \g} = 0; \ \ 
\del^{\b} \del_{\b} H_a = 0.
\ee
 
The resulting supergravity algebra is given in (\ref{bian1})-(\ref{bian4}).
This supergravity algebra is actually invariant under local superscale 
and local $U(1)$ transformations as we now describe. 
To discover these transformations we start from the 
transformation rule of the vierbein
\be
\d E_\a{}^M = {1 \over 2} L E_\a{}^M,
\ee 
where $L$ is a complex parameter. If $L$ is real we are dealing with 
local scale transormations, whereas if it is imaginary with $U(1)$
transformations. The tranformation rules of the connections 
are determined by requiring invariance of the supergravity algebra
under local superscale and/or local $U(1)$ transformations.
The manipulations involved are very similar to the ones we 
performed in section 5.
The resulting transformation rules are given below
\bea
&&\d \del_\a = {1 \over 2} L \del_\a + \del_\b (L + \bar{L}) M_\a{}^\b
-(2 \del_\a L + \del_\a \bar{L}) Y \nonu 
&&\d \del_{\dda} = {1 \over 2} \bar{L} \del_{\dda}
+ \del_{\ddb} (L + \bar{L}) M_{\dda}{}^{\ddb}
+(2 \del_{\dda} \bar{L} + \del_{\dda} L) Y \nonu 
&&\d \del_a = {1 \over 2} (L + \bar{L}) \del_a
+{i \over 2} \del_{\dda} (L + \bar{L}) \del_\a 
+{i \over 2} \del_{\a} (L + \bar{L}) \del_{\dda} \nonu
&&+{i \over 2} \del_{\b} \del_{\dda} (L + \bar{L}) M_\a{}^{\b}
+{i \over 2} \del_{\dda} \del_{\b} (L + \bar{L}) M_\a{}^{\b}
+{i \over 2} (\del_\a \del_{\dda} \bar{L} - \del_{\dda} \del_\a L) Y 
\eea
In addition
\bea
&&\d H_a = {1 \over 2} (L+\bar{L}) H_a
- {i \over 4} [\del_\a, \del_{\dda}](L+ \bar{L})  \nonu
&&\d W_{\a \b \g} = {1 \over 2} (2 \bar{L} + L) W_{\a \b \g}.
\eea
The parameter $L$ satisfies the equation
\be
\del^2 (L+\bar{L}) =0.
\ee
This means $L$ is an imaginary unconstrained superfield
for $U(1)$ transformations and a real linear superfield for scale 
transformations.
From these transformation rules we can read off the relation between the 
conformal weight $d$ and the $U(1)$ charge $w$ of a chiral superfield,
using the fact that 
scale transformations should respect the chirality condition,
$\d (\del_{\dda} \f) = 0$.
This yields 
\be
d + 3 w =0.
\ee 
 
We now return to the antisymmetric tensor Bianchi's.
Having already solved the supergravity Bianchi's it is rather easy to 
solve (\ref{bianchiH2}). Most of the components of $H_{ABC}$ have already 
been determined by the tree-level constraints 
\be
H_{a \tb \tg} = H_{a b \tg} = 0; \ \ 
H_{\a \ddb c} = - i C_{\a \g} C_{\ddb \ddg}.
\ee
Equations  $(\a, \b, \g, \d)$,
$(\dda, \b, \g, \d)$, $(\dda, \ddb, \g, \d)$ and 
$(a, b, \g, \d)$ yield
\bea
&\ & \del_{\a} H_{\b \g \d} + {\rm cyclic \ in} \ \a, \b, \g, \d  = 0, \\
&\ &\del_{\dda} H_{\b \g \d}
+\del_{\b} H_{\dda \g \d} +
\del_{\g} H_{\dda \b \d} + \del_{\d} H_{\dda \b \g} =0, \\
&\ &\del_{\dda} H_{\ddb \g \d}
+\del_{\ddb} H_{\dda \g \d} +
\del_{\g} H_{\dda \ddb \d} + \del_{\d} H_{\dda \ddb \g} =0, \\
&\ & T_{a b}^{\ \ \ \te} H_{\te \g \d} = 0. 
\eea
We conclude that $H_{\a \b \g} = H_{\dda \b \g} = H_{\dda \ddb \g} = 0$.

\end{document}